\documentclass[prb,twocolumn,showpacs,amsmath,amssymb,floatfix,superscriptaddress]{revtex4}
\usepackage{color,graphics,epsfig,rotating}
\usepackage{amsmath,bm}
\usepackage{verbatim}
\usepackage{graphicx,pdflscape}%
\usepackage{color}

\usepackage{hyperref}
\usepackage{color}
\newcommand{\hh}{ {\alpha_0}}
\newcommand{\beq}{\begin{equation}}
\newcommand{\eeq}{\end{equation}}
\newcommand{\barray}{\begin{eqnarray}}
\newcommand{\earray}{\end{eqnarray}}
\newcommand{\disp}[1]{Eq.~(\ref{#1})}

\newcommand{\figdisp}[1]{Fig.~\ref{#1}}

\newcommand{\lsim}{\raisebox{-4pt}{$\,\stackrel{\textstyle <}{\sim}\,$}}

\newcommand{\tJ}{\ $t$-$J$ \ }
\newcommand{\nn}{\nonumber}

\renewcommand{\Re}{\mathrm{Re}}
\renewcommand{\Im}{\mathrm{Im}}

\renewcommand{\emph}{\textit}

\newcommand{\G}{\mathcal{G}}

\newcommand{\GH}{{\bf g}}
\newcommand{\GHI}{\GH^{-1}}

\newcommand{\chem}{ {\mu}}

\newcommand{\sw}{a_\G}
\newcommand{\swo}{1- \frac{n}{2}}

\newcommand{\vk}{\vec{k}}

\usepackage[svgnames]{xcolor}

\newif\ifshowcomments\showcommentstrue

\begin{document}

\title{
Extremely correlated Fermi liquid theory meets Dynamical mean-field theory:\\
Analytical insights into the doping-driven Mott transition}

\author{ R. \v{Z}itko}
\affiliation{Jo\v{z}ef Stefan Institute, Jamova 39, SI-1000 Ljubljana, Slovenia}
\affiliation{Faculty for Mathematics and Physics, University of
Ljubljana, Jadranska 19, SI-1000 Ljubljana, Slovenia}

\author{ D. Hansen}
\affiliation{Physics Department, University of California Santa Cruz, CA 95064, USA}

\author{ E. Perepelitsky}
\affiliation{Physics Department, University of California Santa Cruz, CA 95064, USA}

\author{ J. Mravlje}
\affiliation{Jo\v{z}ef Stefan Institute, Jamova 39, SI-1000 Ljubljana, Slovenia}

\author{A. Georges}
\affiliation{Centre de Physique Th\'eorique, \'Ecole Polytechnique, CNRS, 91128 Palaiseau Cedex, France}
\affiliation{Coll\`ege de France, 11 place Marcelin Berthelot, 75005 Paris, France}
\affiliation{DPMC-MaNEP, Universit\'e de Gen\`eve, CH-1211 Gen\`eve, Switzerland}

\author{  B. S. Shastry}
\affiliation{Physics Department, University of California Santa Cruz, CA 95064, USA}

\date{\today}

\begin{abstract}
We consider a doped Mott insulator in the large dimensionality limit within both 
the recently developed Extremely Correlated Fermi Liquid (ECFL) theory and the Dynamical Mean-Field Theory (DMFT). 
We show that the general structure of the ECFL sheds light on the rich frequency-dependence 
of the DMFT self-energy. Using the leading Fermi-liquid form of the two key auxiliary 
functions introduced in the ECFL theory, we obtain an analytical  ansatz which provides a good quantitative 
description of the DMFT self-energy down to hole doping level $\delta \simeq 0.2$. In
particular, the deviation from 
Fermi-liquid behavior and the corresponding particle-hole asymmetry
developing at a low energy scale 
are well reproduced by this ansatz. 
The DMFT being exact at large dimensionality, our study also provides
a benchmark of the ECFL in this limit. 
We find that the main features of the self-energy and spectral line-shape are 
well reproduced by the ECFL calculations in the $O(\lambda^2)$ `minimal scheme', 
for not too low doping level $\delta\gtrsim 0.3$. 
The DMFT calculations reported here are performed using a state-of-the-art 
numerical renormalization-group impurity solver, which yields accurate 
results down to an unprecedentedly small doping level $\delta\lesssim 0.001$. 
\end{abstract}

\pacs{71.10.Ay, 71.10.Fd, 71.30.+h}

\maketitle
\section{Introduction}

Strong electronic correlations constitute one of the major challenges
in condensed-matter physics and continue to inspire new theoretical
approaches.  In search for novel functionalities, new materials are
being synthesized on a regular basis, giving the field a steady impetus. 
Significant progress in the understanding of electronic correlations has been achieved 
from the Dynamical Mean-Field Theory (DMFT), in which the self-energy 
is assumed to be momentum-independent (see Ref.~\onlinecite{georges_review_dmft} for a review). 
This theory becomes exact in the limit of infinite dimensionality. 

The situation in low dimensions has further  challenges relating to the k dependence of the self energy, and thus new methods for strongly correlated electrons continue to be
developed. One promising approach is Shastry's Extremely Correlated
Fermi Liquid theory (ECFL), developed in a recent series of papers
\cite{ECFL, Monster,Anatomy, Khatami-Shastry}. This theory starts from
the infinite-$U$ limit and is based on the Schwinger equation of
motion for  Gutzwiller projected  electrons, these non-canonical objects requiring special attention. The theory  leads to
a set of analytical expressions that are in principle exact. So far, solutions of  the second order 
expansion of these expressions in a partial projection parameter $\lambda$ are available. They can be
obtained for any lattice by an iterative process analogous to the
skeleton diagram method. The ECFL theory expressions have been
successfully applied to account for the Angle Resolved Photo-Emission Spectroscopy (ARPES) 
line-shapes of cuprate superconductors \cite{Gweon-Shastry, Kazue} in the normal state.

In this work, we perform a comparative study of these two methods. 
We use as a test-bed the single-band doped Hubbard model at strong coupling $U$,
in the limit of large dimensionality. This limit leads to 
simplifications in the ECFL theory which we introduce here (the details of the 
formalism are provided elsewhere\cite{edward-shastry}). 
The comparison focuses on the frequency-dependence of the self-energy and
single-particle spectral line-shapes, 
and their evolution as the Mott 
insulator is approached by reducing the doping level $\delta$ (defined in \disp{deltadef}).   

The first outcome of the present work is that, by looking at the DMFT results 
within an ECFL perspective, we are able to obtain new analytical insights into 
the DMFT description of the doping-driven Mott transition. 
Within the DMFT, the single-particle self-energy $\Sigma(\omega)$ %
displays a rich and complex frequency dependence. 
This has been known for some time (see e.g. Ref.~\onlinecite{deng_2013} for a recent study), but is 
further investigated in the present work down to unprecedentedly low doping levels $\delta\lesssim 0.001$ 
using a state-of-the-art numerical renormalization group (NRG) solution of the DMFT equations. 
Local Fermi-liquid behavior $\mathrm{Im}\Sigma \propto \omega^2 + (\pi T)^2$ is %
obeyed only below a very low energy-scale. Above this energy scale, a marked particle-hole asymmetry 
develops, a feature which is beyond the Fermi liquid theory. 
Furthermore, the strong suppression of spectral weight in the intermediate range of energies separating the 
quasiparticle peak from the lower Hubbard band corresponds to a marked quasi-pole in the 
self-energy. 

We show that all of these features can be well reproduced by constructing an analytical  ansatz for the 
one-particle self-energy which is directly motivated by the ECFL construction. 
The ECFL introduces two key quantities, $\Psi$ and $\chi$, which play the role of auxiliary self-energies in the Schwinger 
construction. The proposed analytical ansatz is obtained by retaining only the dominant Fermi-liquid 
terms in the low-frequency expansion of these auxiliary quantities. This is found to provide a satisfactory 
fit of the DMFT results for doping levels $\delta\gtrsim 0.2$. 
Hence, quite remarkably, the marked deviations from Fermi liquid behavior, and
the particle-hole asymmetry %
in the physical single-particle self-energy, can be accounted for by an
underlying Fermi-liquid form of the ECFL auxiliary quantities.
For very large $U$, and especially for very small doping levels, additional
structures appear in the DMFT results which are not present in this simplest ECFL ansatz, and presumably 
require additional terms beyond the Fermi liquid ones in the auxiliary functions $\Psi$ and $\chi$.   

Another synergistic outcome of our study is that, because the DMFT provides an exact solution 
in the limit of large dimensionality, it can be used to benchmark the ECFL in this limit. 
We present here quantitative results obtained within the `minimal scheme' implementation of 
the ECFL in high dimensions \cite{edward-shastry}, giving rise to an expansion to order $\lambda^2$ in the projection parameter $\lambda$. 
We find that the main features of the self-energy and the spectral line-shape are 
well reproduced by the $O(\lambda^2)$ ECFL calculations, on a semi-quantitative level, 
for not too low doping $\delta\gtrsim 0.3$. Improvement will require further developments of the ECFL approach. 
Since the DMFT is able to handle any finite $U$, while the ECFL construction is motivated by the very large $U$ limit, 
this comparison also sheds light on the adiabatic connection between the regime of moderate and extreme 
correlations. 

We emphasize that ECFL can be used on two different
levels. On one level, it provides a functional form for the physical
Green's function and the corresponding self-energy in terms of the
auxiliary ECFL self-energies $\Psi(\omega)$ and $\chi(\omega)$. By
assuming the simplest Fermi-liquid form for these two self-energies
over a certain frequency range centered around $\omega=0$, we
successfully fit the physical self-energy obtained through DMFT in
this frequency range for $\delta\gtrsim 0.2$. This is remarkable since
the frequency range used is substantially larger than the
characteristic frequency at which the physical self-energy begins to
deviate from Fermi-liquid behavior, and even encompasses the
quasi-pole in the physical self-energy at negative frequencies. This
phenomenological approach to ECFL is the one used in the first five
sections of the paper, and the results of this fit are displayed in
Figs.~\ref{fig:effects}, \ref{fig:re_sigma1}, and \ref{fig:param}. On
the second level, ECFL provides a microscopic theory by which one can
obtain concrete results for $\Psi$ and $\chi$ via an expansion in the
projection parameter $\lambda$. In the remainder of the paper, the
results obtained from the $O(\lambda^2)$ theory are benchmarked
against the results obtained from DMFT, which are exact in the limit
of infinite dimensions. In the long run, further combined use of the
ECFL and the DMFT approaches could lead to a better understanding of
the momentum-dependence of the self-energy that becomes important in
lower dimensions.

The paper is organized as follows. 
After defining the model in Sec.~\ref{sec:model},  
the general structure of the ECFL formalism is reviewed in Sec.~\ref{sec:ecfl_framework}. 
In Sec.~\ref{sec:dmft}, we present detailed DMFT results for the
hole-doped Hubbard model using high-precision Wilson's NRG as a solver. 
In Sec.~\ref{sec:dmft_ecfl}, the DMFT self energies are interpreted in light of the ECFL-motivated 
analytical expressions. The second part of the paper is devoted to the $O(\lambda^2)$ ECFL minimal 
implementation. The basic equations and their simplification in infinite dimensions are established
in Sec.~\ref{sec:ecfl_lambda}, and in Sec.~\ref{sec:lambda_dmft}, a quantitative comparison is made to the DMFT results. 

\section{Model}
\label{sec:model}
We study the Hubbard model defined by the Hamiltonian
\begin{equation}
H=\sum_{k\sigma} \varepsilon_k c^\dag_{k\sigma} c_{k\sigma}
+U \sum_i n_{i\uparrow} n_{i\downarrow},
\end{equation}
where $\varepsilon_k$ is the bare band dispersion relation obtained by Fourier transforming the hopping matrix. 
In this study we consider a doped Hubbard model with nearest-neighbor hopping on a Bethe lattice,  with
semicircular density of states:
\beq
\rho_0(\varepsilon) = \frac{2}{\pi D^2} \sqrt{D^2-\varepsilon^2},
\eeq
where $D$ is the half-bandwidth, and thus any sum over the band energy can be converted to an integral as:
\barray
\frac{1}{N} \sum_k {\cal A}(\varepsilon_k)&  \to &  \int_{-D}^D  \ d\varepsilon \  \rho_{0}(\varepsilon) {\cal A}(\varepsilon). 
\earray
We note that the Fermi energy $\varepsilon_F$ satisfies 
\beq
\sin^{-1}\left(\frac{\varepsilon_F}{D}\right) + \left(\frac{\varepsilon_F}{D}\right) 
\sqrt{ 1-\left(\frac{\varepsilon_F}{D}\right)^2} = - \frac{\pi}{2} (1-n), \label{bare-ef}
\eeq
and vanishes near $n\sim1$ as $\varepsilon_F = - \frac{\pi}{4} (1-n) \ D$.
The hole  doping level $\delta$ is related to the particle density $n$ ($n = N/N_{sites}$) as:
\beq
\delta = 1-n. \label{deltadef}
\eeq
We will use $\rho_Q(\omega)$ as a short-hand notation for the
spectral function associated with any relevant  quantity $Q(i\omega_n)$ (Green's function,
self-energy):
\beq
\rho_Q(\omega) = -\frac{1}{\pi} \Im\left( Q(\omega+ i 0^+) \right).
\eeq

\section{ECFL: General framework}
\label{sec:ecfl_framework}

\subsection{ECFL formalism}

The ECFL methodology has been discussed extensively in recent
literature \cite{ECFL,Monster}; here we highlight only the aspects that
are of relevance to this work. ECFL deals with
Gutzwiller-projected states obtained in the limit of $U \to \infty$,
with the no-double-occupancy constraint built into the electron
operators, leading to the well known \tJ model. This results in a
non-canonical theory, where familiar Feynman diagram methods fail due
to the absence of Wick's theorem. The ECFL formalism is an exact
alternative to the Feynman diagram technique. Instead it works  with
the Schwinger equations of motion for the projected electrons. It
provides results for the electronic Green's functions that describe
the physics of the low-energy sector in the problem, namely the dynamics of 
the quasiparticle (QP) states near the Fermi energy and of the lower Hubbard band (LHB).  

For our purposes, we need to express the ECFL theory in the
large-dimensionality limit.  The related technical problems outlined in
Ref.~\onlinecite{shastry-prl-reply} (paragraph 3) have been recently solved
\cite{edward-shastry} by analyzing the infinite-dimensional limit of
the Schwinger equations of motion in the ECFL\cite{matho}.   
The exact mapping of the momentum-independent self-energy of the infinite-dimensional Hubbard model
onto that of a self-consistent Anderson impurity model\cite{georges_kotliar_dmft} 
provides  a roadmap  for  a suitable formulation  of the ECFL equations in this limit.     

 In the simplest version of the ECFL theory \cite{minimal}, the
 physical (i.e. projected) electronic Green's function is expressed
 as a product of an auxiliary Green's function $\GH(k)$ and a
 caparison term denoted in the present work as $\widetilde{\mu}(k)$. Thus, 
 \beq \G(k)=
 \widetilde{\mu}(k)\times \GH(k) \label{eq7}, \eeq 
 where $k \equiv
 (\vec{k}, i \omega_k)$ and $\omega_k$ is a Fermionic Matsubara
 frequency.  Here $\GH(k)$ is a Fermi-liquid-like Green's function
 \beq \GHI(k)=i\omega_k+\chem- \left( \swo \right) \ \varepsilon_k-
 \Phi(k), \label{eq8} \eeq and $\mu$ is the chemical potential.  The
 factor $\widetilde{\mu}(k)$ (here distinguished from the chemical potential $\mu$ by the tilde), plays the
 role of an adaptive spectral weight, and is given by \beq
 \widetilde{\mu}(k)=\swo+ \Psi(k), \label{eq9} \eeq with $\Psi(k)$
 vanishing at infinite frequency. The functions $\Phi(k)$ and
 $\Psi(k)$ are the twin self-energies in the theory, and are exactly
 defined as the appropriate functional derivatives of $\GHI$ and $\widetilde{\mu}$ respectively \cite{ECFL, Monster}. The term $\widetilde{\mu}(k)$ is
 termed the caparison (i.e., dressing) factor, since it provides a
 second layer of renormalization to the propagator $\GH$, which is
 already dressed by $\Phi$.  Both Green's functions satisfy an
 identical number sum rule $\sum_{k, \omega_k} \G(k) = \frac{n}{2} =
 \sum_{k, \omega_k} \GH(k) $; this enables us to satisfy the
 Luttinger-Ward volume theorem.

In  the large-$d$ limit, a further simplification can be established\cite{edward-shastry}: 
$\Psi$ is independent of $\vec{k}$ and $\Phi$ is decomposable into two $\vec{k}$ independent functions, 
\barray
\Psi(k)&=& \Psi(i \omega_k),  \\
\Phi(k)&=& \chi(i \omega_k) + \varepsilon_k \Psi(i \omega_k),
\earray
i.e., the two frequency-dependent  (but $\vec{k}$ independent) functions $\chi$ and $\Psi$ determine the Green's function.  

The single-electron physical (Dyson) self-energy $\Sigma$ is defined from the single-electron Green's function $\G$ 
in the usual manner, as (using the analytic continuation $i \omega_k \to \omega + i \eta$, $\eta=0^+$):
\beq
\G(k,\omega+ i \eta)=\frac{1}{\omega+ i \eta + \mu - \varepsilon_k - \Sigma(\omega+ i \eta)}.
\eeq
Within the large-dimensional ECFL, the Dyson self-energy $\Sigma$ can be related to $\Psi$ and $\chi$ as follows: 
\beq
\Sigma(\omega+ i \eta) - \chem -\omega= \frac{\chi(\omega+ i \eta) - \chem -\omega}{1-\frac{n}{2}+ \Psi(\omega+ i \eta)}. 
\label{dyson}
\eeq
We see that the Dyson self-energy is manifestly momentum-independent in this limit. 
Note also that, as seen from (\ref{dyson}), its real part grows linearly  with $\omega$ as $\omega \to \infty$. 
This is a consequence\cite{Dyson-Mori} of the Gutzwiller projection in the $U\rightarrow\infty$ limit. At finite $U$, this behavior is regularized at high-enough frequency and $\Sigma$ goes to a constant.   

For a concrete implementation, the ECFL formalism allows for a
perturbative expansion in a projection parameter $\lambda \in [0,1]$,
ultimately identified with the   double occupancy density
\cite{Monster}. The theory to  $O(\lambda^2)$  is expected to be quantitatively accurate for densities  up to $n \lsim 0.7$ \cite{Hansen-Shastry}.  We postpone the description of these equations to Section \ref{second-order}, but note an  important general  insight gained from examining and evaluating such an expansion \cite{ECFL, Monster,Hansen-Shastry}; the two self-energies $\chi$ and $\Psi$   have  simple Fermi-liquid functional forms,  with a  dissipative part  that is quadratic in $\omega$,  at sufficiently low energies (see Figs.~\ref{fig:self_all} in Sec.~\ref{sec:ecfl_lambda}). 
This insight is used in the following to obtain a low-energy expansion for the Green's function.

\subsection{Low-frequency expansion of self-energies and Green's functions} 

In this section we derive the low-frequency behavior of the Green's
function and self-energy within the ECFL. We obtain an analytical expression which will 
be used to interpret and fit the DMFT results in Sec.~\ref{sec:dmft_ecfl}. 
We show, in particular how a characteristic particle-hole asymmetry in the Dyson
self-energy is generated even when the expansion of $\Im \chi$ and $\Im \Psi$ is limited 
to the particle-hole symmetric lowest-order Fermi-liquid  terms. 

Indeed, as mentioned above, the first few terms of a systematic $\lambda$
 expansion of the ECFL equations indicate that the
self-energies $\Psi$ and $\chi$ are very similar functions and
resemble the self-energy of a Fermi liquid at low enough $T,\omega$,
 with suitable  scale constants. 
For low $\omega$ and low $T$, up to a  low-frequency  cutoff scale $\Omega_c$, so that   $|\omega| \leq \Omega_c \ll D$,  we define
(with $k_B=1$)
\beq
{\cal R}(\omega,T)\equiv\pi \left[ \omega^2+ (\pi T)^2 \right], \label{rdef}
\eeq
and write a Fermi-liquid-like  expansion for the  complex ECFL self-energies:
\newcommand{\OmPsi}{\Gamma_\Psi}
 \barray
\Psi(\omega)&\sim& 
\Psi_0  + c_\Psi \ \omega +  \frac{ i}{\OmPsi} \ {\cal R}(\omega,T) +
\Psi_{\mathrm{rem}}(\omega), \label{flpsi-2}\\
\chi(\omega)&\sim& \chi_0 - c_{\chi} \  \omega  - \frac{i
}{\Omega_\chi}  \ {\cal R}(\omega,T)+ \chi_{\mathrm{rem}}(\omega), \label{flpsi-1} 
\earray
where
\barray
c_\Psi&=&  2 \frac{\Omega_c}{\OmPsi}
\quad \text{and} \quad 
c_\chi =  2 \frac{\Omega_c}{\Omega_\chi}.
\label{flpsi}
\earray
$\Omega_\chi$  and $\Gamma_\Psi$    are  parameters  that determine the curvatures of the two imaginary parts, with  
$\Omega_\chi$ having the  dimensions of energy, while $\Gamma_\Psi$  has the dimensions of  the square of energy.
Consequently, $c_\Psi$ has the dimensions of   an inverse energy, while  $c_\chi$ is dimensionless.
The  terms $\Psi_{\mathrm{rem}}(\omega)$ and
$\chi_{\mathrm{rem}}(\omega)$ in
\disp{flpsi-2} represent the remainders containing the leading
corrections to the Fermi-liquid behavior, 
of the type $O(\omega^2)$ for the real part and  $O( \omega^3)$ for the imaginary parts of these functions. 
In the initial analysis below, we simply ignore these terms. 
They can be readily incorporated to find a systematic improvement of the  fits, and lead to further corrections to the low-frequency behavior 
of the imaginary part of the  Dyson self-energy beyond the terms considered here. 
Note that in \disp{flpsi} the constants  $c_\Psi$ and $c_\chi$ also  receive  contributions  from higher terms beyond the quadratic. Hence these approximate relations become  exact  if we retain the imaginary terms only to quadratic order, i.e., assuming  $\rho_\Psi \equiv  - \frac{{\cal R}(\omega, T)}{\pi \OmPsi }$ and  
$\rho_\chi \equiv  \frac{{\cal R}(\omega, T)}{\pi \Omega_\chi }$.
In general, however, $c_\Psi$ and $c_\chi$ can be considered as additional free parameters.
 
Some further remarks about this expansion are called for: 
\begin{enumerate}
\item Expressions \eqref{flpsi-2} and \eqref{flpsi-1} are of the standard Fermi-liquid type (symmetric in $\omega$ for the imaginary parts). Nonetheless,  
when processed  through the ECFL formalism, they lead to important contributions to the imaginary part of the Dysonian self-energy which are {\em anti-symmetric in $\omega$}. 
Revealing the origin of this important asymmetry is one of the main strengths of the ECFL analysis.   

\item We shall find that as we approach half filling, $\Omega_\chi$ and $\OmPsi$ turn out to be similar functions of the electron density, in view of their 
parallel role in the two self-energies within the $\lambda$ expansion.
In the analysis below,
we will find that as $n \to 1$, it is consistent to choose
$\Omega_\chi,\OmPsi \sim \delta$.

\item The energy scale $\Omega_c$, which determines the range of
frequencies where the quadratic behavior of $\Im \ \Psi$ and $\Im \
\chi$ applies, is itself a function of the
density.  It shrinks linearly with $\delta=1-n$ as $n \to 1$, and
therefore $c_\Psi$ and $c_\chi$ are finite as $n \to 1$.  We should
note that these are leading-order assumptions in $\delta=1-n$. 

\end{enumerate} 
Thus we find at low $T, \omega$:
\barray
&&\G(k,\omega+ i \eta)  \\
&&\sim \frac{\hh+c_\Psi \ \omega + \frac{i }{\OmPsi} {\cal R}}
{\omega (1 + c_\chi) + \mu + \frac{i }{\Omega_\chi} {\cal R }  - \varepsilon_k \{\hh+c_\Psi \ \omega + \frac{i }{\OmPsi} {\cal R } \} }, \nn
\earray
where we have introduced
\beq 
\hh \equiv 1-\frac{n}{2} + \Psi_0,
\eeq
and $\chi_0$ has been absorbed into the chemical potential $\chem$.
The entire momentum-dependence is contained in $\varepsilon_k$.  At
$T=0$ and $\omega=0$ we must require $\G^{-1}(k_F,0)=0$, so we need to set
\begin{equation}
\chem=\hh\ \varepsilon_F.
\end{equation}

At  low $\omega+ i \eta$ and a fixed $\vec{k}$, we 
 can write a useful  expression 
\barray
&&\G^{-1}\sim -\varepsilon_k + \frac{\omega (1 + c_\chi) +\hh\ \varepsilon_F+ \frac{i }{\Omega_\chi} {\cal R } }{\hh+c_\Psi \ \omega + \frac{i }{\OmPsi} {\cal R}}
\earray
and therefore
\beq
\Sigma(\omega+ i \eta)\sim\hh\ \varepsilon_F + \omega - \frac{\omega (1 + c_\chi) +\hh\ \varepsilon_F+ \frac{i }{\Omega_\chi} {\cal R } }{\hh+c_\Psi \ \omega + \frac{i }{\OmPsi} {\cal R}}.
\eeq
Note that we adjusted  the self-energy so that $\Re \ \Sigma(0) =
\chem - \varepsilon_F$, thereby placing the zero-energy pole at the Fermi momentum.
 We now extract the wave function renormalization  factor $Z$  from
 $Z^{-1}= \frac{\partial}{\partial \omega}\G^{-1}(k,\omega)|_{\omega=0}$ as
\beq Z=  \frac{\hh}{1+c_\chi- \varepsilon_F \ c_\Psi}. \label{QPweight}\eeq  

\begin{widetext}
Using the above expansion we find the spectral function $
\rho_\G(k,\omega)$ (or equivalently $A(k,\omega)$, as denoted in the
experimental literature), at low $\omega$ and $k \sim k_F$
\barray
A(k,\omega)  &\sim& \left( \frac{\hh^2}{\pi \Omega_\Sigma} \right) 
 \frac{  {\cal R}\ \left( 1- \frac{\omega}{\Delta} \right) }{  \{(1+c_\chi - c_\Psi \ \varepsilon_k) \omega -\hh(\varepsilon_k - \varepsilon_F) \}^2 +  \{ \hh^2 {{\cal R}^2}/{\Omega_\Sigma^2} \} },  \label{eq11}
 \earray
\end{widetext}
 where
 \barray
 \Omega_\Sigma & \equiv &\hh\frac{\Omega_\chi \OmPsi}{\OmPsi- \varepsilon_F \Omega_\chi},\label{omegaSigma}  \\
\Delta &\equiv & \frac{\hh^2 \ \OmPsi \Omega_\chi}{\Omega_\Sigma \{ (1+c_\chi) \Omega_\chi - c_\Psi \OmPsi\}}.\label{asymmDelta}
\earray
In terms of the wave function renormalization factor $Z$,
\barray
A(k, \omega)\! \sim\! \!  \left( \frac{Z^2}{\pi \Omega_\Sigma} \right) 
\frac{  {\cal R}\ \left( 1- \frac{\omega}{\Delta} \right) }{  \{ \omega - Z (\varepsilon_k - \varepsilon_F) \}^2 +  \{ Z^2 {{\cal R}^2}/{\Omega_\Sigma^2} \} }. \nn \\
\label{spectral}
\earray
We thus obtain the following final form for the low-energy expression of the Dysonian self-energy:
\barray
&&\Im  \Sigma(\omega) \sim - \frac{{\cal R}}{\Omega_\Sigma}
\frac{1- \frac{\omega}{\Delta}}{\{1+ \omega \ c_\Psi/\hh  \}^2 + {\cal R}^2/(\hh \OmPsi)^2}, \nn \\
&&\Re  \Sigma(\omega)\sim\hh \varepsilon_F + \omega \nn \\
&& -  \frac{\{\varepsilon_F+ \omega (1 + c_\chi)/\hh  \} (1+  \omega
c_\Psi \ /\hh) +{\cal R}^2/(\hh^2\Omega_\chi\OmPsi)}{\{1+ \omega \ c_\Psi/\hh  \}^2 + {\cal R}^2/(\hh \OmPsi)^2},\nn \\  \label{eq-19}
\earray
where we recall that ${\cal R}(\omega,T)$ is defined in \disp{rdef}.
These expressions, and in particular that of $\Im\Sigma$, are among the key results of the present paper, and will be used below 
in order to fit and interpret the DMFT data.

If we take the $\Psi_{\mathrm{rem}}$ and $\chi_{\mathrm{rem}}$ terms
in \disp{flpsi-2} and \disp{flpsi-1} into account, then \disp{eq-19}
receives higher-order polynomial corrections in $\omega$ in both the
numerator and denominator. Let us also note that  \disp{spectral} is
of the  form of a phenomenological version of  the ECFL theory that
has been recently tested 
against experimental data with considerable success, in some cases after adding a constant times $(\varepsilon_k - \varepsilon_F)$ in the numerator \cite{ECFL,Gweon-Shastry,Kazue}.

\subsection{Low-doping limit $n \to 1$} %

At $T=0$ and  near half-filling we get $\varepsilon_F = - \frac{\pi}{4}  \delta D $ from \disp{bare-ef}. Further, from the single assumption that $\Psi_0 = - \frac{n}{2}$ near half-filling\cite{comment-alt}, we find that the self-energy and wave function renormalization factor scale correctly with $\delta$ as $\delta\to0$. This assumption gives $\hh= \delta$, and a scaling of various energy scales with $\delta$. In particular, we find from the equations that $\Omega_\chi \sim \OmPsi \sim \Omega_c \sim \delta$.  
This, together with \disp{flpsi}, leads to $c_\Psi \sim O(1)$ and $c_\chi \sim O(1)$. This is consistent with the scaling behavior described in remarks 2 and 3 in the previous section. 
Keeping the dominant terms in Eqs. (\ref{QPweight}), (\ref{omegaSigma}), and (\ref{asymmDelta}), we find that 
\barray
 Z& = &  \frac{\hh}{1+c_\chi} \label{eq-n19}, \\
 \Omega_\Sigma & = &\hh \ \Omega_\chi , \label{eq-n20}  \\
\Delta & = & \frac{\hh \ \OmPsi }{ \{ (1+c_\chi) \Omega_\chi - c_\Psi \OmPsi\}}. \label{eq-n21}
 \earray
Near half-filling ($\delta\rightarrow 0$), we define
\barray
Z&=& \delta \ \overline{Z},  \label{eq-20}\\
\Omega_\Sigma&=& \delta^2  \ \overline{\Omega}_\Sigma, \label{eq-21} \\
\Delta &\equiv &\delta \ \overline{\Delta}. \label{eq-22}
\earray 
All objects with an overline, such as $\overline{Q}$, are determined to be finite as $\delta \to 0$. 
\disp{eq-20} is expected on general grounds near the
insulating limit: to leading order $\G^{-1}(k,\omega) =
\varepsilon_F-\varepsilon_k + \frac{\omega}{Z} + O( \omega^2) $,
and therefore the propagating solutions correspond to
quasiparticles with an energy dispersion 
$Z(\varepsilon_k-\varepsilon_F)$ that shrinks to zero at the
insulating point $n=1$. We find here that this occurs as a
linear function of $\delta$.
\disp{eq-21}, together
with \disp{eq-19}, implies that at small $\omega\sim O(\delta)$, the
imaginary part of the self-energy has a finite value. Further combined with a cutoff
$\Omega_c \sim O(\delta)$, it gives $\Re \ \Sigma \sim \omega - 2 \omega \
\frac{ \Omega_c}{\Omega_\Sigma}$, which is then consistent with the
linear vanishing of Z in \disp{eq-20}. \disp{eq-22} shows that the particle-hole asymmetry in the spectral function increases as we approach half-filling. Finally, we see that 
 the spectral density $\Im\Sigma$ becomes a scaling function of
 $\omega/\delta$ at low doping levels.
\section{Doped Mott insulator: single-site DMFT}
\label{sec:dmft}

The dynamical mean-field theory\cite{georges_review_dmft} is based on the fact\cite{metzner_vollhardt} that in
the limit of a large number of dimensions $d$ the self-energy becomes
a momentum-independent local quantity, $\Sigma(\mathbf{k},\omega) \to \Sigma(\omega)$. This
implies that the bulk problem for $d\to\infty$ coincides with the
problem of an interacting impurity embedded in an appropriate
non-interacting bath\cite{georges_kotliar_dmft}. DMFT formulates a practical prescription for
finding this effective impurity problem and the self-consistency
equation.
For the Hubbard model, the corresponding impurity problem is
the single-impurity Anderson model, which can be efficiently solved
with the numerical-renormalization group (NRG) method
\cite{wilson1975,krishna1980a,bulla1999,bulla2008,resolution}.

\subsection{NRG method}

The NRG calculations have been performed with the discretization parameter
$\Lambda=2$ using the discretization scheme with reduced systematic
artifacts described in Ref.~\onlinecite{resolution}. Furthermore, the
twist averaging over $N_z=8$ different discretization meshes has been
used to reduce the oscillatory NRG discretization artifacts
\cite{errors}. The truncation cutoff in the NRG was set in the energy
space at $10\omega_N$ (here $\omega_N$ is the characteristic energy at
the $N$-th NRG step); such results are well converged with respect to
the truncation. The U(1) charge conservation and SU(2) spin rotational
invariance symmetries have been used explicitly. The raw spectral data
(weighted delta peaks) were collected in bins on a logarithmic mesh
with 1000 bins per frequency decade, then the broadening scheme from
Ref.~\onlinecite{weichselbaum2007} with $\alpha=0.2$ has been used to
obtain the continuous representation of the spectral functions. To
calculate the self-energy $\Sigma$, we have used the procedure\cite{bulla1998}  
based on the following exact relation from equations of motion:
\beq
\label{sigmatrick}
\Sigma_\sigma(z) = \frac{\langle\langle [d_\sigma,H_\mathrm{int}];
d_\sigma^\dag \rangle\rangle_z}{\langle\langle d_\sigma;d_\sigma^\dag
\rangle\rangle_z} =
\frac{U \langle\langle n_{\bar\sigma} d_\sigma; d_\sigma^\dag
\rangle\rangle_z}{\langle\langle d_\sigma;d_\sigma^\dag
\rangle\rangle_z}.
\eeq
Here $d_\sigma$ is the impurity annihilation operator, while
$H_\mathrm{int}$ is the interaction part of the Hamiltonian. The two
correlators in this expression were computed using the
full-density-matrix NRG algorithm \cite{peters2006,weichselbaum2007}.
To accelerate the convergence of the DMFT self-consistency loop,
Broyden mixing algorithm has been used \cite{broyden}. This technique
is particularly important to ensure the convergence at small doping as
the Mott transition is approached. The Broyden solver has been used
both to control the chemical potential to obtain the desired band
filling and to apply the DMFT self-consistency equations
\cite{broyden}.

When performing the calculations in the large-$U$ limit, it is
important to note that the upper Hubbard band (UHB) is outside the NRG
discretization energy window. The correlator $F_\sigma(z)
=\langle\langle n_{\bar\sigma} d_\sigma;d^\dag_\sigma
\rangle\rangle_z$ receives a contribution
\begin{equation}
F_\mathrm{UHB}(z) = \frac{w_\mathrm{UHB}}{z-(\epsilon_d+U)}
\mathop{\longrightarrow}_{U \to \infty}
\frac{w_\mathrm{UHB}}{-U}
\end{equation}
from the UHB, where $w_\mathrm{UHB}$ is the total weight of the upper
Hubbard band, which in the $U\to\infty$ limit is equal to $n/2$. The
correlator $F(z)$ in Eq.~\eqref{sigmatrick} is multiplied by a factor
$U$, thus the UHB contribution to the numerator in the $U\to\infty$
limit is equal to $-w_\mathrm{UHB}$. It is crucial to correct the raw
numerical results by making this subtraction when the UHB is outside
the discretization window, otherwise the causality is very strongly
violated. (No such subtraction is necessary for the correlator
$G_\sigma(z) = \langle\langle d_\sigma; d_\sigma^\dag
\rangle\rangle_z$, because the UHB only makes an $\mathcal{O}(1/U)$
contribution to the denominator.)

An analysis of the convergence of the NRG results with respect
to the variation of various parameters in the method is presented in
Appendix~\ref{appB}.

\subsection{DMFT results}

\subsubsection{Scaling of quasiparticle weight $Z$ vs. doping level $\delta$}

\begin{figure*}[htbp!]
\includegraphics[width=1.6\columnwidth,keepaspectratio]{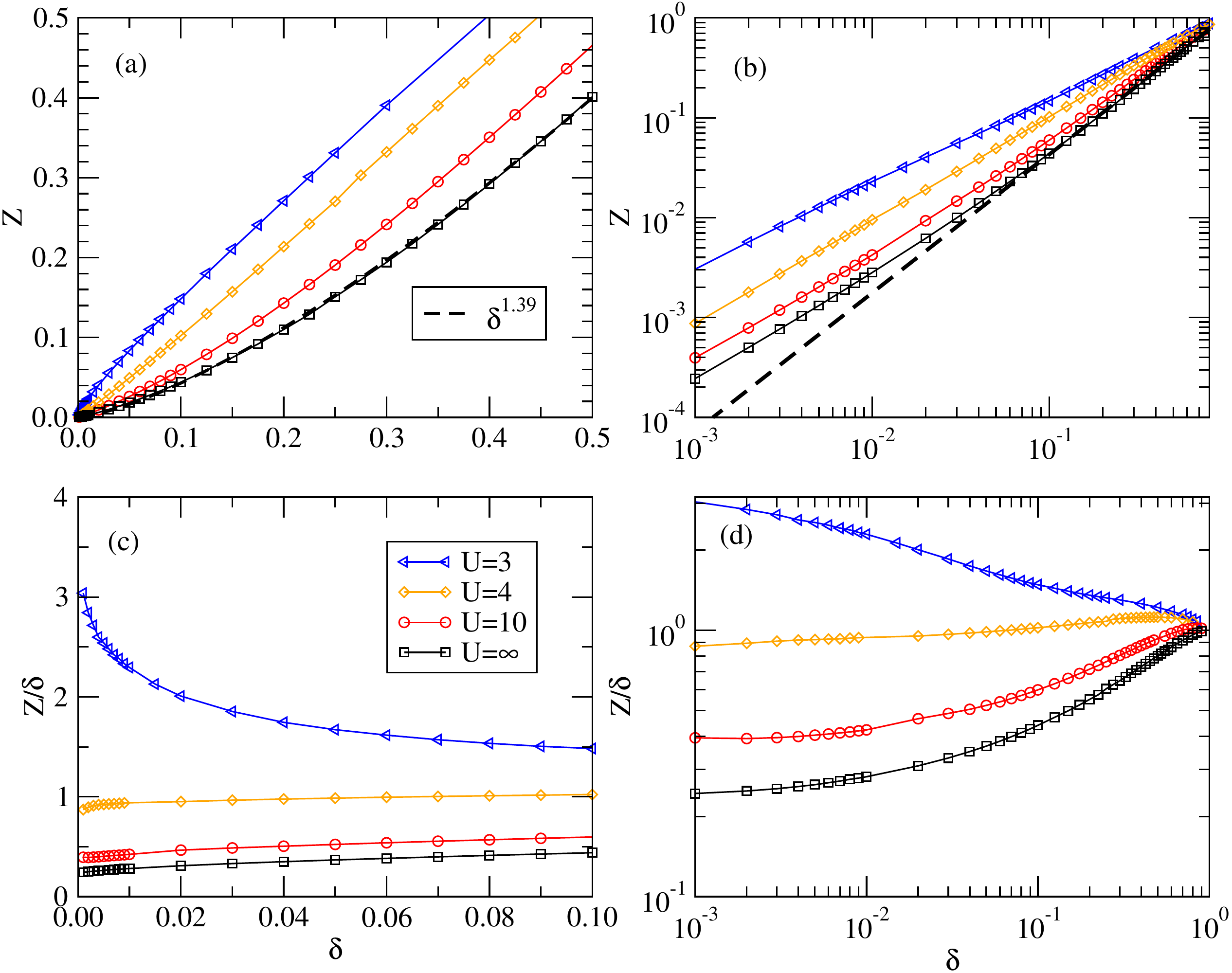}
\caption{\label{fig:z} (Color online) Doping-driven Mott transition
within the DMFT:
approach to the Mott insulating state for $U>U_c=2.918D$ with decreasing
doping $\delta$. Top panels: doping dependence of $Z$ for a range of
$\delta$ on a lin-lin (a) and on a log-log (b) plot. The dashed line
is a fit to a power-law function $Z=\delta^\alpha$ with $\alpha=1.39$.
Bottom panels: $Z/\delta$ vs. $\delta$\ (c), and $Z/\delta$ vs.
$\delta$ on a log-log plot (d). }
\end{figure*}

In this work, we only consider paramagnetic solutions. At low
temperatures, the DMFT equations, depending on the strength of the
interaction $U$ and the doping $\delta$, give either an insulating or
a metallic Fermi-liquid state. The key quantity characterizing the
metallic state is the quasiparticle residue 
\beq
Z = \left( 1-\frac{\partial \mathrm{Re}\Sigma(\omega)}{\partial
\omega}\Bigr|_{\omega=0} \right)^{-1}.
\eeq
At $\delta=0$ (half-filled system), %
a metallic solution is found for $U<U_c$ (this critical value of $U$
is often denoted $U_{c2}$ in the DMFT literature -- the spinodal of
the metallic solution\cite{georges_review_dmft} -- and will be denoted
$U_c$ here for simplicity). 
For $U>U_c$, the DMFT equations only have a unique insulating solution. 
At $U_c/D=2.918$, a Mott metal-insulator transition takes place, with characteristics similar to 
the Brinkman-Rice picture\cite{brinkman_rice_prl_1970} in that the quasiparticle weight $Z$ vanishes continuously and the 
quasiparticle effective mass diverges. 

Away from half filling, i.e., for any $\delta
\neq 0$, the solution is always metallic ($Z>0$); the Mott insulator
($Z=0$) only exists exactly at half-filling 
(for $U>U_{c1}$, the spinodal of the insulator\cite{georges_review_dmft}). 
As $\delta\to 0$ for $U>U_c$, $Z$ diminishes and vanishes at $\delta=0$. 
This doping-driven Mott transition is illustrated in Fig.~\ref{fig:z}. 
In panels a-d) we plot the results of $Z$ vs. $\delta$ for a set of values of $U$. 
It is seen that, when considered over a broad doping range
$\delta\lesssim 0.5$ (panels a and b of Fig.~\ref{fig:z}), the overall doping-dependence of 
$Z$ is fairly linear at intermediate values of $U/D$, while at strong coupling (large $U/D$) a marked curvature is 
seen (approximately fit by a power-law with exponent close to $1.4$). 

A plot of $Z/\delta$ vs. $\delta$ focusing on the low-doping region
(panels c and d) reveals, however, that the asymptotic 
low-$\delta$ behavior is actually linear, $Z\propto\delta$ (except close to the multicritical 
point $U=U_c, \delta=0$ where sizable corrections are found).  
This is indeed the behavior expected within the Gutzwiller 
approach\cite{brinkman_rice_prl_1970,vollhardt_gutzwiller_helium3_rmp_1984,nozieres_cdf_1986}: 
Fig.~\ref{fig:z}c-d thus confirm that 
DMFT obeys this mean-field behavior.  
The prefactor of this linear dependence is also decreasing with $U$ in reasonable agreement 
with the Gutzwiller estimate\cite{nozieres_cdf_1986} $\sim (1-U_c/U)^{-1/2}$. 
Note that the results displayed here extend previous studies to much lower doping levels 
($\delta\lesssim 0.001$) than previously reported in the literature, due to the improvements in 
the NRG methodology. 

\subsubsection{Self-energy and spectral function: overview of the main structures}

We now address the properties of the self-energy $\Sigma(\omega)$ and one-particle spectral function in more detail. 

\begin{figure}[htbp]
\includegraphics[width=\columnwidth,keepaspectratio]{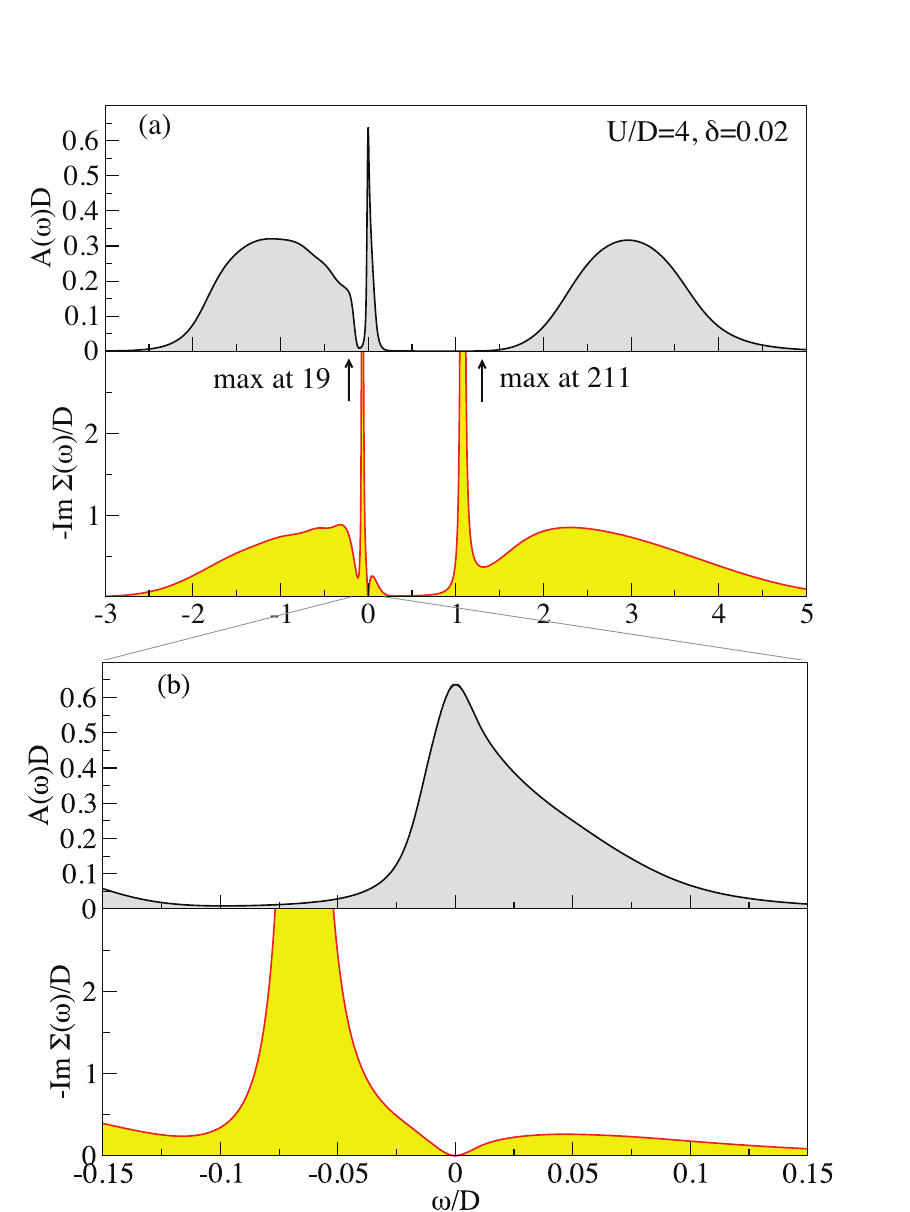}
\caption{\label{figr1} (Color online) (a) Overview plot showing the
full structure of spectra on high frequency scales (lower Hubbard
band, quasiparticle band, upper Hubbard band) at finite $U=4D$. We
show the DMFT local spectral function (top panel) and the imaginary
part of the self-energy (bottom panel). (b) Close-up on the
quasiparticle band at low frequencies.}
\end{figure}

An overview plot in Fig.~\ref{figr1}a shows the main
features in the local spectral function $A(\omega)$ and in the imaginary
part of $\Sigma(\omega)$ in a broad frequency range. 
$\Im\Sigma$ has two very pronounced and sharp resonances (quasi-poles). 
These resonances are responsible for the suppression of the spectral weight in $A(\omega)$ between the 
QP peak and the LHB and UHB, respectively. They are correspondingly positioned close to the
minima of $A(\omega)$. 
In contrast to the half-filled case, where these resonances are symmetrically positioned on each side of 
$\omega=0$ at a scale\cite{rozenberg_prb_1994} $\propto \pm\sqrt{Z}$, their locations in the doped 
case are no longer symmetric and will be discussed below.  
In addition, there are two broad humps in $\Im\Sigma$ in the frequency ranges associated with the two Hubbard
bands. 
As $U$ increases towards very large values at fixed doping, the
UHB moves to higher frequencies, while the LHB and QP band gradually
converge to their high-$U$ asymptotic form. This convergence is,
however, rather slow and the spectra start to very closely agree with
the asymptotic ones only for $U$ on the order of $100D$. 

In Fig.~\ref{figr1}b we plot a close-up on the low-energy structures, i.e., the QP band and its vicinity. 
We notice that the Fermi-liquid quadratic behavior of $\Im\Sigma$ is limited to a very narrow frequency 
interval, much smaller than the width of the QP peak itself.  
We also see (Fig.~\ref{figr1} and Fig.~\ref{fig:im_sigma_U_inf_Z})  that at low doping level, $\Im\Sigma$ develops a 
marked particle-hole asymmetry.
These deviations to Fermi-liquid behavior are discussed below in a more quantitative manner.    

One of the goals of this work is to provide an analytical account of the 
complex frequency dependence of the self-energy that we just summarized.  
 It should be kept in mind that the ECFL theory that we are going to use for this purpose 
 works with the $U=\infty$  model, which begins by throwing out the UHB altogether and deals only with the LHB. 
 Thus the comparison carried out later in this paper refers only to the LHB and QP sector (no double occupancies), 
 containing the interesting low-energy physics of the problem. 

\subsubsection{Dynamical Particle-hole asymmetry}

\begin{figure*}[htbp!]
\includegraphics[width=1.6\columnwidth,keepaspectratio]{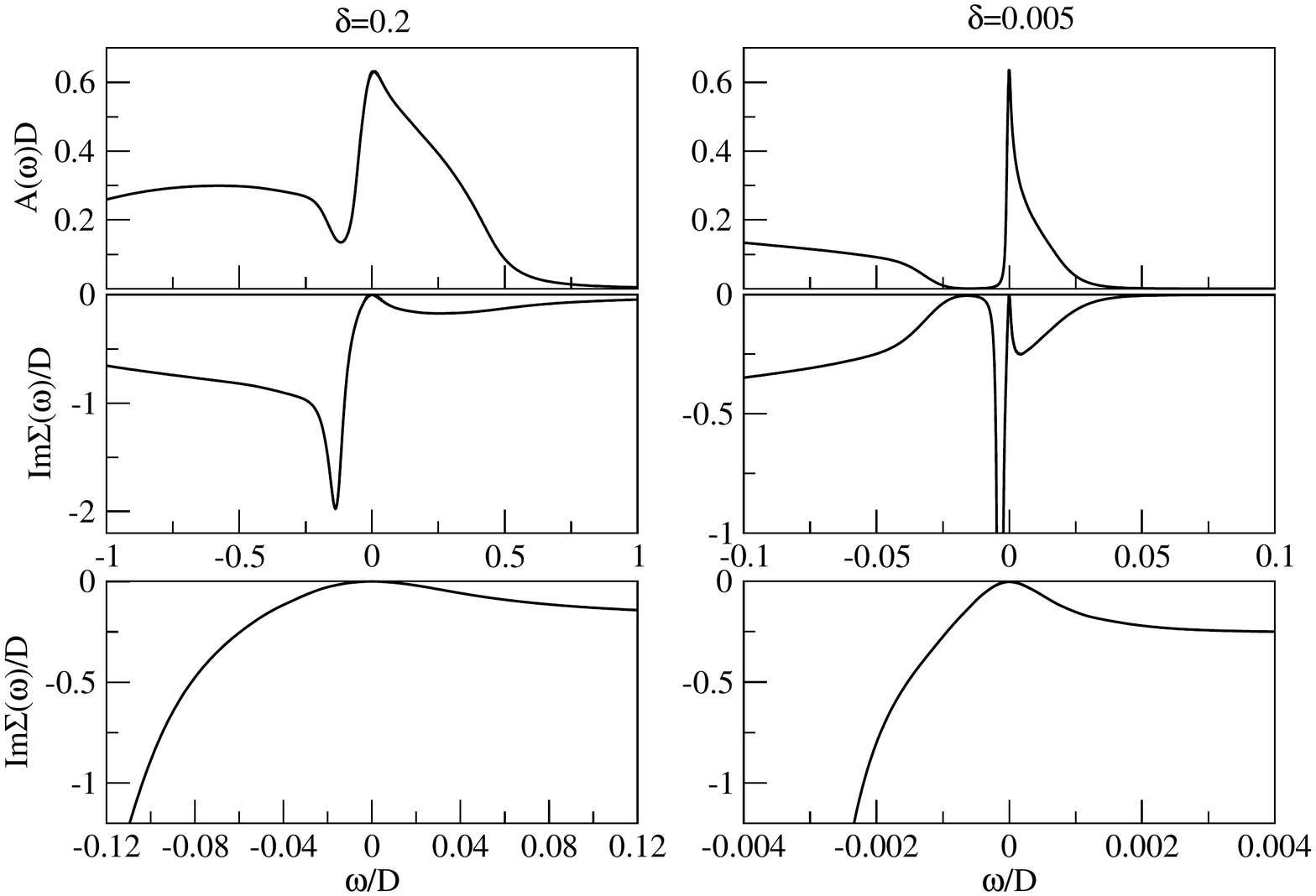}
\caption{\label{fig:3new} (Color online) Local spectral function and the imaginary part of
the self-energy for large doping $\delta=0.2$ (left) and small doping $\delta=0.005$ (right), for $U=\infty$.
}
 \end{figure*}
 
The local spectral function and self-energy are displayed on Fig.~\ref{fig:3new} 
for $U=\infty$, at two different doping levels (a small and large one, for comparison). 
An immediately apparent feature of these plots is the large asymmetry between 
hole-like ($\omega<0$) and particle-like ($\omega>0$) excitations. 

Indeed, for $\omega<0$, $|\Im\Sigma|$ increases rapidly from $\omega=0$ in order to 
connect with the negative-energy quasi-pole. The detailed form of this increase is somewhat different 
depending on the doping level. At large doping it is approximately parabolic, in continuity with the 
low-$\omega$ FL $\sim\omega^2$ dependence. In contrast, at small doping, the low-$\omega$ parabolic 
dependence evolves into a more linear-like increase at higher frequency. The local spectral function also 
displays an almost complete suppression of the spectral weight between the QP peak and the LHB at low doping level, while 
this suppression is only modest at higher doping.  

In contrast, for $\omega>0$, $|\Im\Sigma|$ rapidly flattens out after its initial FL increase. It has a plateau-like 
behavior with a broad maximum at large and intermediate doping level (the maximum is sharper at smaller doping). 
Overall, $|\Im\Sigma|$ remains much smaller at $\omega>0$ than at $\omega<0$. 

This asymmetry also reflects into the QP peak in the local spectral
function $A(\omega)$, i.e., the local $\rho_{\G}(\omega)$, which  has a very asymmetric line-shape. 
The decrease from its $\omega=0$ value $A(0)$ is much faster on the $\omega<0$ side, in accordance with the 
large $|\Im\Sigma|$. The detailed form of the line-shape on the more extended $\omega>0$ side is different at lower and 
higher doping levels, with a convex  and concave shape, respectively.

 \begin{figure*}[htbp!]
\includegraphics[width=1.5\columnwidth,keepaspectratio]{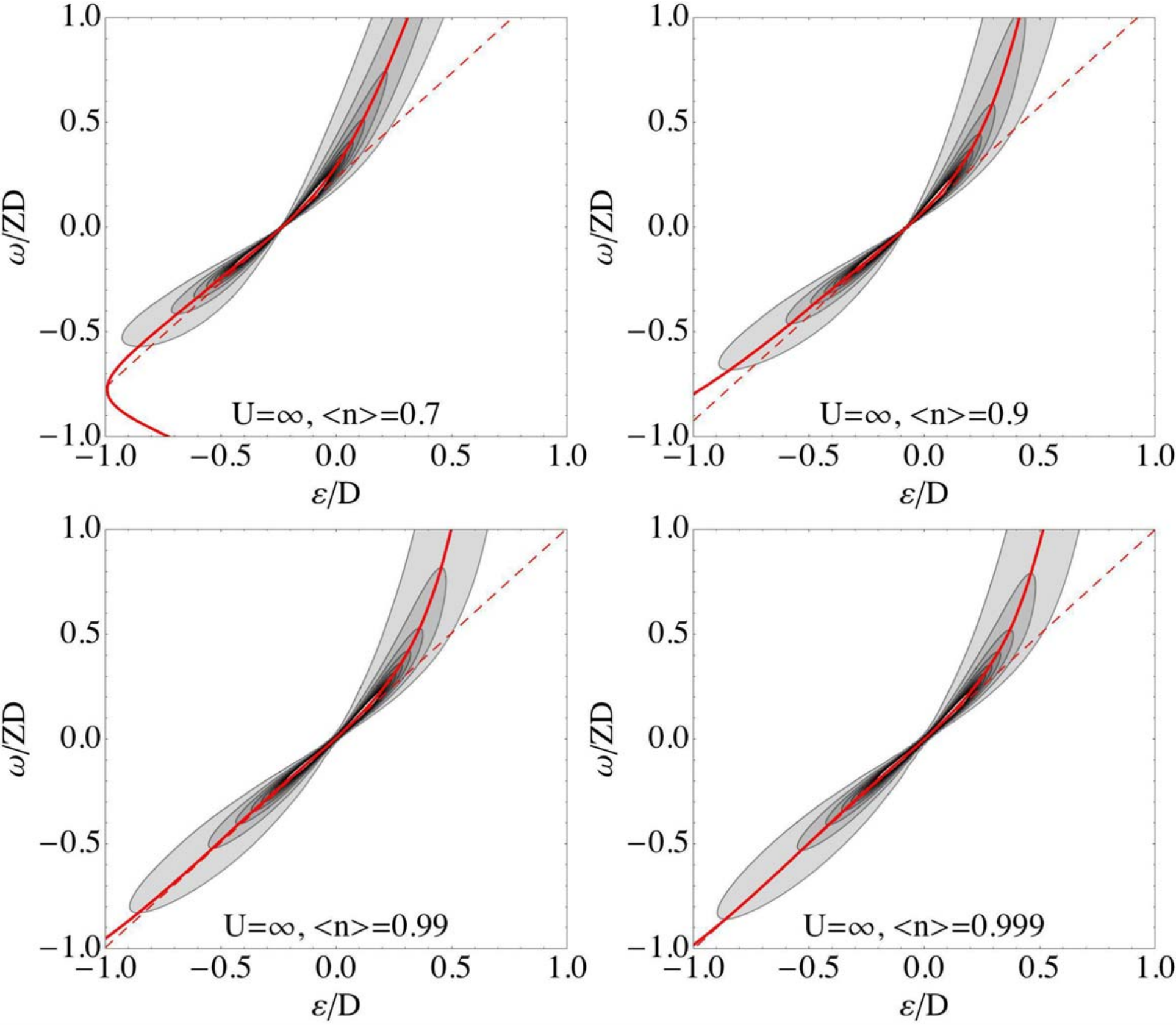}
\caption{\label{fig:eps_resolved_dmft} 
(Color online) Intensity plots of the momentum ($\varepsilon_k$-)
resolved spectral function $A(\varepsilon,\omega)$ for $U=\infty$ at 
four different doping levels, plotted as a function of $\varepsilon/D$
and $\omega/ZD$. The plain line locates the solution of the QP pole
equation $\omega+\mu-\Re\Sigma(\omega)-\varepsilon=0$ (neglecting
$\Im\Sigma$). By definition of the QP excitations, this line has slope
unity (cf. dashed line) at low-$\omega$ when plotted in this manner
since $\omega_{\mathrm{QP}}=Z(\varepsilon-\varepsilon_F)$ (i.e., $v_F^\star=Z
v_F$) within the DMFT. 
}
 \end{figure*}

Finally, the particle-hole asymmetry has a very distinctive signature in the momentum-resolved spectra 
$A(\varepsilon,\omega)$, which are displayed in Fig.~\ref{fig:eps_resolved_dmft}. It is seen there that the dispersion of the 
QP peak deviates from its low-energy form $\omega_{QP}=Z(\varepsilon-\varepsilon_F)$ much more rapidly on the 
$\omega>0$ side, where a stronger dispersion closer to that of the bare band is rapidly found. 
This is mostly due to the distinct behavior of the real part $\Re\Sigma$ for positive and negative frequencies 
(shown later in Fig.~\ref{fig:re_sigma_U_inf_delta}). 
This finding, which is also supported by the ECFL results as discussed below, is one of the main predictions of our work. 
It calls for the development of momentum-resolved spectroscopies for
unoccupied states (the `dark side' that is not 
directly accessible to ARPES). 
The physical significance of this `dark side' has also been recently pointed out in cluster-DMFT 
studies of the two-dimensional Hubbard model\cite{sakai_darkside}.

\subsubsection{$\omega/Z$ scaling}

\begin{figure}[htbp!]
\includegraphics[width=0.8\columnwidth,keepaspectratio]{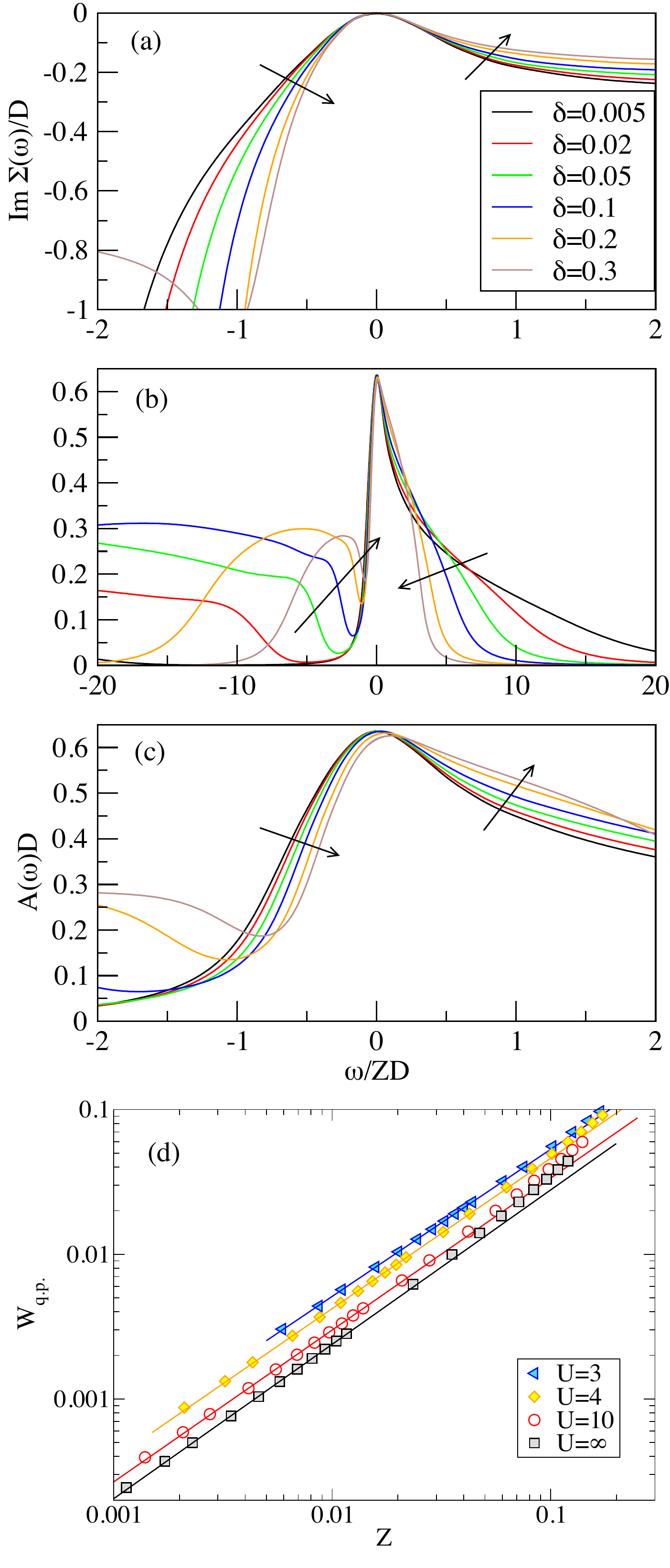}
\caption{\label{fig:im_sigma_U_inf_Z} (Color online) (a) Imaginary
part of the self-energy $\Sigma$ versus the rescaled frequency
$\omega/Z$ for $U=\infty$.  (b,c) Corresponding local spectral
functions.  Note that when plotted vs. $\omega/Z$, the peak related to
the onset of the LHB moves to the left with diminishing $\delta$, as
seen in panel b. The results for finite $U$ are qualitatively very
similar. The arrows indicate the direction of the increasing value of
$\delta$.
}
 \end{figure}

Close to the Mott transition, all low-frequency properties are expected to scale with $Z$, i.e., be described 
by scaling functions\cite{georges_review_dmft,moeller_prl_1995,bulla_scaling_1999} of $\omega/Z$. 
This is indeed the case, as demonstrated in Fig.~\ref{fig:im_sigma_U_inf_Z} in which good data 
collapse is obtained in the lowest frequency range when plotted vs. $\omega/Z$. 
However, we also clearly observe that the scaling is limited to the asymptotic region of very small frequencies. 

On panels b and c of Fig.~\ref{fig:im_sigma_U_inf_Z}, one can compare the evolution of the shape of the 
local spectral function, discussed above, as a function of the doping
level. One sees that the QP peak 
becomes increasingly asymmetric at very low doping.  
We also observe that the LHB has some internal structure, quite similar to that observed at
half-filling as the correlation-driven Mott transition is approached from the metallic 
side \cite{karski2005,karski2008,resolution,pade}.   

In Figs.~\ref{fig:re_sigma_U_inf_delta}, the real and imaginary part of the self-energy are plotted 
against $\omega/\delta D$. The different panels cover different frequency ranges. 
While a better collapse of the different curves at very low frequency was obtained above 
when using $\omega/ZD$ as a scaling variable, it is seen from the plots in a broader frequency range 
that the overall structures of the self-energy obey rather good scaling properties with respect to $\omega/\delta D$. 
For example, the sharp peak (quasi-pole) structure at $\omega<0$ in $\Im\Sigma$ is seen to be 
located at a frequency proportional to doping level
($\omega_{\mathrm{peak}}\simeq -0.7 \delta D$). 
This peak in $\Im\Sigma$ is associated with the suppression of the spectral weight between the 
QP peak and the LHB in the spectral function. 
Correspondingly, it is associated with a resonance-like structure in $\Re\Sigma$. 

 \subsubsection{Deviation from the low-frequency Fermi-liquid behavior}

\begin{figure*}[htbp!]
\includegraphics[width=1.66\columnwidth,keepaspectratio]{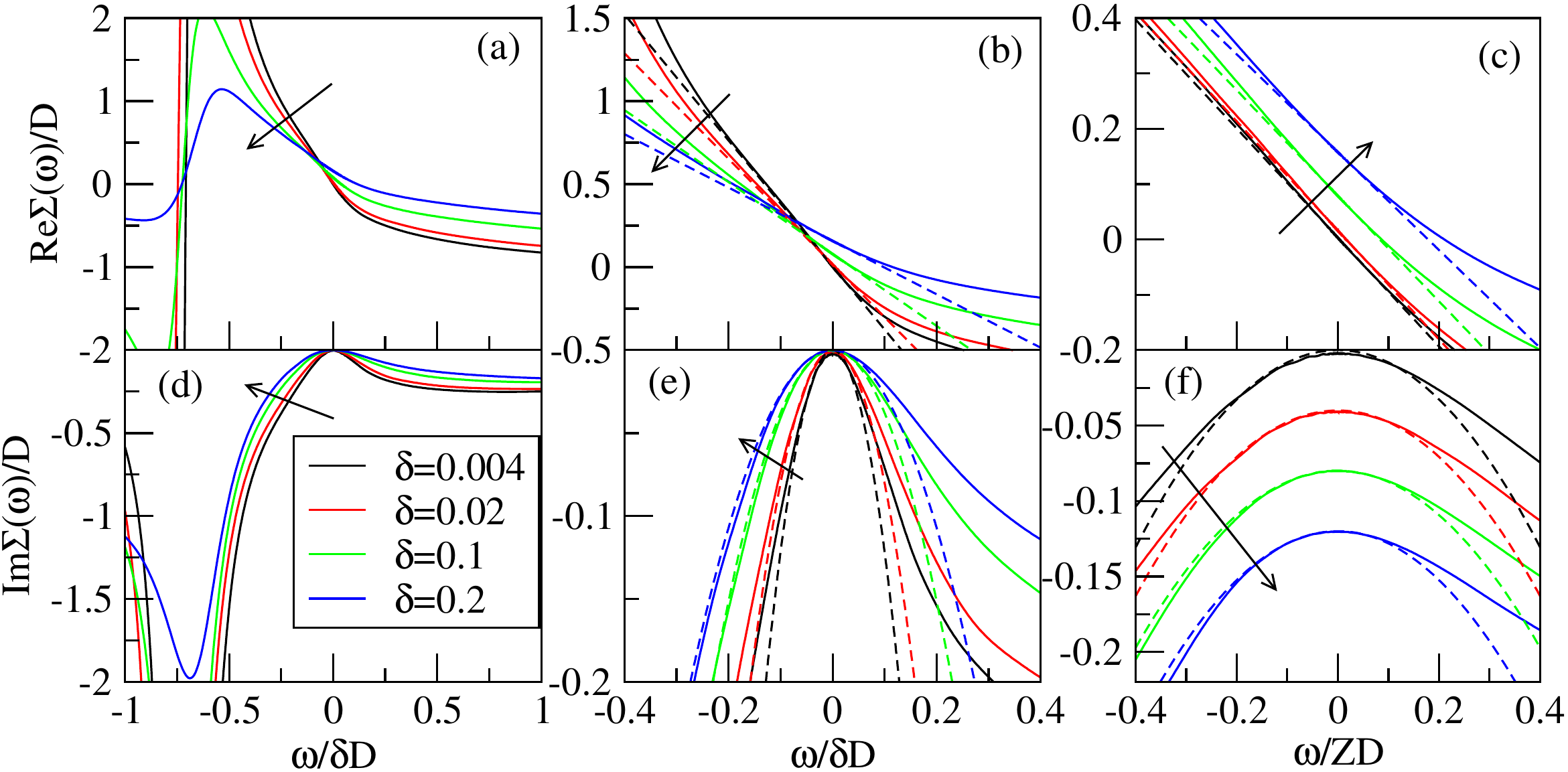}
\caption{\label{fig:re_sigma_U_inf_delta} (Color online) Real and
imaginary parts of the DMFT self-energy $\Sigma(\omega)$ for a range
of doping for $U=\infty$, as a function of $\omega/\delta D$ (left
and centre) or as a function of $\omega/Z D$ (right).
The arrows indicate the direction of the increasing value of $\delta$.
The linear (for real parts) and parabolic (for imaginary parts) are
performed in the frequency range $[-0.05:0.05]\delta D$.
In panel (f), the data is vertically offset for clarity.
}
\end{figure*}
From panels b, c (for $\Re\Sigma$) and e, f (for $\Im\Sigma$) of Fig.~\ref{fig:re_sigma_U_inf_delta}, 
one can visualize the low-frequency deviations from Fermi-liquid behavior. 
The latter is indicated by the dashed straight and parabolic lines on this figure: 
$\Re\Sigma-\Re\Sigma(0)=\omega (1-1/Z)$ and $\Im\Sigma \propto -(\omega/Z)^2$. 

When visualized on an intermediate frequency scale (panels b and e) it is seen that 
deviation from the FL behavior is more apparent on the $\omega>0$ side, in accordance with the 
particle-hole asymmetry discussed above and as pointed out in previous studies\cite{deng_2013,zemljic_07}. 
$\Re\Sigma$ deviates from linearity and 
flattens upwards for $\omega>0$, resulting in the bending of the dispersion 
of $\omega>0$ quasiparticles towards the non-interacting bare dispersion, displayed above 
on Fig.~\ref{fig:eps_resolved_dmft}.
Accordingly, the deviations from parabolic behavior in
$\Im\Sigma(\omega)$ are much more pronounced on the positive frequency side.  

Zooming further on the low-frequency range (panels c and f) allows one to locate more quantitatively the
deviation from the FL behavior. At $U=\infty$, it is seen to occur at 
$\omega_{\mathrm{FL}}^\star\simeq 0.1\,ZD$, which is of order $0.025\delta D$ to $0.05\delta D$ depending on 
$\delta$. In agreement with previous studies\cite{deng_2013} at finite $U$, the scale below which FL 
is found to apply is seen to be a very low one. It is one order of magnitude smaller than the Brinkman-Rice scale 
$\approx \delta D$ which corresponds to scaling the bare bandwidth by the (inverse) of the effective mass. 
When converted to a temperature scale, the Brinkman-Rice scale roughly corresponds to the temperature at 
which QP excitations disappear altogether (and the resistivity approaches the Mott-Ioffe-Regel limit)\cite{deng_2013}, 
but it should not be identified with the much lower scale associated with deviations from FL behavior. 

The low-frequency zooms in panels c and f actually reveal that the deviations from FL behavior 
are seen both on the $\omega<0$ and $\omega>0$ side, at similar scales
$\pm\omega_{\mathrm{FL}}^\star$.
This scale corresponds to a low-energy `kink' in $\Re\Sigma$. The corresponding low-energy kink in 
the QP dispersion\cite{byczuk_kinks_2007,grete_prb_2011,held_2013} 
 is actually visible upon close examination of Fig.~\ref{fig:eps_resolved_dmft}.

As seen on Figs.~\ref{fig:im_sigma_U_inf_Z}b and c, the full collapse
of the data is limited to very low frequencies. Two kinds of
deviations from the universal behavior can be recognized. On negative
frequency side, at moderate doping the deviations occur at the onset
of the Hubbard band as the quasiparticle peak is not clearly separated
from the LHB. On the positive frequency side, the different curves
deviate from each other also in the small doping limit. Comparing the
two lowest dopings, for instance, reveals the excess of the spectral
weight for the lower doping curve. This suggests that the
quasiparticle peak is not fully characterized by the renormalization
factor $Z$ alone: the quasiparticle peak weight $W_{q.p.}$ and $Z$ are
not necessarily simply proportional. This question is best addressed
at very low doping, when the quasiparticle peak is well separated from
the LHB. We can then extract $W_{q.p.}$ by integrating the spectral
function between the two local minima in $A(\omega)$. The results are
plotted in Fig.~\ref{fig:im_sigma_U_inf_Z}d. We find that at low
$\delta$, $W_{q.p.}$ and $Z$ are related by a power-law
$W_{q.p.}=Z^\gamma$, with $\gamma$ close to one, but not exactly 1.
More specifically, $\gamma$ is found to be $U$-dependent:
$\gamma=1.017$ for $U=3$, $\gamma=1.039$ for $U=4$, $\gamma=1.049$ for
$U=10$ and $\gamma=1.067$ for $U=\infty$.

\subsubsection{Charge compressibility: absence of phase separation}

\begin{figure*}[htbp!]
\includegraphics[width=1.9\columnwidth,keepaspectratio]{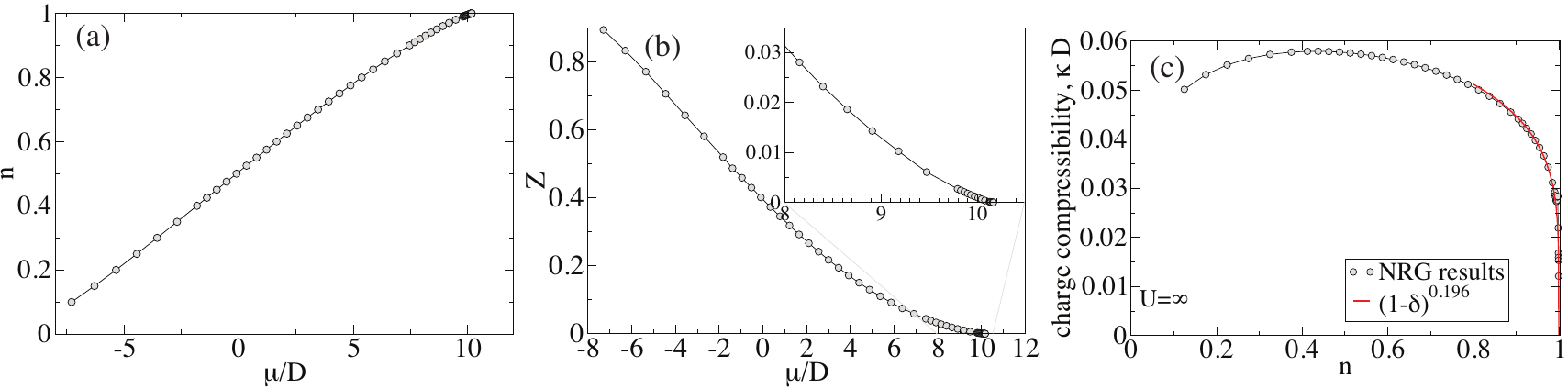}
\caption{a) Band filling $n$ vs. chemical potential $\mu$ for
$U=\infty$. b) Quasiparticle residue $Z$ vs. chemical potential $\mu$
for $U=\infty$. c) Charge compressibility for $U=\infty$.}
\label{fig:kappa}
\end{figure*}

For some types of the (non-interacting) conduction-band density of
states, there can be phase separation near half filling
\cite{Eckstein:2007bi}. We verify that this is not the case for the
Bethe lattice by plotting the band filling $n$ as a function of the
chemical potential $\mu$ in Fig.~\ref{fig:kappa}, panel a) for
$U=\infty$. The dependence is monotonous, thus all solutions are
physically stable with positive charge compressibility
$\kappa=\partial n/\partial \mu$. We also plot the quasiparticle
residue $Z$ as a function of the chemical potential $\mu$ (panel b).
The charge compressibility $\kappa$ as a function of the band filling
(panel c) has a maximum near quarter filling. For smaller $n$, the
decrease is due to the particular form of the non-interacting DOS
(semi-circular function). For larger $n$, $\kappa$ drops to zero as
the Mott transition is approached. The asymptotic behavior is
a power law $\delta^\beta$ with $\beta \approx 1/5$.

\section{Doped Mott insulator: an ECFL perspective on the DMFT}
\label{sec:dmft_ecfl}

In this section we make use of the general structure of the
self-energy resulting from the ECFL in order to interpret, fit, and better
understand the complex frequency-dependence of the DMFT self-energies.
The emphasis will be on the intermediate frequency range which
encompasses both the vicinity of $\omega=0$ and of the quasi-pole
(sharp peak) in $\Im\Sigma$ on the negative frequency side at 
$\omega_{\mathrm{peak}}\simeq -0.7\delta D$.
We focus on intermediate doping levels which
turns out to be the range where the ECFL applies best, rather than on very low doping. 
For these reasons, we can use as a scaling variable:  
\begin{equation}
x \equiv \omega/\delta
\end{equation}

\subsection{ECFL line-shapes: main features} 
 
The low frequency  ECFL analysis from Sec. II gives a simple expression, \disp{eq-19}, for $\Im\Sigma$ at $T=0$. Using the overline convention of  Eqs.~(\ref{eq-20},\ref{eq-21}, \ref{eq-22}) to denote variables that 
remain finite as $\delta \to 0$, e. g. $\overline{P} = P/\delta$,
 we rewrite \disp{eq-19} as
\begin{equation}
\Im \Sigma =
-\frac{x^2}{\overline{\Omega}_1}\frac{1-x/~\overline{\Delta}_1}{(1+x/~\overline{\Delta}_2 )^2+ x^4/~\overline{\Omega}_2^2} \label{eq:eq}.
\end{equation}
This ansatz  function is determined by two  variables with the meaning of
curvature $\overline{\Omega}_{1,2}$ that are simply related to 
$\Omega_\Sigma$ and $\Gamma_\Psi$, respectively, and by two parameters which adjust
the asymmetry $\overline{\Delta}_{1,2}$ that are related to $\Delta$ and $c_\psi$.

\begin{figure}[htbp]
\includegraphics[width=.8\columnwidth,keepaspectratio]{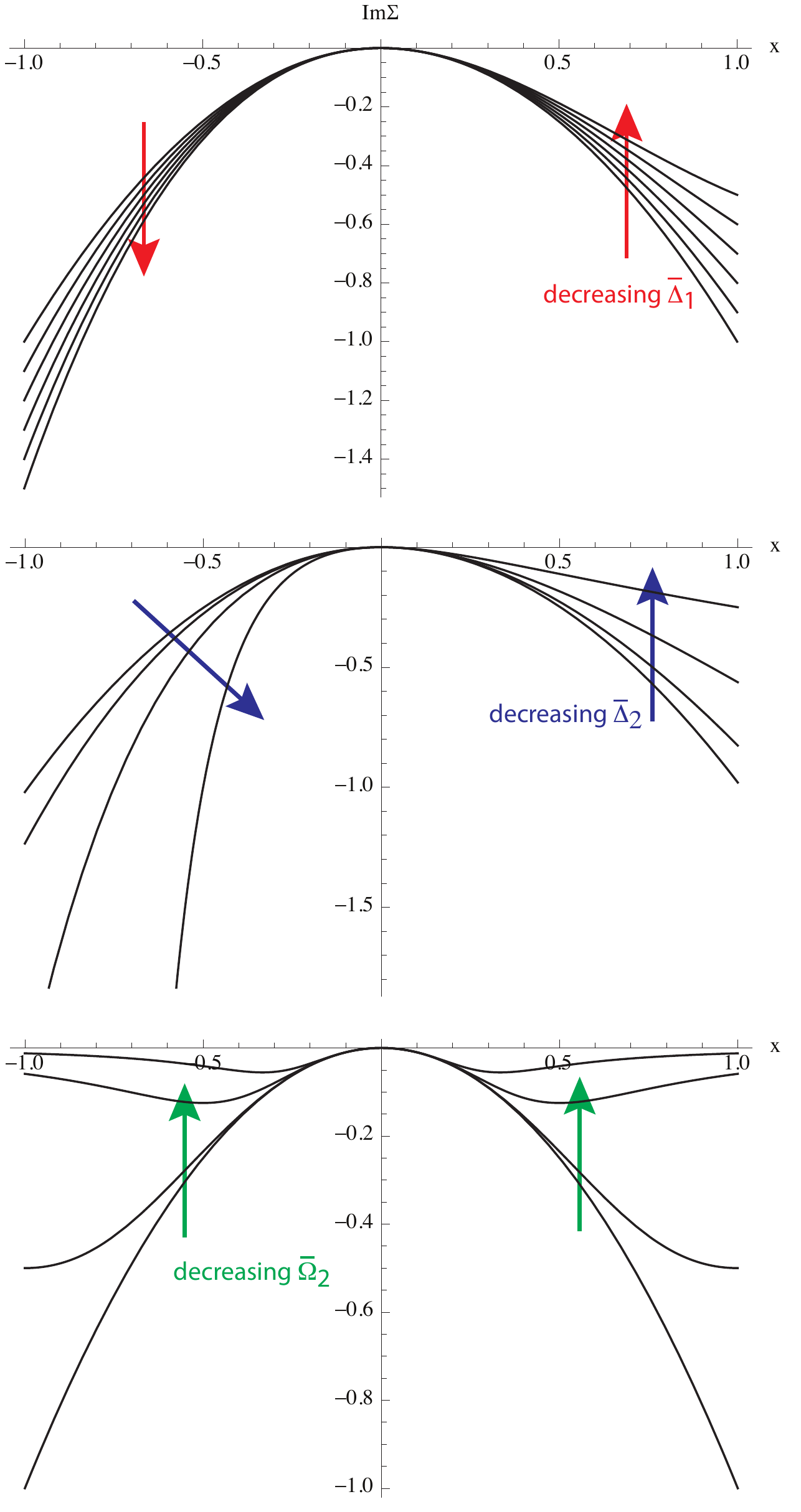}
\caption{(Color online) \label{fig:effects} Role of the parameters in
the ECFL ansatz function, Eq.~\eqref{eq:eq}, use for fitting in
Figs.~\ref{fig:re_sigma1} and \ref{fig:param}.}
\end{figure}

The numerator of the expression describes a parabolic dependence,
with  a cubic correction term. This ansatz function has a peak (quasi-pole) 
at frequency $x= -\overline{\Delta}_2$ for a finite $\overline{\Omega}_2$,  
turning into a true pole when $\overline{\Omega}_2 \to \infty$. The low-frequency asymmetry of the self-energy, important
for the low-temperature thermoelectric properties, arises through a
combination of the terms present in the numerator and denominator. 
Expanded to cubic order in frequency, the ansatz gives 
\begin{equation}
\Im \Sigma= -x^2/\overline{\Omega}_1 [1
  -(1/\overline{\Delta}_1+2/\overline{\Delta}_2)x]. 
\end{equation}
The ansatz function and its evolution as the parameters are varied is illustrated in Fig.~\ref{fig:effects}.

Summarizing,   the ansatz function \disp{eq:eq} contains a parabolic dependence, multiplied by
a function with a sharp peak at negative frequencies; therefore,  it can be
expected to describes the coarse structure found in the DMFT very well
already at this order including {\it only} the Fermi-liquid structure
of the underlying functions $\Psi$ and $\chi$.

\subsection{ECFL fits of the DMFT self-energy}

\begin{figure*}[htbp!]
\includegraphics[width=1.4\columnwidth,keepaspectratio]{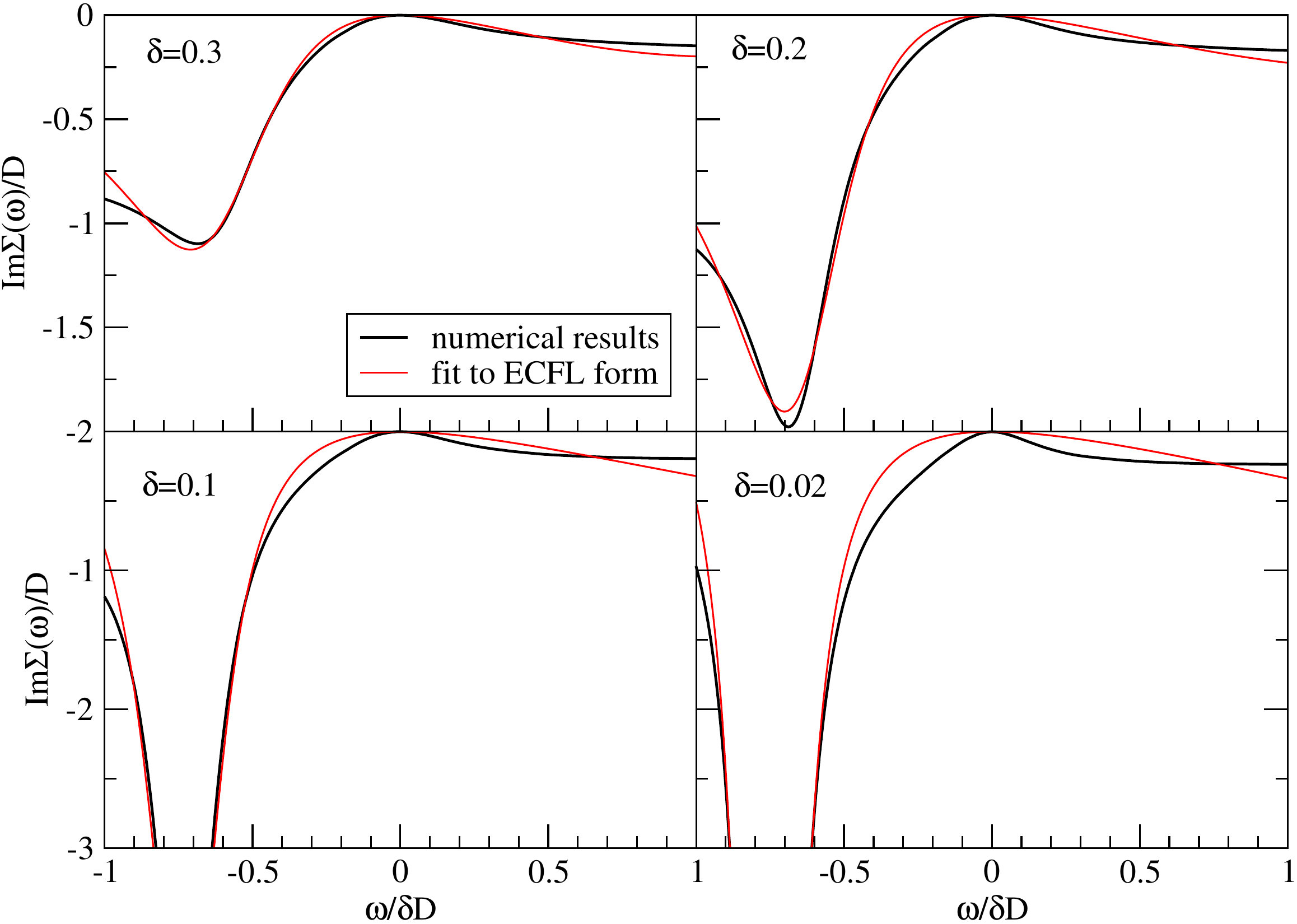}
\caption{\label{fig:re_sigma1} (Color online) $\Im
\Sigma(\omega)$ vs. rescaled frequency $\omega/\delta D$ and
fits to Eq.~\eqref{eq:eq} for $U=\infty$. The ansatz describes the
DMFT self-energy at moderate dopings remarkably well. The nipple
structure that becomes pronounced at small doping signals that $\chi$ and
$\Psi$ develop non-Fermi-liquid corrections.}
 \end{figure*}

The DMFT results for $U=\infty$ self-energy for a range of doping
levels are presented in Fig.~\ref{fig:re_sigma1} together with fits to
Eq.~\eqref{eq:eq}. At large doping, the ansatz function describes the
DMFT data remarkably well: the low-frequency dependence and the main
shape of the self-energy are fully reproduced.

At smaller doping, the quasi-pole at negative frequencies becomes very
sharp and $\Im\Sigma$ in DMFT develops a ``nipple-like'' structure at
low-frequency, with semi-linear frequency dependence at negative
frequencies. 
These two features (quasi-pole and nipple) cannot be simultaneously 
well described by the simplest ECFL ansatz function in a broad frequency range. 
Given that the ansatz has a structure that already contains the pole,
the fits in a broad frequency window
are more meaningful. The DMFT data can still be described successfully,
but terms beyond the lowest-order Fermi-liquid  form in $\Psi$ and $\chi$ need to
be retained. Work along these lines to reproduce the precise shape and
to analyze its physical contents should be possible.

\begin{figure}[htbp!]
\includegraphics[width=.9\columnwidth,keepaspectratio]{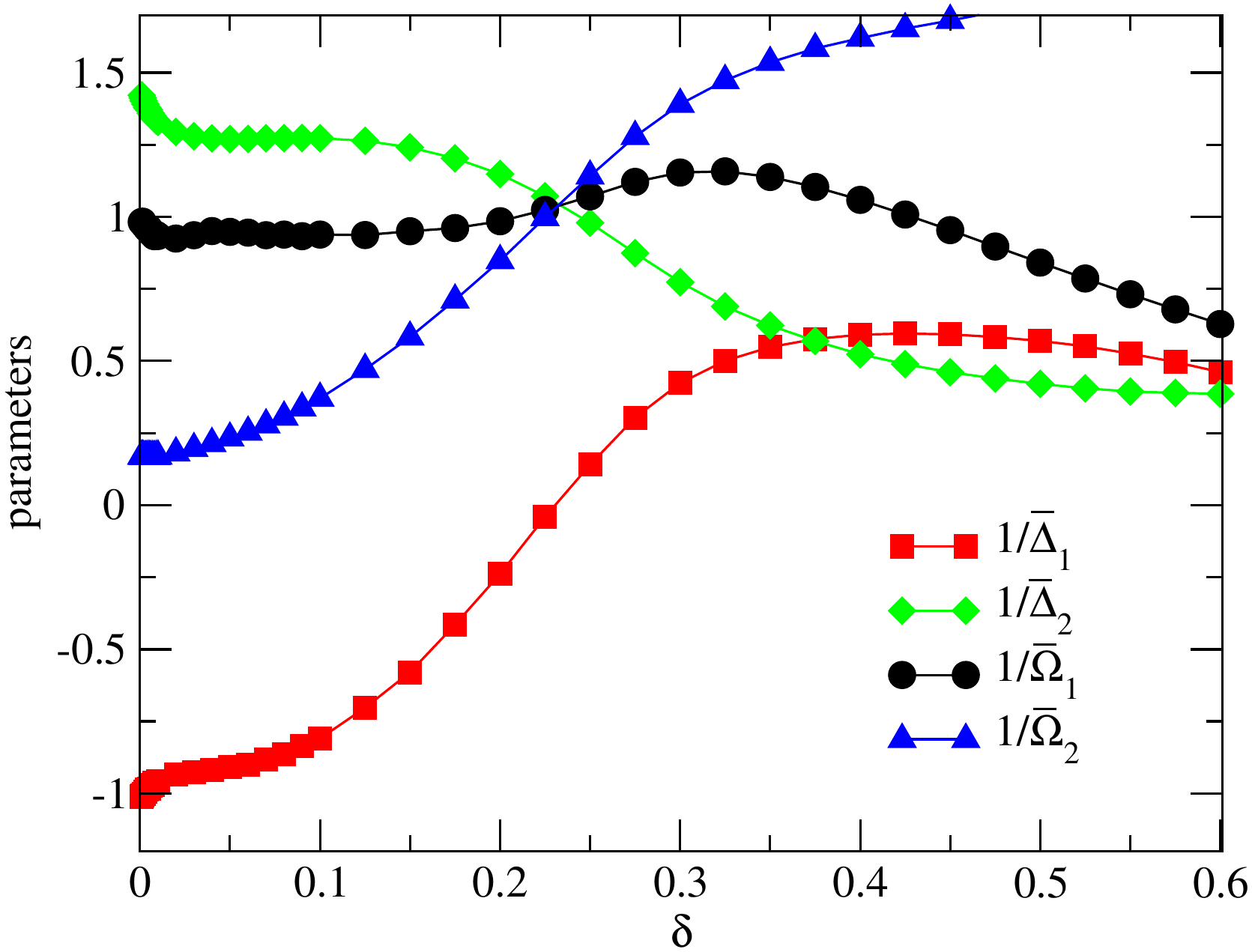}
\caption{\label{fig:param} (Color online) Parameters in the fit
function, Eq.~\eqref{eq:eq}, for a range of doping $\delta$ at
$U=\infty$. For a broad fitting energy range, the fitting parameters
are smooth as a function of $\delta$, thus the fitting procedure is
well defined.  At dopings where the fits work best (that is around
$\delta=0.25$), $\overline{\Omega}_1$ is found to be close to
$\overline{\Omega}_2$.  }
\end{figure}

The evolution with doping of the fitting parameters is shown in Fig.~\ref{fig:param}.
The first observation is that the fitting parameters (except for $\overline{\Delta}_1$
to be discussed below) do not depend substantially on the doping,
hence validating our assumptions stated above. 

The second observation is that $\overline{\Omega}_1$ is usually found to be very close to
$\overline{\Omega}_2$. This supports the conclusions of the $\lambda^2$ analysis (to be
discussed in the next section) which also finds that $\Gamma_\Psi$ is close to
$\Omega_\chi$. 

The third observation is that the bulk of the asymmetry
does actually not come from the explicitly cubic term (parametrized by
$1/\Delta_1$), but rather from $1/\Delta_2$. 
Hence, it is the presence of the quasi-pole at negative frequency which is responsible 
for the strong particle-hole asymmetry. 
This is seen most explicitly in the  doping range 
$\delta=0.2-0.3$, where $1/\Delta_1$ almost vanishes and where,
furthermore, the
data is excellently described by the  ansatz function from \disp{eq:eq}. The physical content
of this observation might be that the low-frequency particle-hole
asymmetry is (at least at not too low dopings) directly related to the
presence of the LHB at much higher frequency scales,
and thus ultimately to the strong-correlation physics. This observation is
consistent with the picture emerging from the ECFL, where the asymmetry is a consequence of the Gutzwiller projection. 
    
\subsection{Summary}

To summarize, the ECFL-derived ansatz, Eq.~\eqref{eq:eq}, which
retains only the lowest-order Fermi-liquid terms in $\Psi$ and $\chi$,
describes the rather complex frequency dependence of the DMFT data
remarkably well at low to intermediate frequencies and for not too
small doping levels.  Importantly, retaining only the Fermi-liquid
(hence particle-hole symmetric) terms in these ECFL self-energies
already yields a marked particle-hole asymmetry in the physical
electron Dysonian self-energy.  The ansatz also describes the pole at
negative frequencies, associated with the onset of the LHB. Whereas
the Fermi-liquid behavior only applies at extremely low frequencies in
the Dysonian self-energy, the Fermi-liquid concepts can still be used
over a much broader frequency range when proper auxiliary quantities
are considered, within the broader framework provided, e.g., by the
ECFL theory. At lower doping levels, however, the DMFT results display
structures (nipple) which signal the increasing importance of
corrections beyond the dominant Fermi-liquid terms in $\Psi,\chi$.

\section{ECFL: expansion to $O(\lambda^2)$ \label{second-order}}
\label{sec:ecfl_lambda}

\subsection{Summary of equations }

We now summarize the results of the $O(\lambda^2)$ expansion of the
ECFL equations in Ref.~\onlinecite{edward-shastry}, which are then
computed and compared with the DMFT results. We note
that the ECFL reformulation of the Dyson self-energy $\Sigma(\omega)$
into the ECFL auxiliary self-energies $\Psi(\omega)$ and
$\chi(\omega)$ is exact. Therefore, if one could perform the $\lambda$
expansion to infinite-order in $\lambda$, one would obtain the exact
answer for these auxiliary self-energies, and consequently the Dyson
self-energy. The resulting Dyson-self energy would then agree exactly
with the one obtained through DMFT for the case of infinite-U. Our aim
here is to benchmark the lowest non-trivial order of the $\lambda$
expansion against the exact DMFT results.
Note that in the $d\rightarrow\infty$ limit, in the paramagnetic phase, 
the single-particle properties of the $t-J$ model are identical to those of the $U=\infty$ Hubbard model. 
In other words,  as long as antiferromagnetic correlations are
short-ranged, $J$ does not enter single-particle 
properties in the $d=\infty$ limit. Accordingly, the only coupling constant entering the simplified ECFL equations 
is the hopping (band dispersion) itself, and not the superexchange. 

In the $O(\lambda^2)$ scheme, the explicit density factors $\swo$ that occur in  Eq.~\eqref{eq8} and \eqref{eq9} are replaced by the rule 
\barray
\swo \to \sw \equiv
1- \lambda \frac{n}{2}+ \lambda^2 \frac{n^2}{4} + O(\lambda^3). \label{adef}
\earray
The second rule is that the explicit self-energy expressions in these equations are multiplied by $\lambda$.
As an illustration of these rules,  we write Eq.~\eqref{eq8} and \eqref{eq9} as
 \barray
\GHI(k) & = & i\omega+\chem- \sw \ \varepsilon_k- \lambda \Phi(k), \label{eq-8prime} \\
\widetilde{\mu}(k)&=& \sw + \lambda \ \Psi(k). \label{eq-9prime}
\earray
The factor $\lambda$ is set  to $1$ before actually computing with these formulas. 
The two self-energy  functions $\Phi(k)$ and $\Psi(k)$ satisfy the equations to second order in $\lambda$:
\barray
 \Psi(i \omega_k) & =& - \lambda  \sum_{pq}(\varepsilon_p+\varepsilon_q-u_0)\GH(p)\GH(q)\GH(p+q-k), \nn \\
 \Phi(k)& =& (\varepsilon_k - \frac{u_0}{2})  \Psi(i \omega_k) +\chi(i \omega_k) - u_0 (\lambda \frac{n^2}{8}   -  \frac{n}{2})\nn \\ && -  \sum_p \varepsilon_p \GH(p) \nn  \\
 \chi(i \omega_k)&=& - \lambda \sum_{pq}(\varepsilon_{p+q-k}-\frac{u_0}{2})(\varepsilon_p+\varepsilon_q-u_0) \times \nn \\
&& \GH(p)\GH(q)\GH(p+q-k) \label{chidef}.
\earray
All equations in \disp{chidef} are implicitly understood to have $O(\lambda^3)$ corrections, so that the $\GH$ and $\widetilde{\mu}$ pieces of $\G$ in \disp{eq7}  are correct to the stated order.
As expected,  the  functions $\Psi,\chi$ depend on the frequency but not the momentum $\vk$.
Both Green's functions satisfy an identical  number sum rule $\sum_{k,
\omega_n} \G(k) = \frac{n}{2} = \sum_{k, \omega_n} \GH(k)  $, and the
theory has two chemical potentials necessary to impose these, namely
$\chem$ and $u_0$. As discussed in Ref.~\onlinecite{Monster}, the
second chemical potential $u_0$ arises from the requirement of
satisfying a ``shift invariance'' in the theory. The shift transformation in the present model  acts as $\varepsilon_p \to \varepsilon_p + c$. This  transformation  shifts the center of gravity of the band; it is absorbable in $u_0$, and thus rendered inconsequential.  
We can easily verify that  $\chi$ and $\Psi$ are independently  shift-invariant.  Combining the expressions, we write
\barray
\chem' &\equiv&\chem +u_0 \left( \lambda^2 \frac{n^2}{8} -  \lambda \frac{n}{2} \right) - \frac{u_0}{2} \sw + \lambda\sum_p \varepsilon_p \GH(p), \nn \\
\GH^{-1}(k)&=&\! i\omega+\chem'-   (\varepsilon_k - \frac{u_0}{2})\left\{\sw+ \lambda \Psi(i \omega_k)  \right\} - \lambda  {\chi}(i \omega_k). \nn \\ \label{gdefred}
\earray
The Green's function is then found by combining \disp{gdefred}, \eqref{chidef} and \eqref{eq-9prime} in the  expression  \disp{eq7}.

\subsection{Setting up the computation }

To set up the computation, we write a local Green's function with weight  $m=0,1,\ldots$ using \disp{gdefred} as  
\barray 
&&\GH_{loc,m}(i\omega_k) \equiv \sum_{\vk} \GH(k) \ (\varepsilon_{\vk})^m \nn \\
&=&   \int_{-D}^D  \! d\varepsilon \  \rho_{0}(\varepsilon) \  \frac{\varepsilon^m}{i\omega+\chem'- (\sw+ \Psi) (\varepsilon - \frac{u_0}{2}) -{\chi}}, \nn \\ \label{gloc}
\earray
where $\chi$ and $\Psi$ are functions of frequency $i \omega_k$ but not the energy $\varepsilon$.
 We find that   both $\GH_{loc,0}$ and $\GH_{loc,1}$  are  needed to
 compute the frequency-dependent self-energy. Similarly, a local $\G$ can be defined, and the number sum rules can be written as $ \frac{n}{2} = \sum_{i\omega} \GH_{loc,0}(i\omega)$
 and $ \frac{n}{2} = \sum_{i\omega} \G_{loc,0}(i\omega)$. 
 Where necessary, the usual convergence factor $ \ e^{i \omega 0^+}$ is inserted.
  The  two $\vk$ independent  functions $\Psi$ and $\chi$ in   \disp{chidef} can be written in a compact way if we first define a function with three indices  $(m_1 m_2 m_3)$ from the weight factors:
\barray
&&I_{m_1 m_2 m_3}(i\omega) =-\frac{1}{\beta^2}\sum_{\nu_1,\nu_2}\ \GH_{loc,m_1}(i\nu_1) \times \nn \\
&& \GH_{loc,m_2}(i\nu_2)\GH_{loc,m_3}(i\nu_1+i\nu_2-i\omega). \label{Isdefined}
\earray
 After continuation $i \omega \to \omega + i \eta$, and for all values of the indices, 
 the  low  frequency and temperature  $I(i\omega)$ is a Fermi-liquid-like self-energy with an imaginary part   $\propto \left(\omega^2+(\pi k_B T)^2\right)$. We can now rewrite \disp{chidef} as:
\barray
\Psi(i\omega)&=&-u_0 I_{000}(i\omega)+2I_{010}(i\omega) \nn \\
\chi(i\omega)&=&-\frac{u_0}{2}\Psi(i\omega)-u_0I_{001}+2I_{011}(i\omega). \label{psicompute}
\earray
Clearly, Eqs.~\eqref{psicompute} and \eqref{gloc} along with the definition \eqref{Isdefined} and the number sum rules 
form a self-consistent set of equations that can be solved  iteratively on a computer. The Dyson self-energy and the spectral function can be computed in terms of these quantities using \disp{dyson}

\subsection{Auxiliary and Dyson self energies to $O(\lambda^2)$}

\begin{figure}[htbp!]
\includegraphics[width=\columnwidth,keepaspectratio]{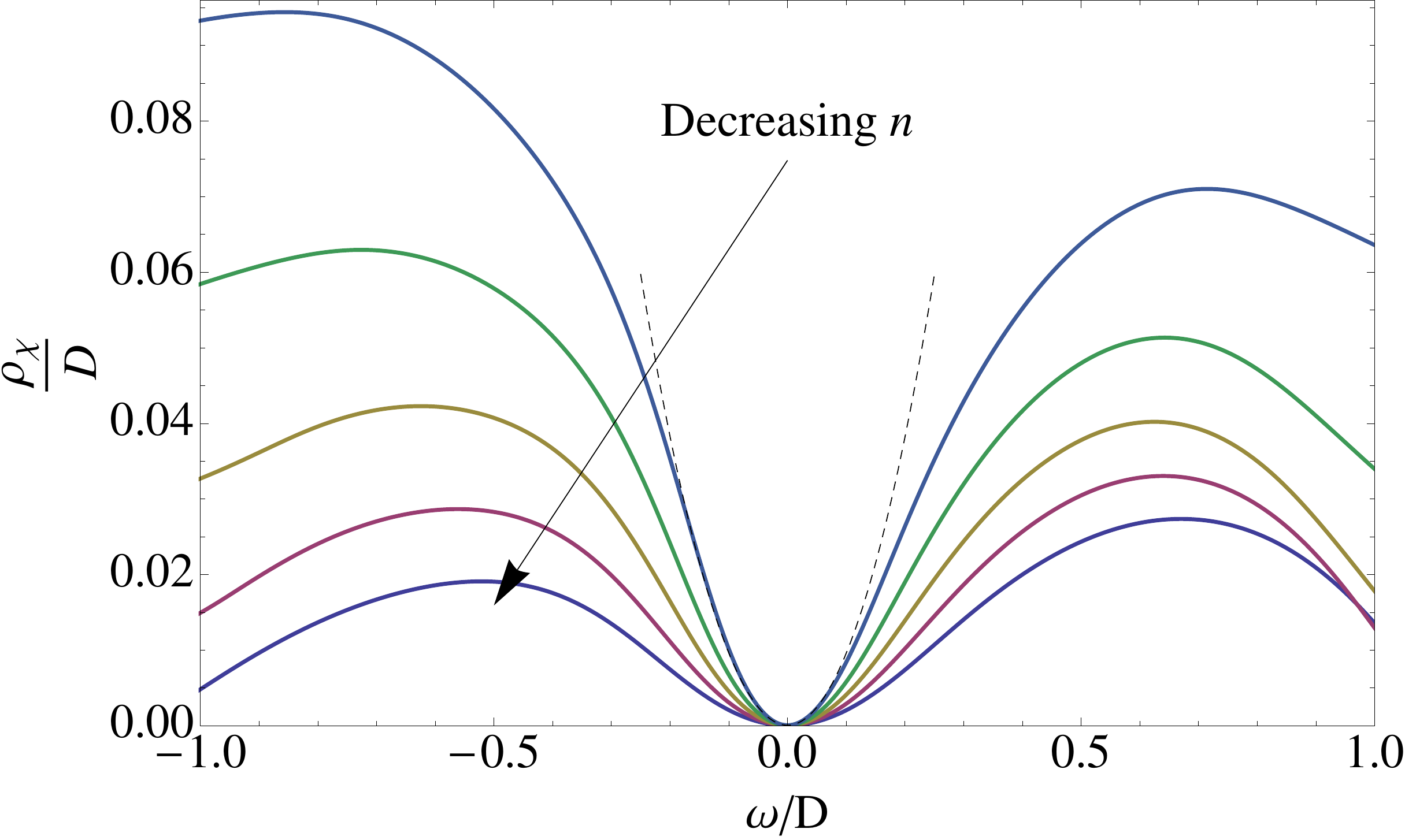}
\includegraphics[width=\columnwidth,keepaspectratio]{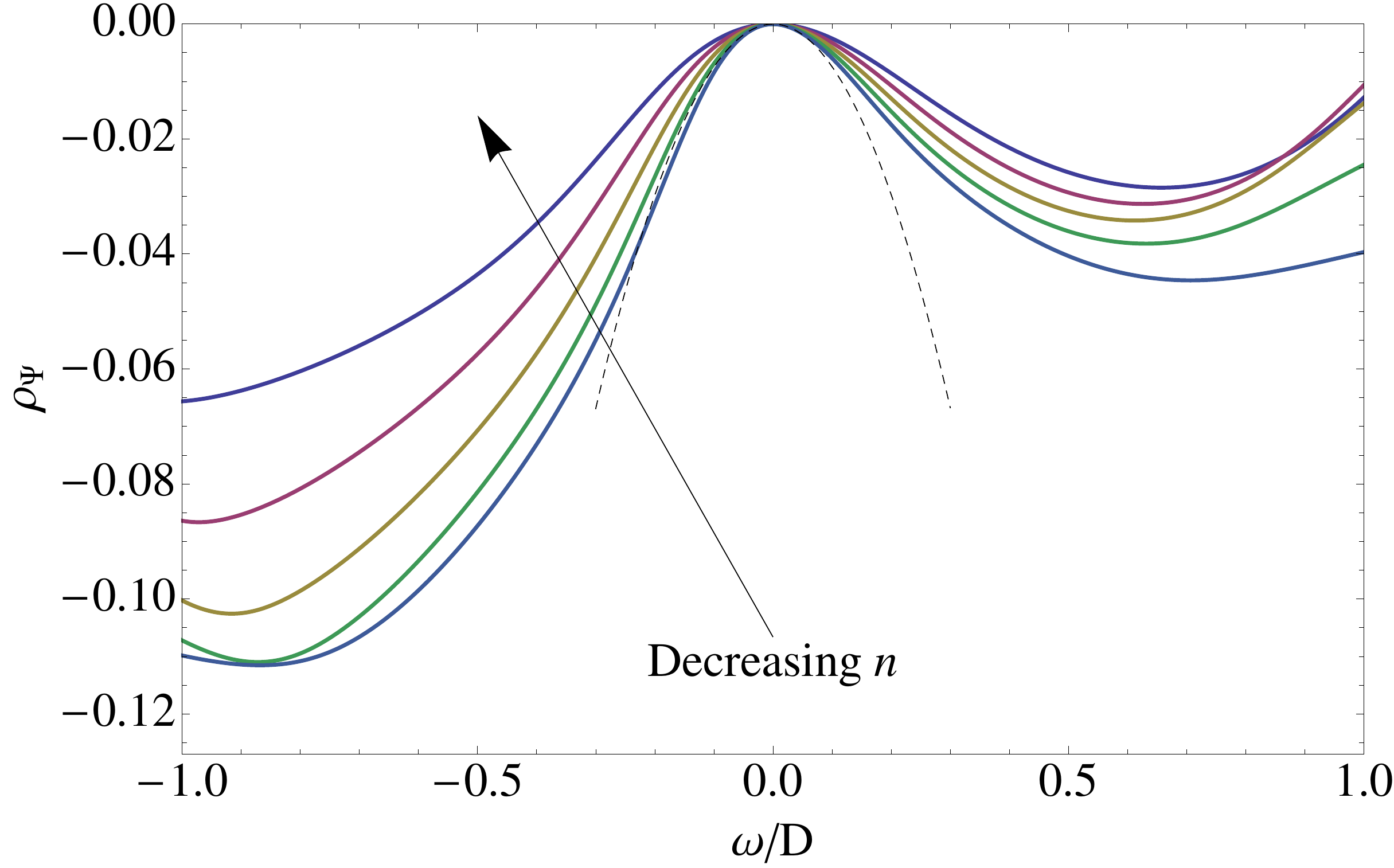}
\includegraphics[width=\columnwidth,keepaspectratio]{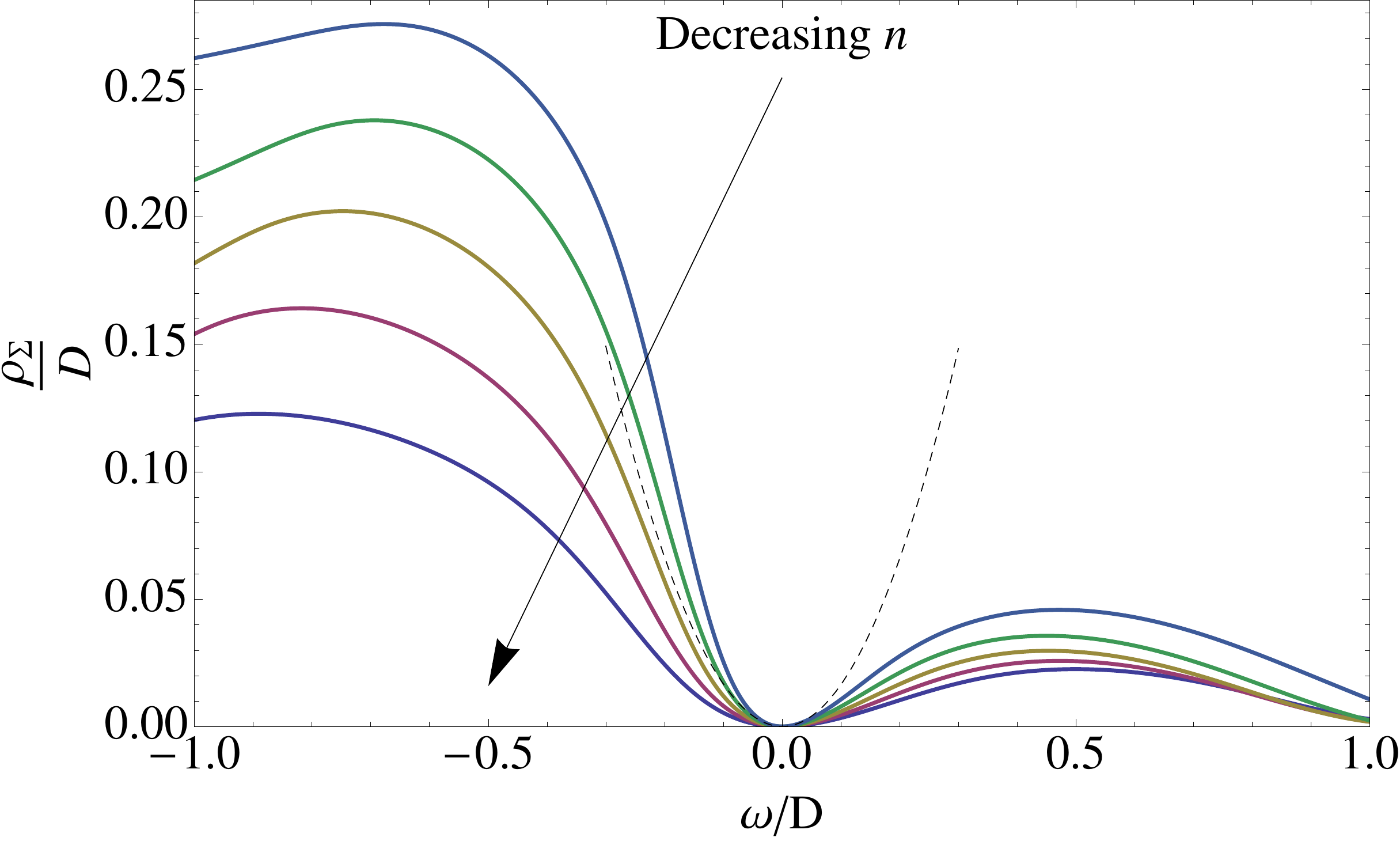}
   \caption{\label{fig:self_all} (Color online)
   Imaginary parts (spectral densities) of the auxiliary self-energies
   $\chi$ and $\Psi$, and the Dyson
   self-energy $\Sigma$ within the $O(\lambda^2)$ ECFL. The dotted lines  are
   parabolic fits at the highest density. Recall that  typical
   Fermi-liquid-type spectral functions exhibit a parabolic and
   therefore particle-hole symmetric behavior over a large energy range.
From these fits   one observes that the auxiliary functions $\Psi$ and $\chi$ have a Fermi-liquid form 
over a wider energy range than the Dyson self-energy $\Sigma$.       }
 \end{figure}

In Fig.~\ref{fig:self_all} we present $\rho_\chi$, $\rho_\psi$, and
$\rho_\Sigma$ (from top to bottom). $\rho_\psi$ and $\rho_\chi$ have
similar frequency-dependence and a Fermi-liquid form is obeyed much
more accurately than what is found for the
Dyson self-energy $\rho_\Sigma$. This supports the ansatz that we
employed above.  In particular, the auxiliary self energies are more
particle-hole symmetric; most of the particle-hole asymmetry follows
from the structure of ECFL equations. This signals that the
Fermi liquid concept has validity outside of the canonical Fermi
liquid behavior.

\section{Detailed Comparison of $O(\lambda^2)$ ECFL results to DMFT} 
\label{sec:lambda_dmft}

\subsection{The effective density of the ECFL spectral functions and its phenomenological adjustment}

The $O(\lambda^2)$ equations of ECFL discussed here give a
high-$\omega$ limiting behavior $\G \sim \frac{\sw}{\omega}$,
differing from the exact form $\G \sim \frac{\swo}{\omega}$ due to the
replacement $\swo \to \sw \equiv 1-\lambda n/2+\lambda^2 n^2/4$ as per
the rules of the calculation. This effect is due to the incomplete
projection of the $O(\lambda^2)$ treatment of the ECFL equation of
motion.  At $n\sim 0.75$ the error in the high-frequency weight is
$22.5\%$. 

A phenomenological scheme for adjusting for this feature defines an
effective density $n_{\mathrm{eff}}$, using the ratio of particle addition and
removal states as the relevant metric, so that
$\frac{n}{1-n+n^2/4}=\frac{n_{\mathrm{eff}}}{1-n_{\mathrm{eff}}}$, thus
yielding
\beq 
n_{\mathrm{eff}}=\frac{n}{1+\frac{n^2}{4}}. \label{neff} \eeq
Clearly higher order calculations would have a corresponding mapping
between the two densities. For several of the comparisons below,
agreement is greatly improved by plotting the results of ECFL as a
function of $n_{\mathrm{eff}}$.

\subsection{Comparison between $O(\lambda^2)$-ECFL and DMFT}

We find that the computed values of the quasiparticle weight $Z$ from
ECFL are close to the $U/D=4$ DMFT curve, we detail this in the
Appendix \ref{AppA}, where the momentum distribution is also shown. 
This is suggestive of an analogy between the two incompletely
projected theories. In particular, making $U$ finite, and truncating
the $\lambda$ expansion at second order, both introduce some double
occupancy into the system. It is therefore not surprising that the
$U/D=4$ DMFT results agree better with the $O(\lambda^2)$-ECFL than
with the $U/D=\infty$ DMFT results.  However the limitations of the
$O(\lambda^2)$ calculation within ECFL preclude obtaining reliable
results for doping levels smaller than $\delta\approx 0.25$. 

\subsubsection{Spectral lineshapes}

\begin{figure*}
\includegraphics[width=\columnwidth]{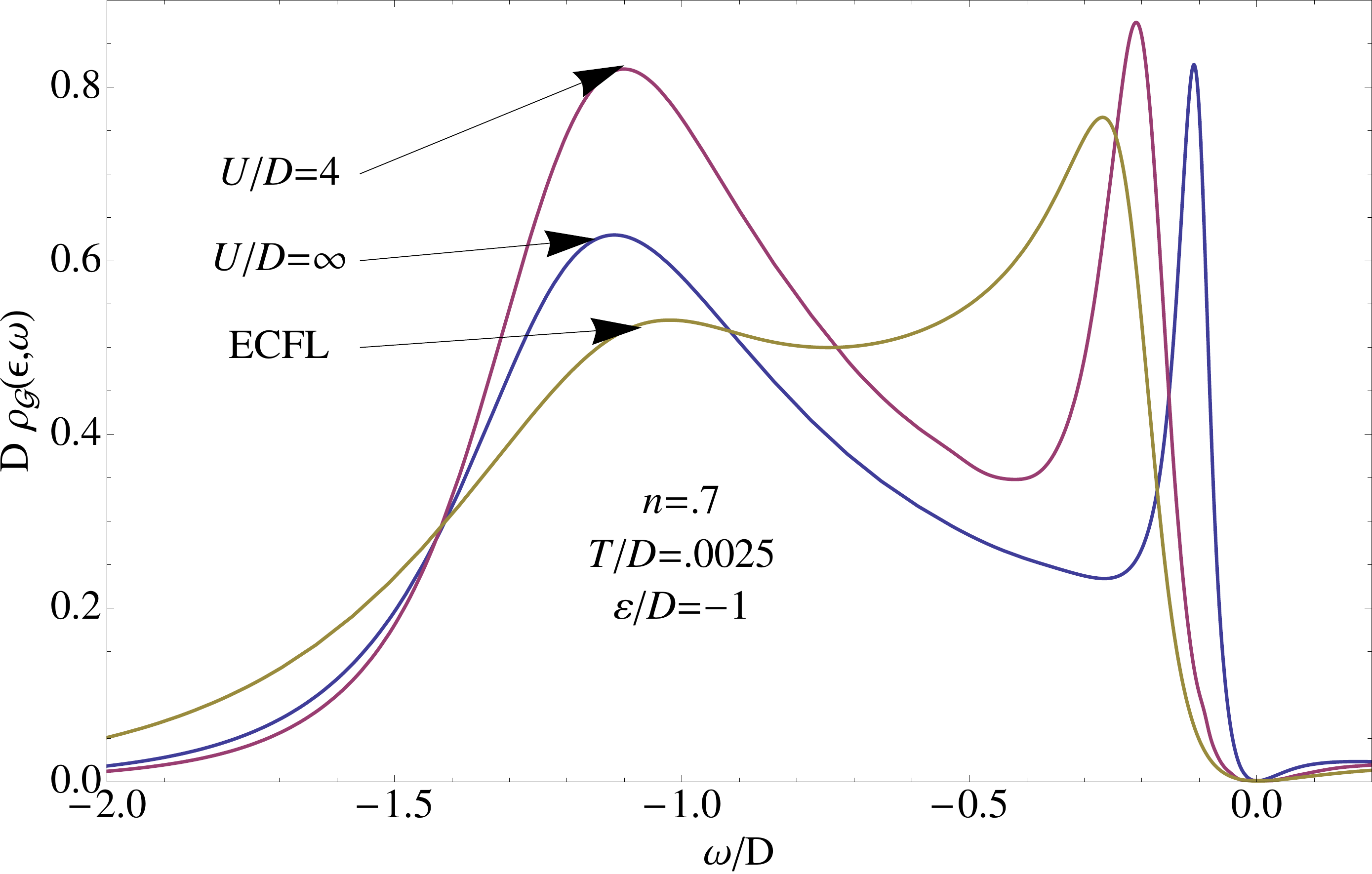}
\includegraphics[width=\columnwidth]{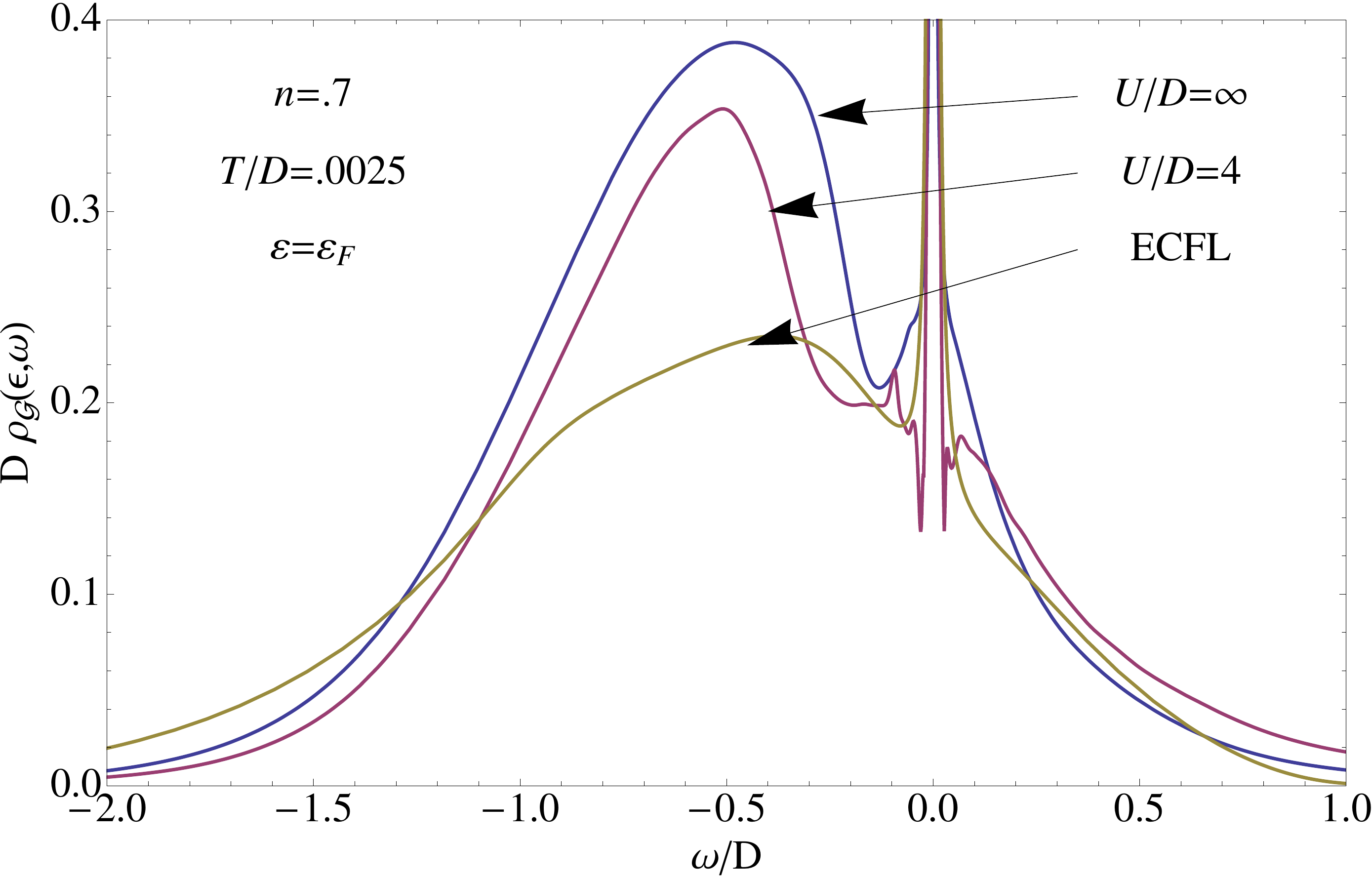}
\caption{(Color online) Spectral functions within the DMFT ($U=\infty$
and $U/D=4$) and ECFL at
two typical energies $\varepsilon=-D$ and 
$\varepsilon=\varepsilon_F$, with $n=0.7$ and $T=0.0025 D$.  The
location of the quasiparticle peak near $\omega \sim 0$ and the broad
secondary peak for $\omega <0$ are common to both calculations. While
there are subtle differences, especially in the magnitudes of the
secondary peaks, the main features of the three calculations match at
high and low frequency. } \label{EDC}
\end{figure*}

\begin{figure*}
\includegraphics[width=.95\columnwidth]{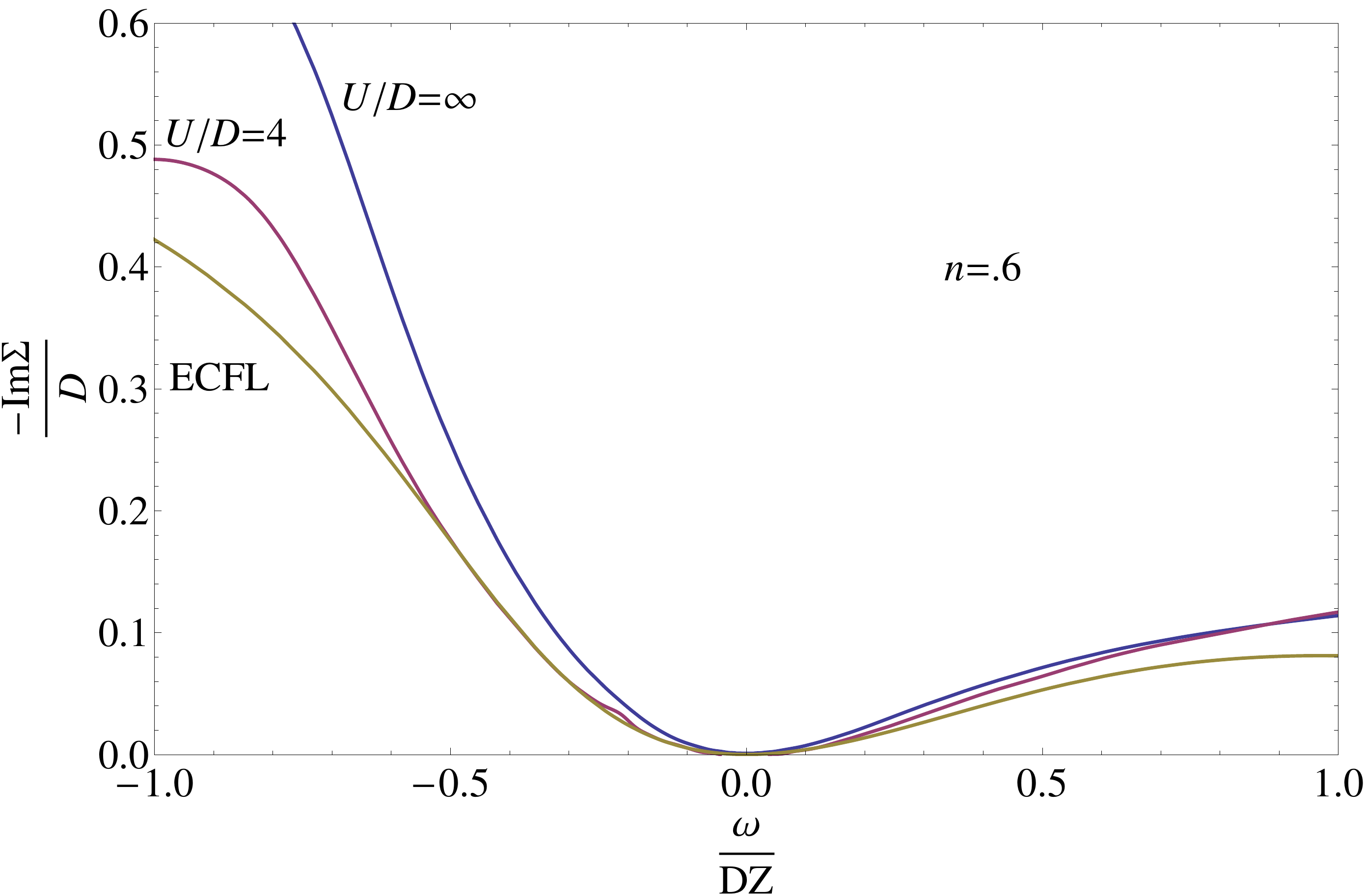}
\includegraphics[width=.95\columnwidth]{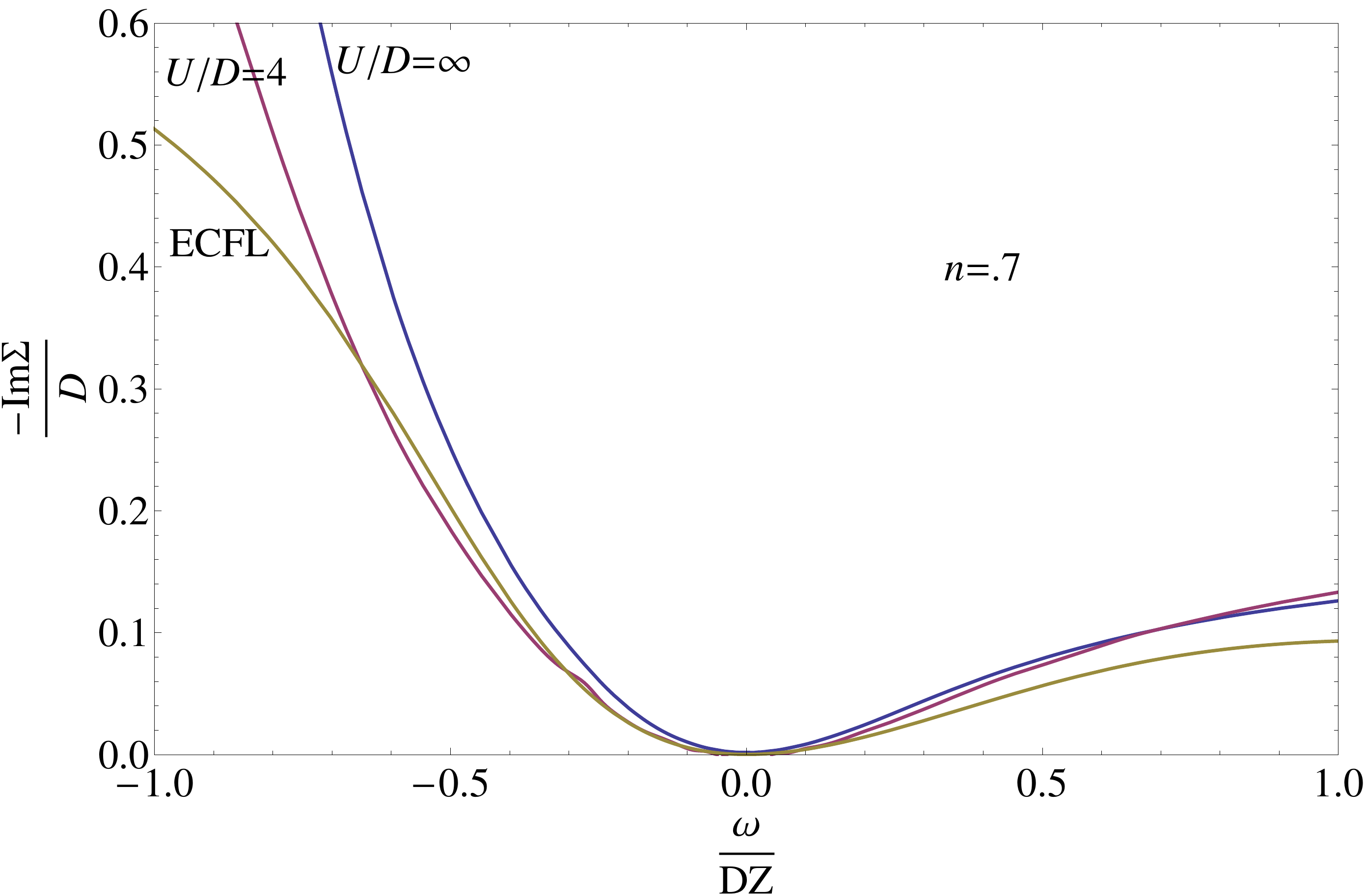}
\includegraphics[width=.95\columnwidth]{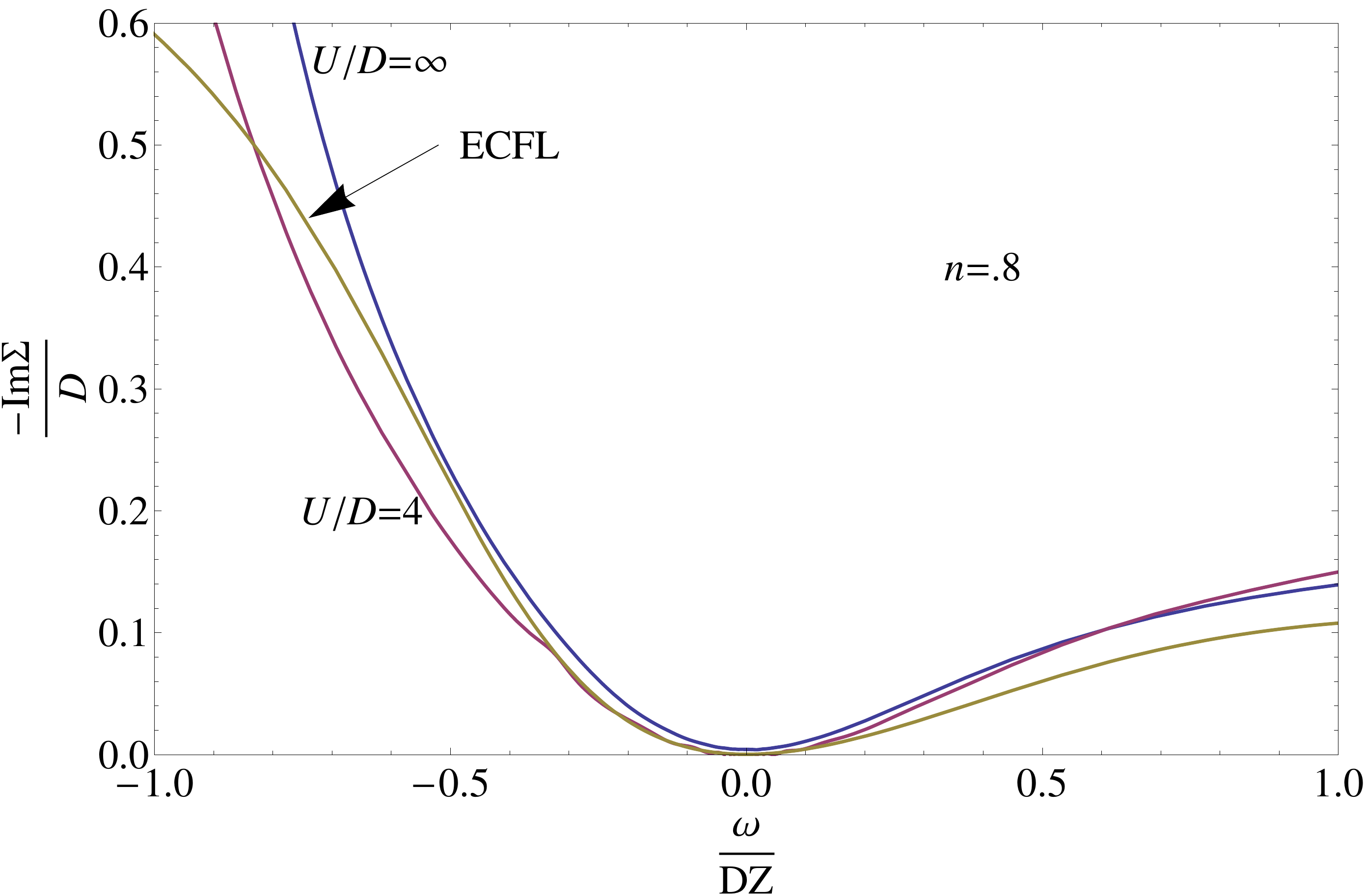}
\includegraphics[width=.95\columnwidth]{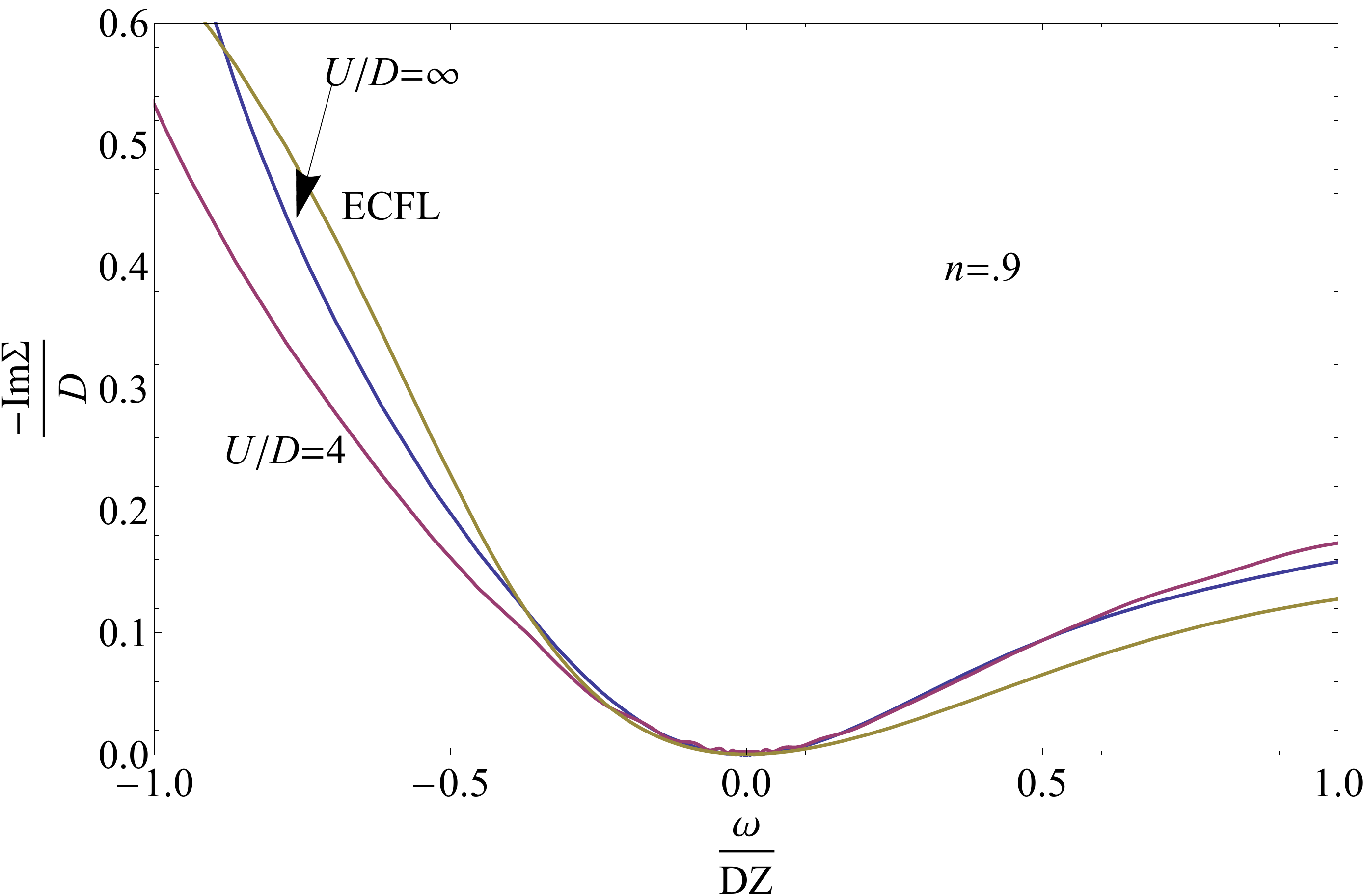}
\caption{(Color online) Spectral function (imaginary part) of the
Dyson self-energy $\Sigma$ versus the scaled variable $\omega/(D Z)$
in the ECFL theory at order $\lambda^2$, and the DMFT at two values of $U$.
The ECFL predicts a value of Z which is too large at low doping, and
significant U dependence creates differences between the $U/D=4$ and
$U=\infty$ results of the DMFT. Nonetheless, all three cases overlap
well at low frequencies when plotted against the scaled frequency.
Surprisingly, this agreement survives to densities far beyond the
expected range of the current version of the  ECFL. } \label{imsigmaZ}
\end{figure*}

In Fig.~\ref{EDC} we compare the ECFL and the DMFT results at $U=4 D$
and $U=\infty$ for the $\epsilon$-resolved spectral functions at two
values of the band energy, $\varepsilon_k = -D$ and
$\varepsilon_k=\varepsilon_F$.  In general, the agreement is
encouraging. At $\varepsilon_k= -D$ (left panel), DMFT has a deeper
minimum between the QP and the secondary feature at high binding
energy than is seen in ECFL, but the position of the ECFL peaks agrees
well with that of the DMFT peaks. At $\varepsilon_F$ (right panel) the
QP are of similar width but have different values of $Z$, as discussed
above. The background of width $\sim D$ lies over essentially the same
frequency range for all three calculations, and has a peak at
$\omega=-0.5D$, approximately the same position for each data set.
However, the height of the peak is less pronounced for the ECFL than
the DMFT. At positive frequencies the spectral functions are in
excellent agreement.  Plotting the spectral function as a function of
the scaled frequency $\omega/ZD$ improves the agreement in the
position and width of the quasiparticle, as illustrated in the more
sensitive self-energy curves in Fig.~\ref{imsigmaZ}. We note that the
scaled ECFL curves agree well with the DMFT curves even for density $n
\sim 0.8 - 0.9$ for scaled frequency $|\omega| \leq 0.5 D Z$. We find
this agreement surprising in view of our criterion discussed above,
placing $n \sim 0.75$ as the limiting density.  
 
\begin{figure*}
\includegraphics[width=.66\columnwidth]{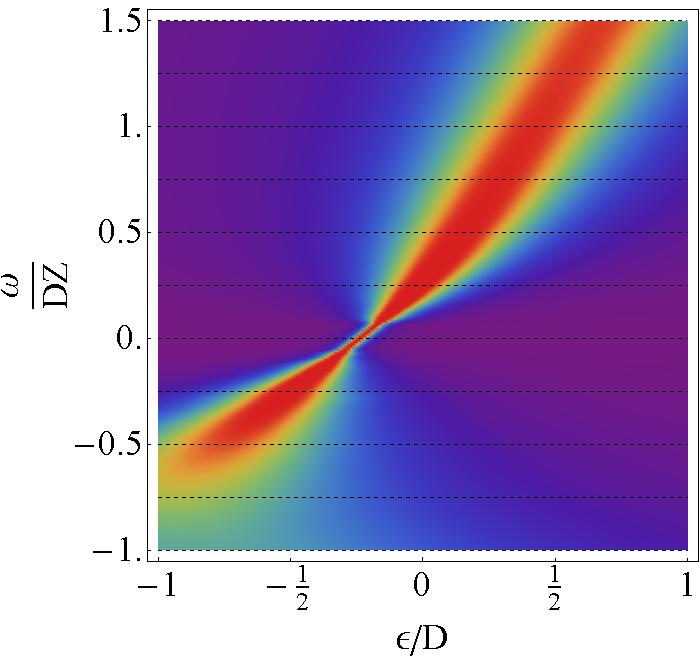}
\includegraphics[width=.66\columnwidth]{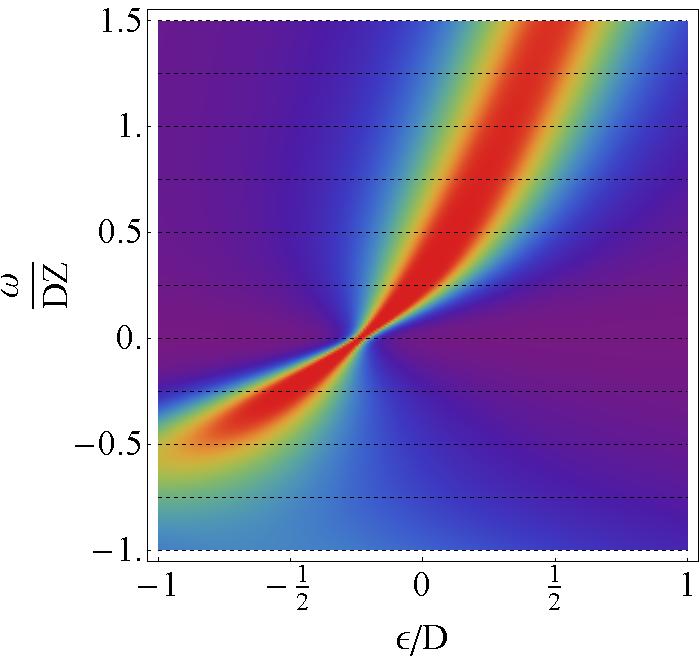}
\includegraphics[width=.66\columnwidth]{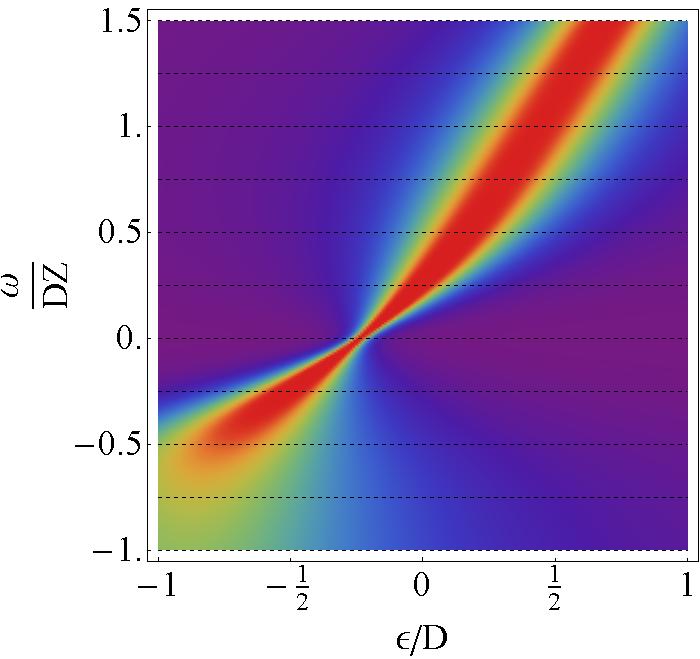}
\caption{(Color online) Physical spectral function $A(\epsilon,\omega)$. From left to
right: $U=4$ DMFT, $U=\infty$ DMFT, and ECFL with $n=0.7$ and
$T/D=0.0025$. Hot colors represent high intensity, while darker blue
represents low intensity. Noting from the left panel of \figdisp{EDC}
that the QP band has a slightly different width in each calculation,
we plot the spectral function here as a function of
$\frac{\omega}{DZ}$. This brings the low energy (QP) features of the
spectral function into impressive agreement, indicating that Z, rather
than $\delta$, is the fundamental energy scale of the extremely
correlated state. } 
\label{specZ}
\end{figure*}

The physical spectral function $A(\omega)$, when displayed as a color intensity
plot using the scaled frequency $\omega/ZD$ as in \figdisp{specZ}, 
further emphasizes this similarity. At this level of description, the
$U= 4D$ DMFT curve and the $O(\lambda^2)$ calculation look almost
identical.
In particular, as clear from this figure, both theories indicate that
the quasiparticle peak becomes rapidly {\it more dispersive} as one
moves to positive energies, corresponding to unoccupied states (i.e.,
the effective Fermi velocity increases as compared to its low-energy
value and becomes closer to the band value).  As discussed above
(Sec.~\ref{sec:dmft}, Fig.~\ref{fig:eps_resolved_dmft}), this is one
of the primary common conclusions of both theories, which could be
tested in future experiments able to probe the unoccupied states in a
momentum-resolved manner. 
 
In view of the remarkable similarity between the different theories,
as seen in \figdisp{imsigmaZ} and \figdisp{specZ}, it appears that the
$O(\lambda^2)$ version of ECFL has the correct {\em shape} of the
spectra built into it, but requires a correction for a too large value
of the QP factor $Z$.  This is the main conclusion of this work
regarding the benchmarking of the ECFL. 
 
\section{Conclusion and prospects}
\label{sec:conc}

In this work we have presented a detailed comparison between the DMFT
and the ECFL theories, applied to the doped Hubbard model at large as well as infinite $U$, 
in the limit of infinite dimensions (Bethe lattice with infinite coordination).    

Our approach here is two-fold. On the one hand, we have used the 
general structure of the Green's function and self-energy in the ECFL theory to 
obtain a useful analytical ansatz which reproduces quite well the rich and complex 
frequency-dependence of the DMFT self-energy at not too low doping level.
This ansatz relies on the lowest-order Fermi-liquid expansion of the two auxiliary 
ECFL self-energies $\Psi$ and $\chi$. 
Quite remarkably, the marked deviations from the Fermi-liquid form and the 
particle-hole asymmetry found in the physical single-particle self-energy can be accounted for 
by this underlying Fermi-liquid form of auxiliary quantities.
In turn, the deviations observed between the DMFT results and this lowest-order ansatz at lower 
doping levels emphasize the need for corrections to FL behavior in
$\Psi,\chi$ within the ECFL.
This part of our study thus provides useful analytical insights into
the DMFT description of the doping-driven Mott transition. 

On the other hand, we have used the DMFT results (obtained here with a high-accuracy NRG solver) 
as a benchmark of the ECFL theory. 
Specifically, we have solved numerically the $O(\lambda^2)$ ECFL equations, 
appropriately simplified in the limit of large dimensions. 
For not too low doping levels, where this $O(\lambda^2)$ scheme is applicable, 
we found that the spectral properties agree well provided the comparison is 
made as a function of the scaled frequency $\omega/ZD$, with $Z$ the quasiparticle weight.  
A similar situation arises in comparing the ECFL method for the Anderson impurity model,  
where $Z$ is rapidly suppressed as the Kondo limit is approached\cite{AIM}. 
This adjustment of the frequency scale compensates the
known weakness of the $O(\lambda^2)$ theory in obtaining $Z$ quantitatively, and
enables, to some extent, a preview of the results of the planned higher-order calculations in the ECFL 
projection parameter $\lambda$. 

From a physics point of view, we now summarize the most significant insights provided by our study. 

Doped Mott insulators are found to be characterized  by a marked particle-hole {\it dynamical asymmetry}, as emphasized 
in recent ECFL\cite{Asymmetry} and DMFT\cite{deng_2013} studies. 
In the case of hole-doping, particle-like ($\omega>0$) excitations are longer lived than hole-like ($\omega<0$) ones, leading to 
more `resilient' electron-like quasiparticles\cite{deng_2013}.   
This dynamical asymmetry has physical implications for the spectral lineshapes\cite{Asymmetry,deng_2013} as 
well as thermopower\cite{Haule_proc09,deng_2013}. 
The asymmetric terms in the low-frequency expansion of the self-energy signal deviations 
from the Fermi-liquid theory which are usually ignored in weak-coupling studies. They become 
large at low  hole doping and strong coupling, as demonstrated here in considerable detail, thus 
confirming the proposal made originally in Ref.~\onlinecite{Asymmetry}.

Due to the importance of this asymmetry, we found that the energy vs. momentum dispersion of the 
quasiparticle state quickly deviates on the $\omega>0$ side from its low-energy value (associated with 
the renormalized effective Fermi velocity). The deviation is towards a weaker dispersion, closer to the 
bare band value. This is a prediction of both ECFL and DMFT which could be tested experimentally 
once momentum-resolved spectroscopies are developed in order to address unoccupied states 
(the `dark side' for photoemission). 

Regarding ARPES lineshapes, we also emphasize that the recent successful comparison\cite{Gweon-Shastry,Kazue}  
between the ECFL and the experimental ARPES lineshapes in the optimally doped and overdoped cuprates along the nodal
direction  can just as well be interpreted as the similar success of the DMFT interpretation of these lineshapes.  
The adjustment of the momentum dependence of the caparison
factor for different systems in Refs.~\onlinecite{Gweon-Shastry,Kazue}
hints at the importance of the momentum-dependence of the self-energy. This 
momentum-dependence is already present in the ECFL in two dimensions, 
and also emerges from cluster DMFT calculations. 

Further comparison between 
the nature of the momentum dependence in both theories is to be addressed in future work. 
More generally, we believe that this work lays the foundation of a useful
program where the momentum-dependent self-energies can be reliably
computed and expressed in simple analytic forms. While cluster DMFT methods can already provide some answers
to this important problem, the ECFL theory readily treats low
dimensions and the momentum dependence. In order to get further solid
results, the current limitation of the ECFL  to the somewhat overdoped
regime needs to be overcome.
This limitation  arises from the low order of the expansion in $\lambda$, and 
 brute-force higher order calculations  in $\lambda$ are planned. In this task, the insights gained from the present comparison with DMFT,  are invaluable.
 
\acknowledgments{
A.G., J.M. and B.S.S. acknowledge useful discussions with Xiaoyu Deng at the initial stage of this project. 
The work at UCSC was  supported by  DOE-BES under Grant No. DE-FG02-06ER46319. 
A.G. and J.M. acknowledge the support of Ecole Polytechnique and Coll\`ege de France.  
R. \v{Z}. and J. M. acknowledge the support of the Slovenian Research Agency (ARRS)
under Program P1-0044.
}

\appendix
\section{Quasiparticle occupation and $Z_k$ in the $O(\lambda^2)$ ECFL. \label{AppA}}

\begin{figure}
\includegraphics[width=.85\columnwidth]{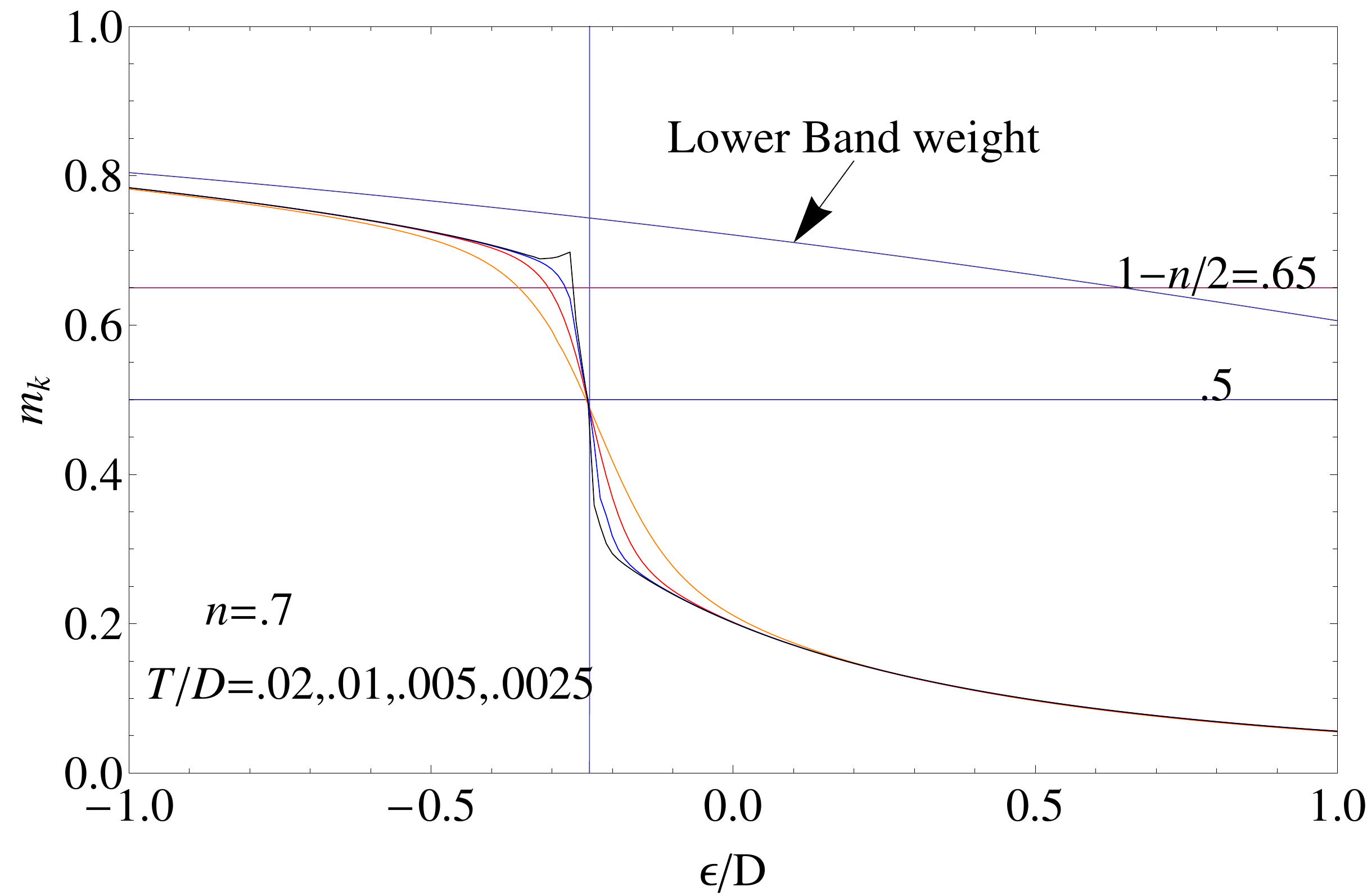}
\includegraphics[width=.85\columnwidth]{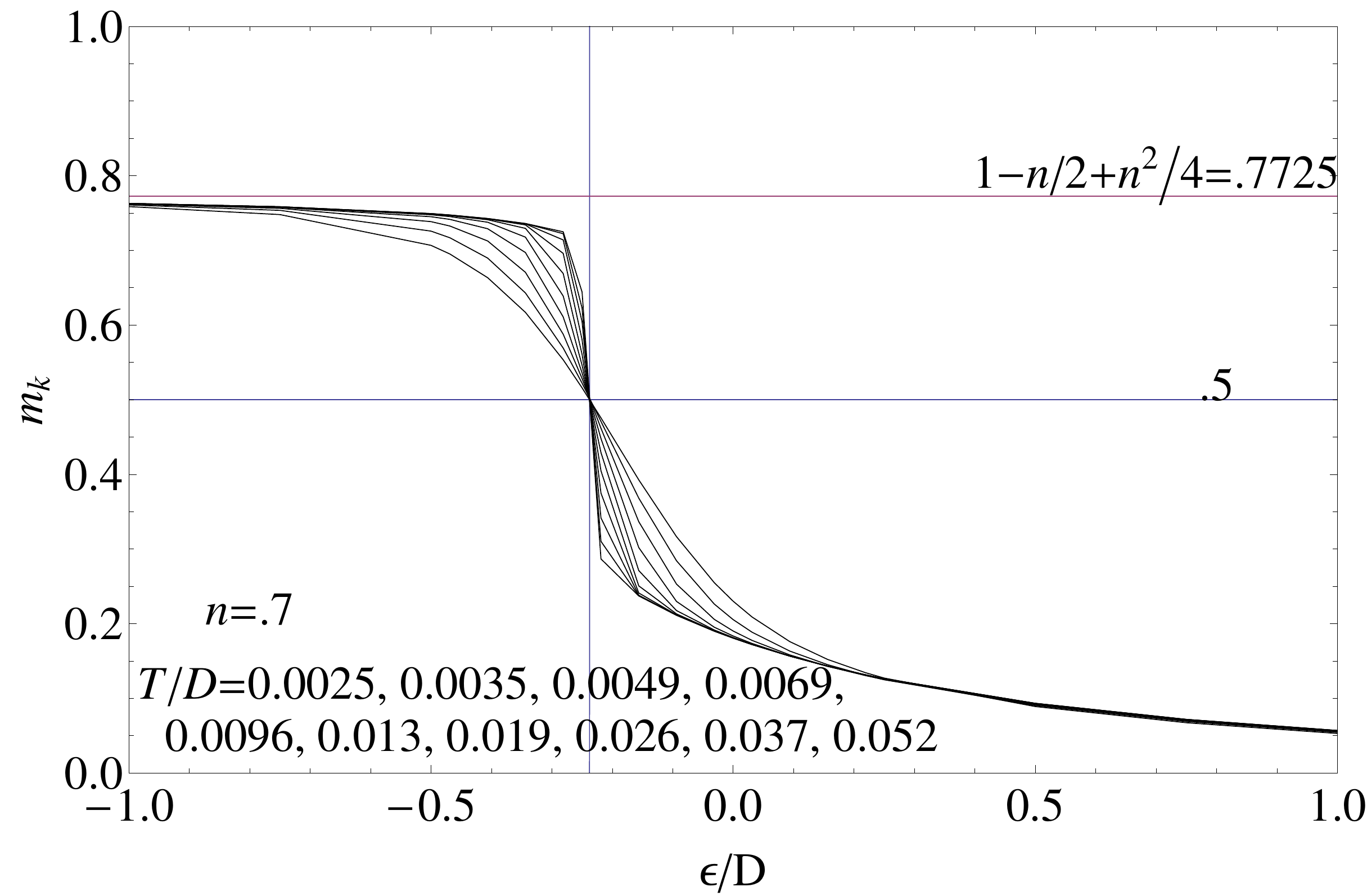}
\caption{ (Color online) Momentum occupation versus $\varepsilon$
within the DMFT at $U/D=4$ (top) and the ECFL (bottom).  } \label{mk}
\end{figure}

 The
momentum distribution function, Fig.~\ref{mk}, shows good agreement
with the DMFT at a density $n=.7$, and the large spillover for $k > k_F$ is of the same
scale in both sets of calculation. Its importance in estimating the
background spectrum in ARPES is well known, so this is already a
reasonably reliable common result. It is also interesting that at the
Fermi momentum, the magnitude of the distribution function is close to
$\frac{1}{2}$ in both calculations, as argued in the literature
\cite{Singh,Hansen-Shastry}. 

\begin{figure}[ht]
\includegraphics[width=0.9\columnwidth]{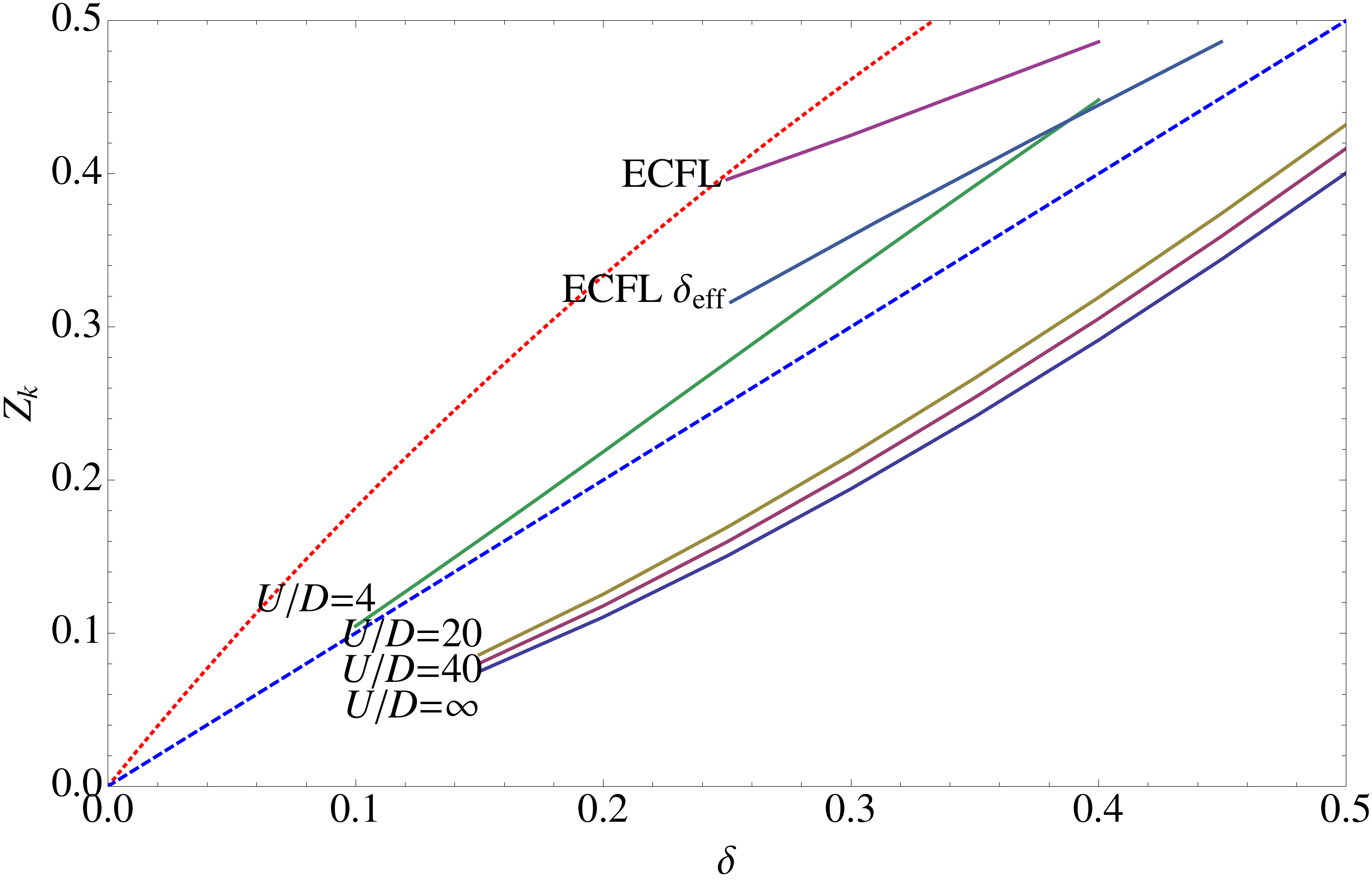}
\caption{(Color online) Quasiparticle weight $Z$ as a function of hole
doping $\delta$ (or $\delta_{\mathrm{eff}}=1-n_{\mathrm{eff}}$, the
effective hole doping) from the $O(\lambda^2)$ version of ECFL, and
from DMFT for various values of U. More detailed DMFT results are in
\figdisp{fig:z}. The blue dashed line represents $Z=\delta$, the
simplest $U=\infty$ slave-boson estimate, as a guide to the eye. The
dotted red line represents the Gutzwiller approximation result
$Z=\frac{2\delta}{1+\delta}$.} 
\label{zkfig}
\end{figure}

In Fig.~\ref{zkfig} we compare the quasiparticle weight $Z$ in ECFL
and DMFT.  The $O(\lambda^2)$ ECFL result has some similarity to the
Gutzwiller approximation \cite{Rice-Ueda} result $2 \delta/(1+\delta)$
in the limited density range of validity.  However it does not seem to
vanish in any obvious way, if we extrapolate by eye to higher density
$n$, highlighting its main weakness in the current state of
development, but plotted against the effective density it becomes
comparable to the $U/D=4$ DMFT curve over a limited range.

\section{NRG impurity solver convergence at small doping}
\label{appB}

\begin{figure}[htbp!]
\includegraphics[width=0.8\columnwidth,keepaspectratio]{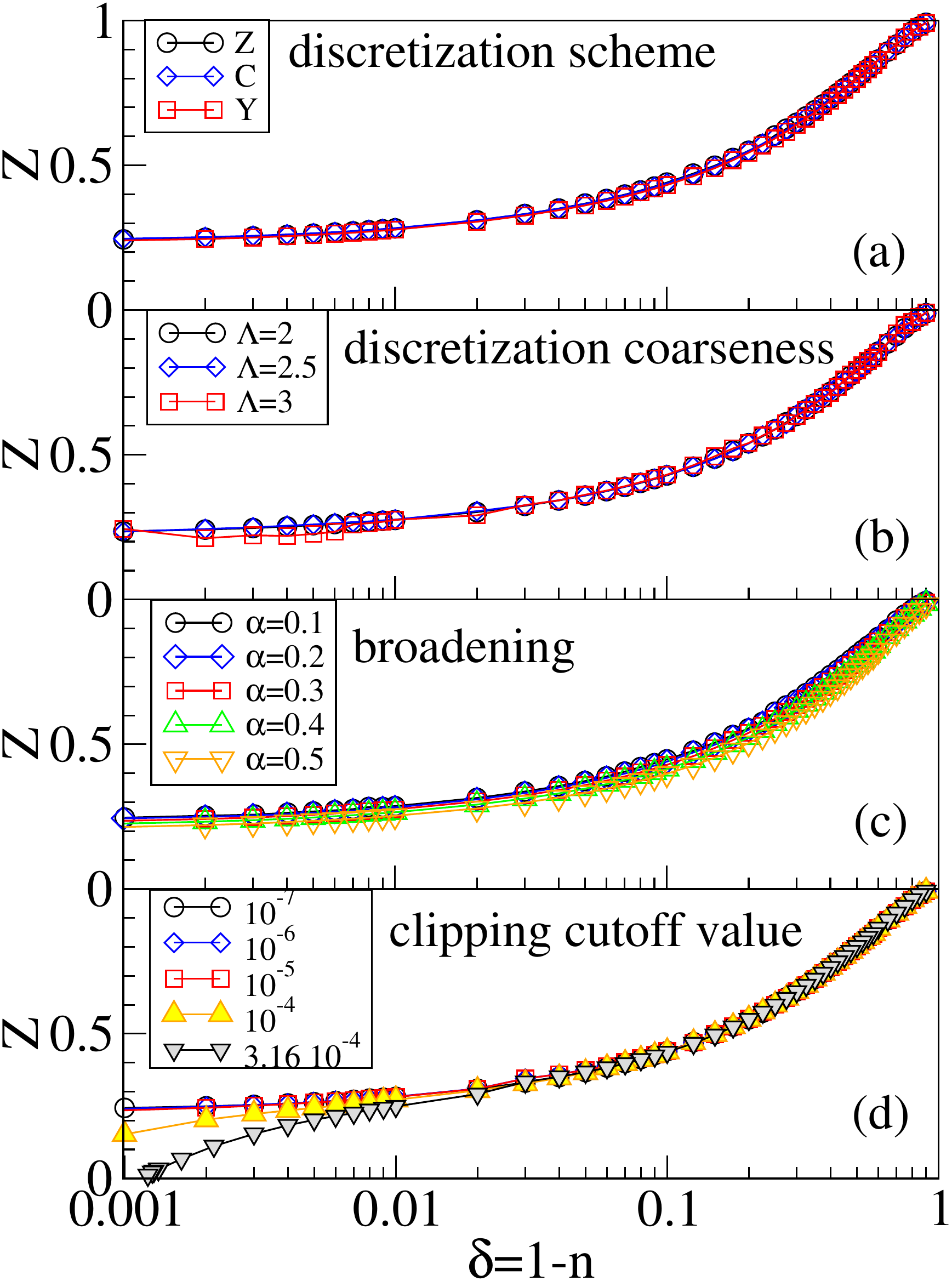}
   \caption{(Color online) \label{fig:nrg_comp1} Quasiparticle residue $Z$ for
   the $U=\infty$ Hubbard model as a function of doping $\delta$. We
   compare the DMFT results for different choices of the NRG impurity solver parameters.}
\end{figure}

\begin{figure}[htbp!]
\includegraphics[width=0.8\columnwidth,keepaspectratio]{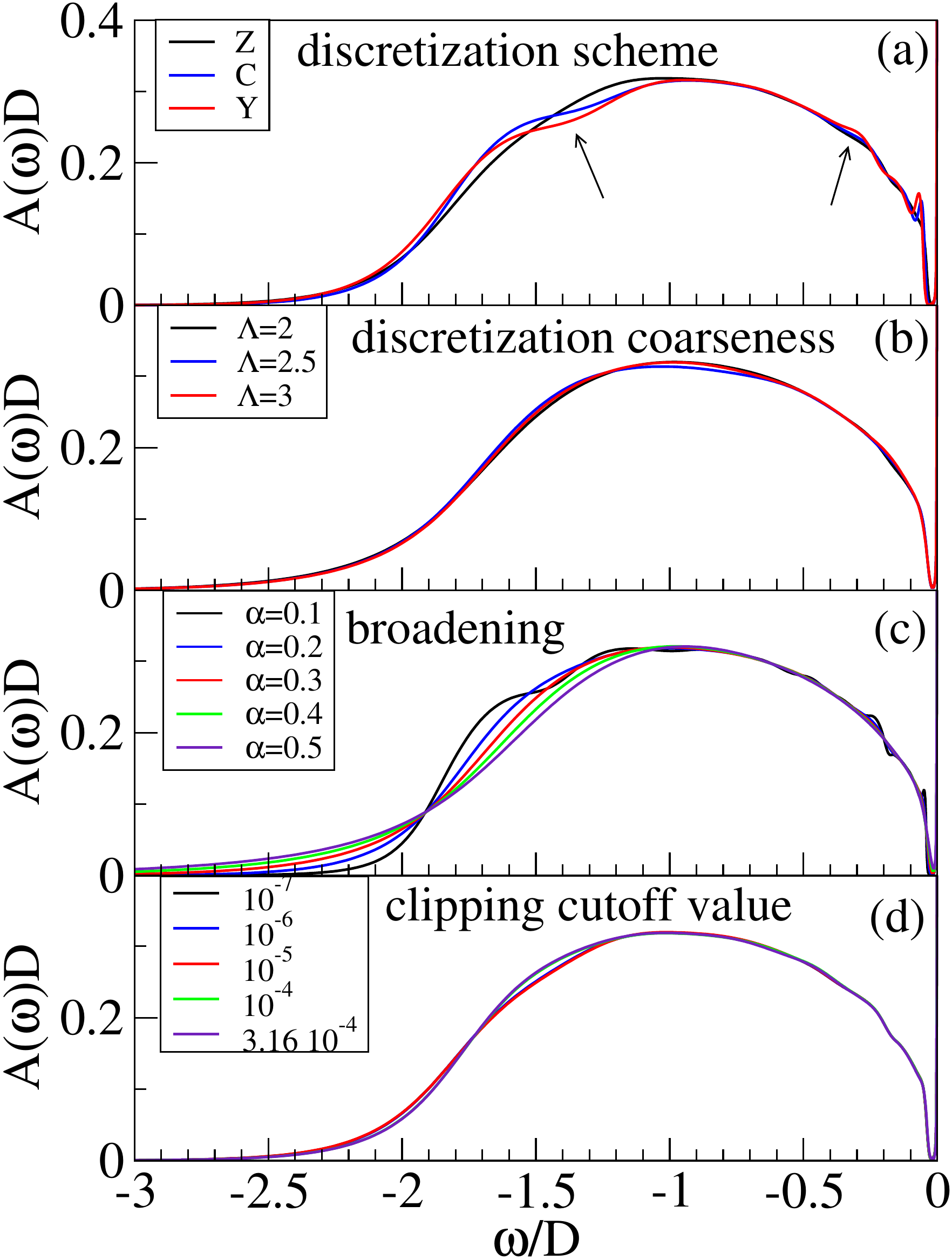}
   \caption{(Color online) \label{fig:nrg_comp2} Lower Hubbard band part of the
   local spectral function $A(\omega)$ for the $U=\infty$ Hubbard
   model at low doping, $\delta=0.01$. In (a), the schemes C and Y
   produce spurious features at the outer band edge and at
   $\omega=0.3D$ (indicated by arrows). In addition, the (expected)
   feature at the inner band edge at $\omega=0.1D$ is overemphasized
   in C and Y schemes. In (d), we note that the curves for clipping
   values $10^{-7}$, $10^{-6}$ and $10^{-5}$ nearly overlap, while
   those for $\geq 10^{-4}$ exhibit some deviations.}
\end{figure}

In order to obtain well-converged spectral functions using the NRG
impurity solver at low doping $\delta$, several parameters in the
method need to be apropriately tuned. Their choice affects both
low-frequency and high-frequency parts of the spectral functions. In
addition, it also significantly affects the numerical requirements --
both the duration of each NRG calculation and the number of the DMFT
cycles until self-consistency. Very close to the Mott transition,
obtaining fully converged results becomes computationally very
expensive (several hundreds of DMFT cycles) even with Broyden
acceleration \cite{resolution}. In this section, we explore the
effects of different choices on the quasiparticle residue $Z$
(low-frequency property), Fig.~\ref{fig:nrg_comp1}, and on the shape of
the LHB (high-frequency property), Fig.~\ref{fig:nrg_comp2}.

We first explore the choice of the discretization scheme (i.e., how
the coefficients of the Wilson chain are computed based from the input
hybridization function). We compare the discretization scheme (denoted
as Z) proposed by R. \v{Z}itko and Th. Pruschke in
Ref.~\onlinecite{resolution} which corrects the systematic
discretization errors near band edges present both in the conventional
discretization scheme (Y), Ref.~\onlinecite{yoshida1990}, and in the
improved scheme by V. Campo and L. Oliveira (C),
Ref.~\onlinecite{campo2005}. At low frequencies, one observes
excellent overlap of the results, as indicated in
Fig.~\ref{fig:nrg_comp1}a. This is in line with the common wisdom that
the NRG is a reliable method for low-frequency properties, having good
spectral resolution in the vicinity of the Fermi level where the
discretization grid is condensed. For this reason, the choice of the
discretization scheme has little effect on $Z$. At high frequencies,
however, one can clearly observe the systematic artifacts present in
schemes Y and C: the LHB presents spurious (non-physical) structure at
the outer edge which is not present in the results of scheme Z, see
Fig.~\ref{fig:nrg_comp2}a. Recent comparisons of the NRG (using scheme
Z) and continuous-time quantum Monte Carlo at finite temperatures have
established that the NRG (using scheme Z) is, in fact, a rather
reliable method also for high-frequency/finite-temperature properties.
On the other hand, the NRG using schemes Y or C is expected to exhibit
more pronounced systematic errors at high-frequencies and at finite
temperatures.

The second important choice concerns the value of the discretization
parameter $\Lambda$, which controls the coarseness of the logarithmic
grid. The standard choice is $\Lambda=2$, which is suitable to obtain
well converged results at both low and high frequencies, see
Figs.~\ref{fig:nrg_comp1}b and \ref{fig:nrg_comp2}b. In fact, the
results do not change much even when going to somewhat higher
$\Lambda=2.5$, while for $\Lambda=3$ we start to observe some
systematic deviations at very low doping $\delta$. We have also
performed some test calculations for smaller values $\Lambda=1.9, 1.8,
1.7$; the results differ little from those for $\Lambda=2$ while being
significantly more computationally expensive to produce.

We now consider the broadening parameter $\alpha$ which controls how
the raw spectral function in the form of a set of weighted delta peaks
is processed to obtain a smooth continuous representation. Too small
values lead to spurious oscillations, too high values to
overbroadening. These effects are nicely illustrated by the results
for the LHB part of the spectral funciton in
Fig.~\ref{fig:nrg_comp2}c. The long high-frequency tail of the LHB for
increasing $\alpha$ is a clear over-broadening effect, while the
oscillatory features for $\alpha=0.1$ are a discretization artifact.
At low frequencies, the QP residue $Z$ converges as $\alpha$ is
decreased, see Fig.~\ref{fig:nrg_comp1}c. We find that for $\alpha
\leq 0.1$, the results practically overlap, while for $\alpha=0.2$
(the value used for most calculations in this work), the deviation
from the asymptotic value is of order one percent. For large values of
broadening (as commonly done in NRG calculations), $Z$ is
underestimated. This is because the spectral weight is more spread
around as $\alpha$ increases, thus less weight remains in the QP peak.
Based on these results, we find that $\alpha=0.2$ is a good
compromise.

Finally, we discuss a subtle issue which becomes important at very
small dopings. The NRG discretization has difficulties if in the
hybridization function there are extended regions of very low values.
In particular, this leads to very slow approach to the
self-consistency. For this reason, it is convenient to use a small,
but finite cut-off value for the hybridization function to clip the
input hybridization function to some minimum value at all frequencies.
It is important, however, to choose this value so that the results are
not perturbed. We find that using too high cut-off leads to incorrect
$Z$ vs. $\delta$ behavior at low doping (a down-turn), see
Fig.~\ref{fig:nrg_comp1}d. The effect is thus similar to
overbroadening, since the spectral weight shifts from the
quasiparticle peak to the region between the LHB and the QP peak,
where the clipping is applied (for small $\delta$, where the LHB and
the QP peak no longer overlap, but rather the QP becomes an isolated
specgtral peak in the gap). There is also some effect of clipping on
the LHB itself, Fig.~\ref{fig:nrg_comp2}d. Again, this effect is
analogous to overbroadening.

\bibliography{paper}

\begin{thebibliography}{49}
\expandafter\ifx\csname natexlab\endcsname\relax\def\natexlab#1{#1}\fi
\expandafter\ifx\csname bibnamefont\endcsname\relax
  \def\bibnamefont#1{#1}\fi
\expandafter\ifx\csname bibfnamefont\endcsname\relax
  \def\bibfnamefont#1{#1}\fi
\expandafter\ifx\csname citenamefont\endcsname\relax
  \def\citenamefont#1{#1}\fi
\expandafter\ifx\csname url\endcsname\relax
  \def\url#1{\texttt{#1}}\fi
\expandafter\ifx\csname urlprefix\endcsname\relax\def\urlprefix{URL }\fi
\providecommand{\bibinfo}[2]{#2}
\providecommand{\eprint}[2][]{\url{#2}}

\bibitem[{\citenamefont{Georges et~al.}(1996)\citenamefont{Georges, Kotliar,
  Krauth, and Rozenberg}}]{georges_review_dmft}
\bibinfo{author}{\bibfnamefont{A.}~\bibnamefont{Georges}},
  \bibinfo{author}{\bibfnamefont{G.}~\bibnamefont{Kotliar}},
  \bibinfo{author}{\bibfnamefont{W.}~\bibnamefont{Krauth}}, \bibnamefont{and}
  \bibinfo{author}{\bibfnamefont{M.~J.} \bibnamefont{Rozenberg}},
  \bibinfo{journal}{Reviews of Modern Physics} \textbf{\bibinfo{volume}{68}},
  \bibinfo{pages}{13} (\bibinfo{year}{1996}).

\bibitem[{\citenamefont{Shastry}(2011{\natexlab{a}})}]{ECFL}
\bibinfo{author}{\bibfnamefont{B.~S.} \bibnamefont{Shastry}},
  \bibinfo{journal}{Phys. Rev. Lett.} \textbf{\bibinfo{volume}{107}},
  \bibinfo{pages}{056403} (\bibinfo{year}{2011}{\natexlab{a}}),
  \urlprefix\url{http://link.aps.org/doi/10.1103/PhysRevLett.107.056403}.

\bibitem[{\citenamefont{Shastry}(2013)}]{Monster}
\bibinfo{author}{\bibfnamefont{B.~S.} \bibnamefont{Shastry}},
  \bibinfo{journal}{Phys. Rev. B} \textbf{\bibinfo{volume}{87}},
  \bibinfo{pages}{125124} (\bibinfo{year}{2013}),
  \urlprefix\url{http://link.aps.org/doi/10.1103/PhysRevB.87.125124}.

\bibitem[{\citenamefont{Shastry}(2011{\natexlab{b}})}]{Anatomy}
\bibinfo{author}{\bibfnamefont{B.~S.} \bibnamefont{Shastry}},
  \bibinfo{journal}{Phys. Rev. B} \textbf{\bibinfo{volume}{84}},
  \bibinfo{pages}{165112} (\bibinfo{year}{2011}{\natexlab{b}}),
  \urlprefix\url{http://link.aps.org/doi/10.1103/PhysRevB.84.165112}.

\bibitem[{\citenamefont{Khatami et~al.}(2013)\citenamefont{Khatami, Hansen,
  Perepelitsky, Rigol, and Shastry}}]{Khatami-Shastry}
\bibinfo{author}{\bibfnamefont{E.}~\bibnamefont{Khatami}},
  \bibinfo{author}{\bibfnamefont{D.}~\bibnamefont{Hansen}},
  \bibinfo{author}{\bibfnamefont{E.}~\bibnamefont{Perepelitsky}},
  \bibinfo{author}{\bibfnamefont{M.}~\bibnamefont{Rigol}}, \bibnamefont{and}
  \bibinfo{author}{\bibfnamefont{B.~S.} \bibnamefont{Shastry}},
  \bibinfo{journal}{Phys. Rev. B} \textbf{\bibinfo{volume}{87}},
  \bibinfo{pages}{161120} (\bibinfo{year}{2013}),
  \urlprefix\url{http://link.aps.org/doi/10.1103/PhysRevB.87.161120}.

\bibitem[{\citenamefont{Gweon et~al.}(2011)\citenamefont{Gweon, Shastry, and
  Gu}}]{Gweon-Shastry}
\bibinfo{author}{\bibfnamefont{G.-H.} \bibnamefont{Gweon}},
  \bibinfo{author}{\bibfnamefont{B.~S.} \bibnamefont{Shastry}},
  \bibnamefont{and} \bibinfo{author}{\bibfnamefont{G.~D.} \bibnamefont{Gu}},
  \bibinfo{journal}{Phys. Rev. Lett.} \textbf{\bibinfo{volume}{107}},
  \bibinfo{pages}{056404} (\bibinfo{year}{2011}),
  \urlprefix\url{http://link.aps.org/doi/10.1103/PhysRevLett.107.056404}.

\bibitem[{Kaz()}]{Kazue}
\bibinfo{note}{Kazue Matsuyama and G.-H. Gweon, arXiv:1212.0299}.

\bibitem[{edw()}]{edward-shastry}
\bibinfo{note}{E. Perepelitsky and B. S. Shastry, ``Extremely Correlated Fermi
  Liquids in the limit of infinite dimension'', Annals of Physics (to appear)
  (2013), arXiv:}.

\bibitem[{\citenamefont{Deng et~al.}(2013)\citenamefont{Deng, Mravlje,
  \ifmmode~\check{Z}\else \v{Z}\fi{}itko, Ferrero, Kotliar, and
  Georges}}]{deng_2013}
\bibinfo{author}{\bibfnamefont{X.}~\bibnamefont{Deng}},
  \bibinfo{author}{\bibfnamefont{J.}~\bibnamefont{Mravlje}},
  \bibinfo{author}{\bibfnamefont{R.}~\bibnamefont{\ifmmode~\check{Z}\else
  \v{Z}\fi{}itko}}, \bibinfo{author}{\bibfnamefont{M.}~\bibnamefont{Ferrero}},
  \bibinfo{author}{\bibfnamefont{G.}~\bibnamefont{Kotliar}}, \bibnamefont{and}
  \bibinfo{author}{\bibfnamefont{A.}~\bibnamefont{Georges}},
  \bibinfo{journal}{Phys. Rev. Lett.} \textbf{\bibinfo{volume}{110}},
  \bibinfo{pages}{086401} (\bibinfo{year}{2013}),
  \urlprefix\url{http://link.aps.org/doi/10.1103/PhysRevLett.110.086401}.

\bibitem[{\citenamefont{Shastry}(2012{\natexlab{a}})}]{shastry-prl-reply}
\bibinfo{author}{\bibfnamefont{B.~S.} \bibnamefont{Shastry}},
  \bibinfo{journal}{Phys. Rev. Lett.} \textbf{\bibinfo{volume}{108}},
  \bibinfo{pages}{029702} (\bibinfo{year}{2012}{\natexlab{a}}),
  \urlprefix\url{http://link.aps.org/doi/10.1103/PhysRevLett.108.029702}.

\bibitem[{mat()}]{matho}
\bibinfo{note}{An early comment on Ref.~\onlinecite{ECFL} (K.~Matho, Phys. Rev.
  Lett. {\bf 108}, 029702 (2012)) and the response to this comment
  (B.S.~Shastry, Phys. Rev. Lett. {\bf 108}, 029702 (2012)) are noteworthy. In
  these works the momentum dependence of the Dyson self energy in the limit of
  large d was discussed, before the exact simplifications arising from that
  limit were known precisely. As shown in Ref.~\onlinecite{edward-shastry} and
  in the present work, ECFL does provide a momentum-independent Dyson self
  energy in this limit.}

\bibitem[{\citenamefont{{Georges} and {Kotliar}}(1992)}]{georges_kotliar_dmft}
\bibinfo{author}{\bibfnamefont{A.}~\bibnamefont{{Georges}}} \bibnamefont{and}
  \bibinfo{author}{\bibfnamefont{G.}~\bibnamefont{{Kotliar}}},
  \bibinfo{journal}{Phys. Rev. B} \textbf{\bibinfo{volume}{45}},
  \bibinfo{pages}{6479} (\bibinfo{year}{1992}).

\bibitem[{min()}]{minimal}
\bibinfo{note}{The minimal version of ECFL is obtained by setting to zero all
  correlation functions satisfying the Gutzwiller constraint- whereas the
  symmetrized version adds in some terms that are known to be zero, in order to
  make the effective Hamiltonian manifestly Hermitean \cite[p 11]{Monster}}.

\bibitem[{Dys()}]{Dyson-Mori}
\bibinfo{note}{To avoid the linear growth with $\omega$ of the self energy, one
  may define the Dyson Mori self energy, as in \cite{Anatomy} with a factor
  $\swo$ in the numerator of the Greens function. However in the present work,
  since DMFT directly yields the Dyson self energy, we simplify by quoting the
  results for this object directly.}

\bibitem[{\citenamefont{Hansen and Shastry}(2013)}]{Hansen-Shastry}
\bibinfo{author}{\bibfnamefont{D.}~\bibnamefont{Hansen}} \bibnamefont{and}
  \bibinfo{author}{\bibfnamefont{B.~S.} \bibnamefont{Shastry}},
  \bibinfo{journal}{Phys. Rev. B} \textbf{\bibinfo{volume}{87}},
  \bibinfo{pages}{245101} (\bibinfo{year}{2013}),
  \urlprefix\url{http://link.aps.org/doi/10.1103/PhysRevB.87.245101}.

\bibitem[{com()}]{comment-alt}
\bibinfo{note}{A more complete discussion of the approach to the insulating
  limit within ECFL is not yet available. It is possible to also achieve a
  vanishing quasiparticle weight near half-filling in other ways as well, such
  as letting $c_\chi$ diverge as $\frac{1}{\delta}$. We prefer the current
  scheme for its simplicity and economy.}

\bibitem[{\citenamefont{Metzner and Vollhardt}(1989)}]{metzner_vollhardt}
\bibinfo{author}{\bibfnamefont{W.}~\bibnamefont{Metzner}} \bibnamefont{and}
  \bibinfo{author}{\bibfnamefont{D.}~\bibnamefont{Vollhardt}},
  \bibinfo{journal}{Phys. Rev. Lett.} \textbf{\bibinfo{volume}{62}},
  \bibinfo{pages}{324} (\bibinfo{year}{1989}).

\bibitem[{\citenamefont{Wilson}(1975)}]{wilson1975}
\bibinfo{author}{\bibfnamefont{K.~G.} \bibnamefont{Wilson}},
  \bibinfo{journal}{Rev. Mod. Phys.} \textbf{\bibinfo{volume}{47}},
  \bibinfo{pages}{773} (\bibinfo{year}{1975}).

\bibitem[{\citenamefont{Krishna-murthy
  et~al.}(1980)\citenamefont{Krishna-murthy, Wilkins, and
  Wilson}}]{krishna1980a}
\bibinfo{author}{\bibfnamefont{H.~R.} \bibnamefont{Krishna-murthy}},
  \bibinfo{author}{\bibfnamefont{J.~W.} \bibnamefont{Wilkins}},
  \bibnamefont{and} \bibinfo{author}{\bibfnamefont{K.~G.}
  \bibnamefont{Wilson}}, \bibinfo{journal}{Phys. Rev. B}
  \textbf{\bibinfo{volume}{21}}, \bibinfo{pages}{1003} (\bibinfo{year}{1980}).

\bibitem[{\citenamefont{Bulla}(1999)}]{bulla1999}
\bibinfo{author}{\bibfnamefont{R.}~\bibnamefont{Bulla}},
  \bibinfo{journal}{Phys. Rev. Lett.} \textbf{\bibinfo{volume}{83}},
  \bibinfo{pages}{136} (\bibinfo{year}{1999}).

\bibitem[{\citenamefont{Bulla et~al.}(2008)\citenamefont{Bulla, Costi, and
  Pruschke}}]{bulla2008}
\bibinfo{author}{\bibfnamefont{R.}~\bibnamefont{Bulla}},
  \bibinfo{author}{\bibfnamefont{T.}~\bibnamefont{Costi}}, \bibnamefont{and}
  \bibinfo{author}{\bibfnamefont{T.}~\bibnamefont{Pruschke}},
  \bibinfo{journal}{Rev. Mod. Phys.} \textbf{\bibinfo{volume}{80}},
  \bibinfo{pages}{395} (\bibinfo{year}{2008}).

\bibitem[{\citenamefont{\v{Z}itko and Pruschke}(2009)}]{resolution}
\bibinfo{author}{\bibfnamefont{R.}~\bibnamefont{\v{Z}itko}} \bibnamefont{and}
  \bibinfo{author}{\bibfnamefont{T.}~\bibnamefont{Pruschke}},
  \bibinfo{journal}{Phys. Rev. B} \textbf{\bibinfo{volume}{79}},
  \bibinfo{pages}{085106} (\bibinfo{year}{2009}).

\bibitem[{\citenamefont{\v{Z}itko}(2011)}]{errors}
\bibinfo{author}{\bibfnamefont{R.}~\bibnamefont{\v{Z}itko}},
  \bibinfo{journal}{Phys. Rev. B} \textbf{\bibinfo{volume}{84}},
  \bibinfo{pages}{085142} (\bibinfo{year}{2011}).

\bibitem[{\citenamefont{Weichselbaum and von Delft}(2007)}]{weichselbaum2007}
\bibinfo{author}{\bibfnamefont{A.}~\bibnamefont{Weichselbaum}}
  \bibnamefont{and} \bibinfo{author}{\bibfnamefont{J.}~\bibnamefont{von
  Delft}}, \bibinfo{journal}{Phys. Rev. Lett.} \textbf{\bibinfo{volume}{99}},
  \bibinfo{pages}{076402} (\bibinfo{year}{2007}).

\bibitem[{\citenamefont{Bulla et~al.}(1998)\citenamefont{Bulla, Hewson, and
  Pruschke}}]{bulla1998}
\bibinfo{author}{\bibfnamefont{R.}~\bibnamefont{Bulla}},
  \bibinfo{author}{\bibfnamefont{A.~C.} \bibnamefont{Hewson}},
  \bibnamefont{and} \bibinfo{author}{\bibfnamefont{T.}~\bibnamefont{Pruschke}},
  \bibinfo{journal}{J. Phys.: Condens. Matter} \textbf{\bibinfo{volume}{10}},
  \bibinfo{pages}{8365} (\bibinfo{year}{1998}).

\bibitem[{\citenamefont{Peters et~al.}(2006)\citenamefont{Peters, Pruschke, and
  Anders}}]{peters2006}
\bibinfo{author}{\bibfnamefont{R.}~\bibnamefont{Peters}},
  \bibinfo{author}{\bibfnamefont{T.}~\bibnamefont{Pruschke}}, \bibnamefont{and}
  \bibinfo{author}{\bibfnamefont{F.~B.} \bibnamefont{Anders}},
  \bibinfo{journal}{Phys. Rev. B} \textbf{\bibinfo{volume}{74}},
  \bibinfo{pages}{245114} (\bibinfo{year}{2006}).

\bibitem[{\citenamefont{\v{Z}itko}(2009)}]{broyden}
\bibinfo{author}{\bibfnamefont{R.}~\bibnamefont{\v{Z}itko}},
  \bibinfo{journal}{Phys. Rev. B} \textbf{\bibinfo{volume}{80}},
  \bibinfo{pages}{125125} (\bibinfo{year}{2009}).

\bibitem[{\citenamefont{Brinkman and Rice}(1970)}]{brinkman_rice_prl_1970}
\bibinfo{author}{\bibfnamefont{W.~F.} \bibnamefont{Brinkman}} \bibnamefont{and}
  \bibinfo{author}{\bibfnamefont{T.~M.} \bibnamefont{Rice}},
  \bibinfo{journal}{Phys. Rev. B} \textbf{\bibinfo{volume}{2}},
  \bibinfo{pages}{4302} (\bibinfo{year}{1970}),
  \urlprefix\url{http://link.aps.org/doi/10.1103/PhysRevB.2.4302}.

\bibitem[{\citenamefont{Vollhardt}(1984)}]{vollhardt_gutzwiller_helium3_rmp_1984}
\bibinfo{author}{\bibfnamefont{D.}~\bibnamefont{Vollhardt}},
  \bibinfo{journal}{Rev. Mod. Phys.} \textbf{\bibinfo{volume}{56}},
  \bibinfo{pages}{99} (\bibinfo{year}{1984}),
  \urlprefix\url{http://link.aps.org/doi/10.1103/RevModPhys.56.99}.

\bibitem[{\citenamefont{Nozi\`eres}(1986)}]{nozieres_cdf_1986}
\bibinfo{author}{\bibfnamefont{P.}~\bibnamefont{Nozi\`eres}},
  \emph{\bibinfo{title}{La m\'ethode de Gutzwiller, Lecture Notes at Coll\`ege
  de France}} (\bibinfo{year}{1986}), \bibinfo{note}{unpublished}.

\bibitem[{\citenamefont{Rozenberg et~al.}(1994)\citenamefont{Rozenberg,
  Kotliar, and Zhang}}]{rozenberg_prb_1994}
\bibinfo{author}{\bibfnamefont{M.~J.} \bibnamefont{Rozenberg}},
  \bibinfo{author}{\bibfnamefont{G.}~\bibnamefont{Kotliar}}, \bibnamefont{and}
  \bibinfo{author}{\bibfnamefont{X.~Y.} \bibnamefont{Zhang}},
  \bibinfo{journal}{Phys. Rev. B} \textbf{\bibinfo{volume}{49}},
  \bibinfo{pages}{10181} (\bibinfo{year}{1994}),
  \urlprefix\url{http://link.aps.org/doi/10.1103/PhysRevB.49.10181}.

\bibitem[{\citenamefont{{Sakai} et~al.}(2012)\citenamefont{{Sakai}, {Blanc},
  {Civelli}, {Gallais}, {Cazayous}, {Measson}, {Wen}, {Xu}, {Gu}, {Sangiovanni}
  et~al.}}]{sakai_darkside}
\bibinfo{author}{\bibfnamefont{S.}~\bibnamefont{{Sakai}}},
  \bibinfo{author}{\bibfnamefont{S.}~\bibnamefont{{Blanc}}},
  \bibinfo{author}{\bibfnamefont{M.}~\bibnamefont{{Civelli}}},
  \bibinfo{author}{\bibfnamefont{Y.}~\bibnamefont{{Gallais}}},
  \bibinfo{author}{\bibfnamefont{M.}~\bibnamefont{{Cazayous}}},
  \bibinfo{author}{\bibfnamefont{M.-A.} \bibnamefont{{Measson}}},
  \bibinfo{author}{\bibfnamefont{J.~S.} \bibnamefont{{Wen}}},
  \bibinfo{author}{\bibfnamefont{Z.~J.} \bibnamefont{{Xu}}},
  \bibinfo{author}{\bibfnamefont{G.~D.} \bibnamefont{{Gu}}},
  \bibinfo{author}{\bibfnamefont{G.}~\bibnamefont{{Sangiovanni}}},
  \bibnamefont{et~al.}, \bibinfo{journal}{ArXiv e-prints}
  (\bibinfo{year}{2012}), \eprint{1207.5070}.

\bibitem[{\citenamefont{Moeller et~al.}(1995)\citenamefont{Moeller, Si,
  Kotliar, Rozenberg, and Fisher}}]{moeller_prl_1995}
\bibinfo{author}{\bibfnamefont{G.}~\bibnamefont{Moeller}},
  \bibinfo{author}{\bibfnamefont{Q.}~\bibnamefont{Si}},
  \bibinfo{author}{\bibfnamefont{G.}~\bibnamefont{Kotliar}},
  \bibinfo{author}{\bibfnamefont{M.}~\bibnamefont{Rozenberg}},
  \bibnamefont{and} \bibinfo{author}{\bibfnamefont{D.~S.}
  \bibnamefont{Fisher}}, \bibinfo{journal}{Phys. Rev. Lett.}
  \textbf{\bibinfo{volume}{74}}, \bibinfo{pages}{2082} (\bibinfo{year}{1995}),
  \urlprefix\url{http://link.aps.org/doi/10.1103/PhysRevLett.74.2082}.

\bibitem[{\citenamefont{Bulla et~al.}(1999)\citenamefont{Bulla, Pruschke, and
  Hewson}}]{bulla_scaling_1999}
\bibinfo{author}{\bibfnamefont{R.}~\bibnamefont{Bulla}},
  \bibinfo{author}{\bibfnamefont{T.}~\bibnamefont{Pruschke}}, \bibnamefont{and}
  \bibinfo{author}{\bibfnamefont{A.~C.} \bibnamefont{Hewson}},
  \bibinfo{journal}{Physica B Condensed Matter} \textbf{\bibinfo{volume}{259}},
  \bibinfo{pages}{721} (\bibinfo{year}{1999}).

\bibitem[{\citenamefont{Karski et~al.}(2005)\citenamefont{Karski, Raas, and
  Uhrig}}]{karski2005}
\bibinfo{author}{\bibfnamefont{M.}~\bibnamefont{Karski}},
  \bibinfo{author}{\bibfnamefont{C.}~\bibnamefont{Raas}}, \bibnamefont{and}
  \bibinfo{author}{\bibfnamefont{G.~S.} \bibnamefont{Uhrig}},
  \bibinfo{journal}{Phys. Rev. B} \textbf{\bibinfo{volume}{72}},
  \bibinfo{pages}{113110} (\bibinfo{year}{2005}).

\bibitem[{\citenamefont{Karski et~al.}(2008)\citenamefont{Karski, Raas, and
  Uhrig}}]{karski2008}
\bibinfo{author}{\bibfnamefont{M.}~\bibnamefont{Karski}},
  \bibinfo{author}{\bibfnamefont{C.}~\bibnamefont{Raas}}, \bibnamefont{and}
  \bibinfo{author}{\bibfnamefont{G.~S.} \bibnamefont{Uhrig}},
  \bibinfo{journal}{Phys. Rev. B} \textbf{\bibinfo{volume}{77}},
  \bibinfo{pages}{075116} (\bibinfo{year}{2008}).

\bibitem[{\citenamefont{\v{Z}iga Osolin and \v{Z}itko}(2013)}]{pade}
\bibinfo{author}{\bibnamefont{\v{Z}iga Osolin}} \bibnamefont{and}
  \bibinfo{author}{\bibfnamefont{R.}~\bibnamefont{\v{Z}itko}},
  \bibinfo{journal}{Phys. Rev. B} \textbf{\bibinfo{volume}{87}},
  \bibinfo{pages}{245135} (\bibinfo{year}{2013}).

\bibitem[{\citenamefont{Zemlji\ifmmode~\check{c}\else \v{c}\fi{}
  et~al.}(2008)\citenamefont{Zemlji\ifmmode~\check{c}\else \v{c}\fi{},
  Prelov\ifmmode~\check{s}\else \v{s}\fi{}ek, and Tohyama}}]{zemljic_07}
\bibinfo{author}{\bibfnamefont{M.~M.}
  \bibnamefont{Zemlji\ifmmode~\check{c}\else \v{c}\fi{}}},
  \bibinfo{author}{\bibfnamefont{P.}~\bibnamefont{Prelov\ifmmode~\check{s}\else
  \v{s}\fi{}ek}}, \bibnamefont{and}
  \bibinfo{author}{\bibfnamefont{T.}~\bibnamefont{Tohyama}},
  \bibinfo{journal}{Phys. Rev. Lett.} \textbf{\bibinfo{volume}{100}},
  \bibinfo{pages}{036402} (\bibinfo{year}{2008}),
  \urlprefix\url{http://link.aps.org/doi/10.1103/PhysRevLett.100.036402}.

\bibitem[{\citenamefont{Byczuk et~al.}(2007)\citenamefont{Byczuk, Kollar, Held,
  Yang, Nekrasov, Pruschke, and Vollhardt}}]{byczuk_kinks_2007}
\bibinfo{author}{\bibfnamefont{K.}~\bibnamefont{Byczuk}},
  \bibinfo{author}{\bibfnamefont{M.}~\bibnamefont{Kollar}},
  \bibinfo{author}{\bibfnamefont{K.}~\bibnamefont{Held}},
  \bibinfo{author}{\bibfnamefont{Y.-F.} \bibnamefont{Yang}},
  \bibinfo{author}{\bibfnamefont{I.~A.} \bibnamefont{Nekrasov}},
  \bibinfo{author}{\bibfnamefont{T.}~\bibnamefont{Pruschke}}, \bibnamefont{and}
  \bibinfo{author}{\bibfnamefont{D.}~\bibnamefont{Vollhardt}},
  \bibinfo{journal}{Nat Phys} \textbf{\bibinfo{volume}{3}},
  \bibinfo{pages}{168} (\bibinfo{year}{2007}), ISSN \bibinfo{issn}{1745-2473},
  \urlprefix\url{http://dx.doi.org/10.1038/nphys538}.

\bibitem[{\citenamefont{Grete et~al.}(2011)\citenamefont{Grete, Schmitt, Raas,
  Anders, and Uhrig}}]{grete_prb_2011}
\bibinfo{author}{\bibfnamefont{P.}~\bibnamefont{Grete}},
  \bibinfo{author}{\bibfnamefont{S.}~\bibnamefont{Schmitt}},
  \bibinfo{author}{\bibfnamefont{C.}~\bibnamefont{Raas}},
  \bibinfo{author}{\bibfnamefont{F.~B.} \bibnamefont{Anders}},
  \bibnamefont{and} \bibinfo{author}{\bibfnamefont{G.~S.} \bibnamefont{Uhrig}},
  \bibinfo{journal}{Phys. Rev. B} \textbf{\bibinfo{volume}{84}},
  \bibinfo{pages}{205104} (\bibinfo{year}{2011}),
  \urlprefix\url{http://link.aps.org/doi/10.1103/PhysRevB.84.205104}.

\bibitem[{\citenamefont{Held et~al.}(2013)\citenamefont{Held, Peters, and
  Toschi}}]{held_2013}
\bibinfo{author}{\bibfnamefont{K.}~\bibnamefont{Held}},
  \bibinfo{author}{\bibfnamefont{R.}~\bibnamefont{Peters}}, \bibnamefont{and}
  \bibinfo{author}{\bibfnamefont{A.}~\bibnamefont{Toschi}},
  \bibinfo{journal}{Phys. Rev. Lett.} \textbf{\bibinfo{volume}{110}},
  \bibinfo{pages}{246402} (\bibinfo{year}{2013}).

\bibitem[{\citenamefont{Eckstein et~al.}(2007)\citenamefont{Eckstein, Kollar,
  Potthoff, and Vollhardt}}]{Eckstein:2007bi}
\bibinfo{author}{\bibfnamefont{M.}~\bibnamefont{Eckstein}},
  \bibinfo{author}{\bibfnamefont{M.}~\bibnamefont{Kollar}},
  \bibinfo{author}{\bibfnamefont{M.}~\bibnamefont{Potthoff}}, \bibnamefont{and}
  \bibinfo{author}{\bibfnamefont{D.}~\bibnamefont{Vollhardt}},
  \bibinfo{journal}{Physical Review B} \textbf{\bibinfo{volume}{75}},
  \bibinfo{pages}{125103} (\bibinfo{year}{2007}).

\bibitem[{AIM()}]{AIM}
\bibinfo{note}{B. S. Shastry, E. Perepelitsky and A. C. Hewson,
  arXiv:1307.3492}.

\bibitem[{\citenamefont{Shastry}(2012{\natexlab{b}})}]{Asymmetry}
\bibinfo{author}{\bibfnamefont{B.~S.} \bibnamefont{Shastry}},
  \bibinfo{journal}{Phys. Rev. Lett.} \textbf{\bibinfo{volume}{109}},
  \bibinfo{pages}{067004} (\bibinfo{year}{2012}{\natexlab{b}}),
  \urlprefix\url{http://link.aps.org/doi/10.1103/PhysRevLett.109.067004}.

\bibitem[{\citenamefont{Haule and Kotliar}(2009)}]{Haule_proc09}
\bibinfo{author}{\bibfnamefont{K.}~\bibnamefont{Haule}} \bibnamefont{and}
  \bibinfo{author}{\bibfnamefont{G.}~\bibnamefont{Kotliar}}, in
  \emph{\bibinfo{booktitle}{Properties and Applications of Thermoelectric
  Materials}}, edited by
  \bibinfo{editor}{\bibfnamefont{V.}~\bibnamefont{Zlatic}} \bibnamefont{and}
  \bibinfo{editor}{\bibfnamefont{A.~C.} \bibnamefont{Hewson}}
  (\bibinfo{publisher}{Springer Netherlands}, \bibinfo{year}{2009}), NATO
  Science for Peace and Security Series B: Physics and Biophysics, pp.
  \bibinfo{pages}{119--131}.

\bibitem[{\citenamefont{Singh and Glenister}(1992)}]{Singh}
\bibinfo{author}{\bibfnamefont{R.~R.~P.} \bibnamefont{Singh}} \bibnamefont{and}
  \bibinfo{author}{\bibfnamefont{R.~L.} \bibnamefont{Glenister}},
  \bibinfo{journal}{Phys. Rev. B} \textbf{\bibinfo{volume}{46}},
  \bibinfo{pages}{14313} (\bibinfo{year}{1992}),
  \urlprefix\url{http://link.aps.org/doi/10.1103/PhysRevB.46.14313}.

\bibitem[{\citenamefont{Rice and Ueda}(1985)}]{Rice-Ueda}
\bibinfo{author}{\bibfnamefont{T.~M.} \bibnamefont{Rice}} \bibnamefont{and}
  \bibinfo{author}{\bibfnamefont{K.}~\bibnamefont{Ueda}},
  \bibinfo{journal}{Phys. Rev. Lett.} \textbf{\bibinfo{volume}{55}},
  \bibinfo{pages}{995} (\bibinfo{year}{1985}).

\bibitem[{\citenamefont{Yoshida et~al.}(1990)\citenamefont{Yoshida, Whitaker,
  and Oliveira}}]{yoshida1990}
\bibinfo{author}{\bibfnamefont{M.}~\bibnamefont{Yoshida}},
  \bibinfo{author}{\bibfnamefont{M.~A.} \bibnamefont{Whitaker}},
  \bibnamefont{and} \bibinfo{author}{\bibfnamefont{L.~N.}
  \bibnamefont{Oliveira}}, \bibinfo{journal}{Phys. Rev. B}
  \textbf{\bibinfo{volume}{41}}, \bibinfo{pages}{9403} (\bibinfo{year}{1990}).

\bibitem[{\citenamefont{Campo and Oliveira}(2005)}]{campo2005}
\bibinfo{author}{\bibfnamefont{V.~L.} \bibnamefont{Campo}} \bibnamefont{and}
  \bibinfo{author}{\bibfnamefont{L.~N.} \bibnamefont{Oliveira}},
  \bibinfo{journal}{Phys. Rev. B} \textbf{\bibinfo{volume}{72}},
  \bibinfo{pages}{104432} (\bibinfo{year}{2005}).

\end{thebibliography}

\end{document}